\documentclass{pasa}%
\title[Australian Aboriginal Astronomy and Navigation]{Dawes Review 5: Australian Aboriginal Astronomy and Navigation}
\author[Ray P. Norris]{Ray P. Norris$^{1,2}$\\
\affil{$^1$  Western Sydney University, Locked Bag 1797, Penrith South, NSW 1797, Australia}
\affil{$^2$ CSIRO Astronomy \& Space Science, PO Box 76, Epping, NSW 1710, Australia}}
\jid{PASA}
\doi{10.1017/pas.\the\year.xxx}
\jyear{\the\year}

\usepackage[authoryear]{natbib}
\bibpunct{(}{)}{;}{a}{}{,}
\usepackage{aas_macros}
\usepackage{hyperref} 
\hypersetup{colorlinks,citecolor=blue,linkcolor=blue,urlcolor=blue}

\usepackage{times,epsfig}
\usepackage{graphics,graphicx}
\usepackage{amsmath}
\usepackage{amsfonts}
\usepackage{psfig}

\usepackage{mathrsfs}
\usepackage{amssymb}
\usepackage{bm}
\usepackage{graphicx}

\newcommand{\kms}{\mbox{km\,s$^{-1}$}}
\def\kms {\ifmmode{{\rm ~km~s}^{-1}}\else{~km~s$^{-1}$}\fi}

\def\lsun {\ifmmode{{\rm ~L}_\odot}\else{~L$_\odot$}\fi}

\def\deg {^{\circ} }

\newbox\grsign \setbox\grsign=\hbox{$>$} \newdimen\grdimen \grdimen=\ht\grsign
\newbox\simlessbox \newbox\simgreatbox
\setbox\simgreatbox=\hbox{\raise.5ex\hbox{$>$}\llap
 {\lower.5ex\hbox{$\sim$}}}\ht1=\grdimen\dp1=0pt
\setbox\simlessbox=\hbox{\raise.5ex\hbox{$<$}\llap
 {\lower.5ex\hbox{$\sim$}}}\ht2=\grdimen\dp2=0pt

\def\lsim{\mathrel{\rlap{\lower4pt\hbox{\hskip1pt$\sim$}}
    \raise1pt\hbox{$<$}}}                
\def\gsim{\mathrel{\rlap{\lower4pt\hbox{\hskip1pt$\sim$}}
    \raise1pt\hbox{$>$}}}                

\def\apjl {{\it Ap.~J.\ (Letters)}}
\def\apjs {{\it Ap.~J.\ Suppl.}}

\def\aap {{\it Astr.~Ap.}}

\def\baas {{\it Bull.~AAS}}
\def\jahh {{\it J. Astr. History \& Heritage}}

\def\nat {{\it Nature}}

\def\apjl{ApJ}                
\def\apjs{ApJS}               
\def\ao{Appl.~Opt.}           
\def\aap{A\&A}                
\def\baas{BAAS}               
\def\jrasc{JRASC}             

\def\nat{Nature}              

\def\grtsim{\mathrel{\hbox{\rlap{\hbox{\lower2pt\hbox{$\sim$}}}\raise2pt\hbox{$>$}}}}
\def\lesssim{\mathrel{\hbox{\rlap{\hbox{\lower2pt\hbox{$\sim$}}}\raise2pt\hbox{$<$}}}}

\def\lsim{\,\lower2truept\hbox{${<\atop\hbox{\raise4truept\hbox{$\sim$}}}$}\,}
\def\gsim{\,\lower2truept\hbox{${> \atop\hbox{\raise4truept\hbox{$\sim$}}}$}\,}
\def\simlt{\mathrel{\rlap{\lower 3pt\hbox{$\sim$}}
        \raise 2.0pt\hbox{$<$}}}
\def\simgt{\mathrel{\rlap{\lower 3pt\hbox{$\sim$}}
        \raise 2.0pt\hbox{$>$}}}

\begin{document}
\begin{abstract}
The traditional cultures of Aboriginal Australians include a significant astronomical component, perpetuated through oral tradition, ceremony, and art. This astronomical knowledge includes a deep understanding of the motion of objects in the sky, which was used for practical purposes such as constructing calendars and for navigation. There is also evidence that traditional Aboriginal Australians made careful records and measurements of cyclical phenomena, recorded unexpected phenomena such as eclipses and meteorite impacts, and could determine the cardinal points to an accuracy of a few degrees. Putative explanations of celestial phenomena appear throughout the oral record, suggesting traditional Aboriginal Australians sought to understand the natural world around them, in the same way as modern scientists, but within their own cultural context. There is also a growing body of evidence for sophisticated navigational skills, including the use of astronomically based songlines. Songlines are effectively oral maps of the landscape, and are an efficient way of transmitting oral navigational skills in cultures that do not have a written language. The study of Aboriginal astronomy has had an impact extending beyond mere academic curiosity, facilitating cross-cultural understanding, demonstrating the intimate links between science and culture, and helping students to engage with science.
\end{abstract}
\maketitle%

\section*{Preface }
The Dawes Reviews are substantial reviews of topical areas in astronomy, published by authors of international standing at the invitation of the PASA Editorial Board. The reviews recognise William Dawes (1762--1836), second lieutenant in the Royal Marines and the astronomer on the First Fleet. Dawes was not only an accomplished astronomer, but spoke five languages, had a keen interest in botany, mineralogy, engineering, cartography and music, compiled the first Aboriginal-English dictionary, and was an outspoken opponent of slavery.

\section*{Warning to Aboriginal Readers }
This paper may contain names and images of people who are deceased.

\section{Introduction }
\label{intro}
\subsection{The European History of Aboriginal Astronomy}
Lt. William Dawes arrived in Sydney Cove with the First Fleet  on 26 January 1788. He took greater interest in the local Darug people  than  any other officer from the First Fleet, and  became an authority on their language and culture. His notebooks  \citep{dawes} are the earliest detailed description of Aboriginal culture. He became close friends, and perhaps partners, with a Darug girl (Patyegarang) and learnt some of her language. Unfortunately he clashed with Governor Phillip, and was therefore unable to stay in the colony as he wished at the end of his three-year term.

Dawes was also an astronomer, and naturally took an interest in Darug astronomy. His notebooks contain many Darug words for stars and planets but, curiously, contain no mention of Darug knowledge of the sky.  It seems inconceivable that Dawes, a man curious about both astronomy and  Darug culture, didn't ask questions about the Darug understanding of the motion of celestial bodies, and yet he left us no record of the answers.  His notebooks were lost for many years, but were recently found and are now available on-line  \citep{dawes}, but are thought to be incomplete.
It is still possible that one day his missing notebooks will be found, and will contain an account of Darug astronomy.

His fellow-officer  Lieutenant-Colonel David Collins wrote of the Aboriginal inhabitants of Sydney:  `Their acquaintance with astronomy is limited to the names of the sun and moon, some few stars, the Magellanic clouds, and the Milky Way. Of the circular form of the earth they have not the smallest idea, but imagine that the sun returns over their heads during the night to the quarter whence he begins his course in the morning.'  \citep{collins98}. Even this fragment contains useful information, telling us that the Eora people believed that the Sun returned to the East {\it over} their heads rather than under the ground, as believed by most other groups. Another early account of Aboriginal astronomy was from the escaped convict William Buckley, who lived with the Wathaurung people from 1803 to 1835 \citep{morgan52}.

The first published substantive accounts of  Australian Aboriginal astronomy were by  \cite{stanbridge57, stanbridge61}, who wrote about the sky knowledge of the Boorong people   \citep[see also][]{hamacherfrew10}. 
This was followed by  accounts  \citep[e.g.][]{bates44, maegraith32, mathews94, parker05} of the sky knowledge of other groups, and also derivative and interpretational works  \citep{griffin23, kotz11, macpherson81}.  Apart from incidental mentions of sky knowledge by other writers  \citep[e.g.][]{bates44, berndt48, mathews94,tindale25}, there was then no substantial discussion of Aboriginal astronomy until two monumental works by Mountford.  One of these volumes  \citep{mountford56} was based on the 1948 Arnhem Land expedition, that he led, and established Aboriginal Astronomy as a serious subject. Unfortunately, the second volume  \citep{mountford76} included accounts of sacred material that should not have been published. Following a court case, this volume was partially withdrawn from sale, and the resulting distrust of researchers by Aboriginal elders has significantly impeded subsequent work in this field. 

The studies by the eccentric Irishwoman Daisy Bates are potentially very important, because she wrote extensively on Aboriginal sky-beliefs, based on her immersion in an Aboriginal community  \citep{bates25, bates44}, but her writings are not often cited because errors of fact and judgement  \citep{devries10} mar her writings, marking her as an unreliable witness. Nevertheless, there have been valuable attempts to deconvolve her errors from her writings  \citep[e.g.][]{fredrick08, leaman}.

Following Mountford's work, although there were several popular accounts  \citep[e.g.][]{isaacs80, wells64},  there were no significant research studies until   \cite{cairns88} suggested that Sydney rock art contained references to Aboriginal astronomy,  \cite{clarke90} examined the Aboriginal astronomy of the people around Adelaide, and  \cite{haynes90, haynes92} wrote extensively on the subject, suggesting that Australian Aboriginal people may be `the worldÕs first astronomers'. Subsequently,  \citet{johnson98} wrote an authoritative academic monograph of the astronomy of several language groups. 
However, most of these works were based on the interpretation of earlier academic texts, and few presented any new data obtained from Aboriginal sources or from fieldwork.

Although evidence of astronomical knowledge has been found in many Aboriginal cultures, the best-documented example is undoubtedly that of the Wardaman people, largely because of Wardaman senior elder Bill Yidumduma Harney's enthusiasm to share his traditional knowledge with the wider world. In particular, a significant step forward was the publication of the book Dark Sparkers  \citep{DS}  which contains an extremely  detailed study of the  sky knowledge and astronomical lore of the Wardaman people, at a level of detail which is unprecedented in the literature. Unfortunately only a small fraction of this knowledge can be presented in this review.

\begin{figure}[h]
\begin{center}
\includegraphics[width=8cm]{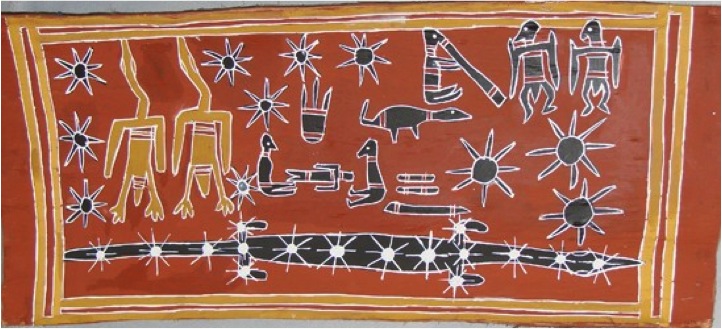} 
\caption{A bark painting by Yolngu artist Dhuwarriny Yunupingu. All elements of it refer to stories involving astronomical constellations. The object in the centre may depict a comet.The crocodile at the bottom is the constellation Scorpius, and some other elements in a very similar painting are discussed by  \citet[][Plate188]{grogerwurm73}. This particular painting is modern ($\sim$2000 AD) but follows a traditional design. }
\label{art}
\end{center}
\end{figure}

By 2005, these detailed published accounts demonstrated that there was a significant body of Aboriginal Astronomical knowledge, and that the art (e.g. Fig. \ref{art}), ceremonies, and oral traditions of many traditional Aboriginal cultures contain references to celestial bodies such as the Sun, Moon, and stars.
However, very few peer-reviewed articles had been written. As a result, most archaeologists and anthropologists still regarded the area as a `fringe' area, and it received little attention in mainstream archaeology or anthropology.

\begin{figure}[h]
\begin{center}
\includegraphics[width=8cm]{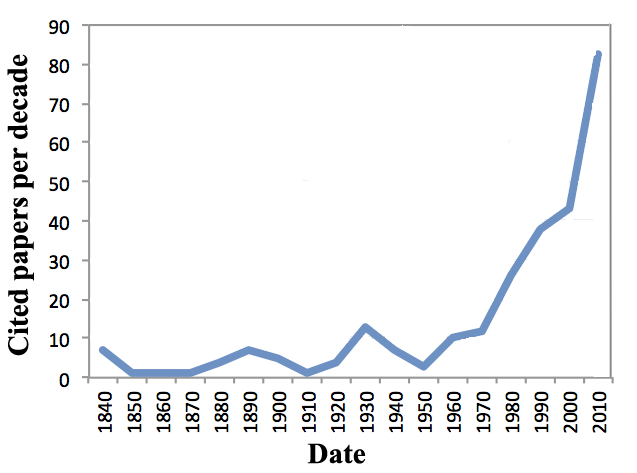} 
\caption{The growth of Aboriginal astronomy literature. The plot shows the number of papers per decade cited in this review. The number for the current decade has been multiplied by 1.4 to correct for the incomplete decade.}
\label{papers}
\end{center}
\end{figure}

From 2005 onwards, there was a rapid growth of articles on Aboriginal astronomy (see Fig. \ref{papers}), reflecting a rapid increase not only in interest in the subject but in rigorous scholarly research.The International Year of Astronomy in 2009 saw an explosion of published literature on Aboriginal Astronomy  \cite[e.g.][]{ED, n230}, including new information from fieldwork as well as the mining of existing texts.  This increase in scholarly work was accompanied by a rapid growth of Aboriginal Astronomy outreach activities. These included an ABC TV `Message Stick' program devoted to Aboriginal Astronomy, public displays  \citep[e.g.][]{goldsmith11}, the `First Astronomers' stage show which started at the Darwin festival and then toured several Arts Festivals  \citep{harney09b}, and, during 2009 alone, over 100 public talks and media interviews on Aboriginal Astronomy  \citep{n254}. A highlight was a one-day symposium on Aboriginal Astronomy at the Australian Institute of Aboriginal and Torres Strait Islanders Studies (AIATSIS) on 27 November 2009. This well-attended symposium at Australia's peak body on Aboriginal studies marked the beginning of acceptance of Aboriginal Astronomy as a serious research area by mainstream academia. It also marked the launch of the `Ilgajiri' art exhibition of Indigenous paintings of sky-knowledge, which later blossomed into the `Shared Skies' exhibition, funded by the Square Kilometre Array organisation, and which has toured Australia, Europe, and South Africa, featuring Australian and South African paintings of traditional sky knowledge  \citep{goldsmith14a}.

Three PhD degrees have been awarded in this field \citep{kotz11, hamacher11a, goldsmith14b}, and three Masters degrees \citep{morieson96, fredrick08, fuller14d}, and there are now several graduate students enrolled at the Universities of New South Wales (UNSW) and Western Sydney (WSU).
Nevertheless, the subject field is currently small enough that this review can aim  to summarise {\bf every} peer-reviewed paper on Aboriginal Astronomy, and also all non-peer-reviewed publications containing significant original research. 

Most early publications on Aboriginal Astronomy were primarily concerned  with ceremonial aspects, although it was well-established that the stars were used for calendars and timing of harvesting food sources.  \cite{haynes92} first suggested that the Australian Aboriginal people may be the world's first astronomers. But `astronomer' implies more than just telling stories about the sky - it implies a quest to understand the phenomena in the sky, develop a self-consistent model of how the world works, and apply this knowledge to practical applications such as calendars and navigation. From 2005 onwards, there was a growing trend to test this hypothesis  \citep[e.g.][]{n214, n230}, and an increasing focus on ethno-science - the idea that traditional Australians were trying to understand their world, in terms of their traditional culture, 
much as science does nowadays.

As a result of this explosion in the field, it is now clear that there is far more depth to the subject than realised by early researchers such as Mountford or Elkin  \citep{n324}. Pre-contact Aboriginal Australians not only knew songs and stories about the sky, but they had a  deep understanding of the positions and motions of celestial bodies, and used that knowledge to construct calendars, and to navigate. Evidence of this knowledge may be found in their songs and stories  \citep{DS}, their art  \citep{n255}, and their navigational skills  \citep{n315}. Perhaps just as importantly, that knowledge was integrated into a world-view which incorporated models of how the world works  \citep{n324}.

The breadth of knowledge has also increased. We now have detailed studies of the astronomical components of several cultures, most notably the  Wardaman  \citep[e.g.][]{DS}, but also Yolngu  \citep{ED}, 
Kamilaroi \& Euahlayi\footnote{The distinction between Kamilaroi and Euahlayi traditions is often blurred. To avoid incorrect attribution, I use `Kamilaroi' throughout this review to mean `either Kamilaroi or Euahlayi', with apologies to Euahlayi readers.} 
 \citep{n311},
and South Australian groups  \citep{clarke97a}. The field is rapidly broadening, with current studies looking at the astronomy of the Wiradjuri  \citep{leaman16a, leaman16b}, and Torres Strait Islanders  \citep{hamacher13c}. However, these still represent only a tiny fraction of the 300-odd Aboriginal language groups, and it is likely we have only scratched the surface. At the same time, the senior elders who possess much of this knowledge are ageing, and passing on, with relatively few new elders rising up to replace them. So recording this astronomical knowledge is something of a race against time. On the other hand, groups such as Kamilaroi  are rebuilding their culture, with younger people learning language and ceremony, and the study of Aboriginal astronomy can lay a significant role in that rebuilding process  \citep{n311}. 
It is hopefully not far-fetched to imagine a future Australia in which Aboriginal languages are taught as second languages in Australian high schools (in the same way that Welsh is taught in  schools in Wales) and that Aboriginal ethnoscience is part of the Australian science curriculum. Aboriginal Astronomy is starting to be a significant and effective component of education and outreach activities  \citep{bhathal08, bhathal11a,hollow08, pring02,  pring06a, pring06b, wyatt14}.  \citet{n324} has argued that Indigenous Astronomy can be used to teach students the basic motivation of science.

This revolution in knowledge of Aboriginal astronomy can also be viewed in the context of a rapidly changing view of other aspects of Indigenous culture. Contrary to widespread opinions by earlier researchers  \citep{n324}, we now know that  pre-contact Indigenous people practised  agriculture  \citep{gammage11,pascoe14}, and possessed sophisticated navigational skills  \citep{kerwin10}. 

In the last ten years the nature of the study of Aboriginal astronomy has also changed. Ten years ago researchers would approach elders and communities, clipboards in hand metaphorically if not literally, and ask for information about stories. The work of Fuller et al.  \citep{n311,n322,n318} is completely different, and is better characterised as a collaboration between the research group (Fuller et al.) and the Kamilaroi and Euahlayi groups (Michael Anderson et al). The project had  clear benefits to  both scholarly achievement and  to rebuilding the Kamilaroi and Euahlayi cultures, marked by a `giving back ceremony' at which teaching materials were given back to the community to help educate high-school students about Kamilaroi and Euahlayi culture. It also triggered a feature film about Euahlayi sky knowledge \citep{ellie}. It is hoped that future projects in this field will adopt this collaborative approach.

Another important approach is that of \cite{nakata14} who propose to develop software, archives, and web interfaces to allow Indigenous communities to share their astronomical knowledge with the world on their terms and in a culturally sensitive manner.

\subsection{The Aboriginal People}

The ancestors of Aboriginal Australians left central Africa around 100,000 BC, and passed through the Middle East about 70,000 BC.  DNA sequencing  \citep{rasmussen} shows that they were closely related to the proto-Europeans who emerged from Africa at around the same time.  
The proto-Australians followed the coast of India and China, crossing through Papua New Guinea, rapidly spread across the Pleistocene continent of Sahul (Australia, New Guinea, and Tasmania),  and arrived in Northern Australia, probably in a single wave   \citep{hudjashov}, 
at least 40,000 years ago  \citep{oconnell04}, probably before their counterparts reached Europe.  Radiocarbon dating of Mungo Man in the Willandra Lakes region of New South Wales (NSW), 
\footnote{Throughout this review I use the following abbreviations for the states and territories of Australia: NSW: New South Wales; Qld: Queensland; Vic: Victoria; SA: South Australia; WA: Western Australia; NT: Northern Territory; ACT: Australian Capital Territory; Tas: Tasmania} 
showed that they had reached NSW by 40,000 BC  \citep{bowler03}.


From 40,000 BC onwards they enjoyed a continuous, unbroken culture, with very little contact with outsiders, other than annual visits over the last few hundred years from Macassan trepang-collectors  \citep{orchiston16}. This isolation continued, with no cultural discontinuities,  until the arrival of the British in 1788, making Aboriginal Australians among the oldest continuous cultures in the world  \citep{mcniven05}. At the time of British invasion, Australia's population was about 300,000  \citep{jupp01}, divided into about 250 distinct Aboriginal language groups with nearly 750 dialects  \citep[e.g][]{walsh91}. Each had its own culture and  language, although many shared common threads, such as the belief that the world was created in the `Dreaming' \index{Dreaming} by an\-cestral spirits.  Some languages were closely related (as close as Italian and Spanish), while others were as different from each other as Italian is from Chinese. 

Many Aboriginal groups divide their world into two moieties: every person and every object has the characteristics of one or other of these moieties. For example, every Yolngu person is either {\it dua} or {\it yirritja}, and this is such a fundamental difference that it affects their language (with different verb endings for dua and yirritja), whom they associate with,  and whom they can  marry. Their sky is also divided into the two moieties.
 \cite{mountford56} notes that Groote Eylandt people divide the stars in the Milky Way into two moieties, while  the central desert people  \citep{mountford76} associate the summer constellations with one moiety, and the winter constellations with the other. Even within one community, the two moieties typically also have different stories, with further variation between different clans or groups. As a result, there are often several different versions of any particular story in a  community, with further (secret) versions reserved for the various levels of initiated men and women.

Most Aboriginal Australians were nomadic hunter/ gatherers, moving in an annual cycle of seasonal camps and hunting-places within the land that they owned, taking advantage of seasonal food sources. They practised careful land management to increase the food yield of their land  \citep{gammage11}, including `firestick--farming' in which the land was burnt in a patchwork pattern to increase food production and reduce the severity of bush-fires. Some groups built stone traps for fish farming, planted crops such as yams, or built stone dwellings  \citep{clark94, gammage11}.

Many Aboriginal cultures were severely damaged by the arrival of the British in 1788, and by the consequent disease, reduced access to food, and, in some cases, genocide. The total Aboriginal population decreased from over 300,000 in 1788 to about 93,000 in 1900. In south-east Australia, entire cultures were destroyed. However,  the language and culture of some groups in the north and centre of Australia are still essentially intact, and initiation ceremonies are still conducted at which  knowledge is passed from one generation to the next. It is from these groups, particularly the Yolngu and Wardaman people, that we have obtained our most detailed information. 

The people who came to Australia over 40,000 years ago were biologically identical to modern humans, and may well have had extensive knowledge of the sky. However, there is no dated record of any Aboriginal astronomy until the invasion of Australia by the British in 1788, at which time astronomical knowledge appears to have deeply embedded in many Aboriginal cultures, and was presumably already ancient. Certainly it was an important part of several Aboriginal cultures in 1788, and was `considered one of their principal branches of education. ... it is taught by men selected for their intelligence and information'  \citep{clarke09a,dawson81}.

 \cite{mountford76}
reported that some Aboriginal people knew every star as faint as fourth magnitude, and knew myths associated with most of those stars. Similarly, elders such as Harney can name most  stars in the sky visible to the naked eye, and understand intimately  how the whole pattern rotates over their heads from east to west during the night, and how it shifts over the course of a year  \citep{DS}. 
To name most of the $\sim$ 3000 stars visible to the naked eye from Northern Australia is a memory feat that rivals winners of the World Memory Championships  \citep[e.g][]{foer11} and which must have taken years of learning.

 \cite{maegraith32} says that `The most interesting fact about Aboriginal astronomy is that all the adult males of the tribe are fully conversant with all that is known, while no young man of the tribe knows much about the stars until after his initiation is complete ... The old men also instruct the initiated boys in the movements, colour and brightness of the stars.' 

Stars are a central part of Aboriginal sky-lore, and are often associated with creator spirits. 
Aboriginal men are also familiar with the changing position of the stars throughout the night and throughout the year, as described below in \S\ref{navigation}. Curiously, the stars regarded as the most important were often not the brightest, but instead importance seemed to depend on factors such as colour and the relationship to other stars  \citep[e.g][]{haynes90, maegraith32}.

The first definite evidence of astronomy in the world is probably either Stonehenge  \citep{pearson13} or the older but less-well-known Warren Field  \citep{gaffney13}, which was built around 8000 BC. Given the continuity of Indigenous Australian culture for at least 40,000 years with little contact with the outside world, it is plausible that Aboriginal astronomical knowledge predates these British sites. This is the basis of the statement that  `the Australian Aborigines  were arguably the world's first astronomers'  \citep{haynes96}, but we currently have no firm evidence to support or refute that hypothesis.

\subsection{Coverage and Limitations of this Review}

To the best of the author's knowledge, this review paper summarises and cites {\bf all} published peer-reviewed literature on Australian Aboriginal astronomy, together with significant non-peer reviewed material that presents original research and cites sources. This paper therefore marks a watershed, in that it is unlikely that a review paper written in the future could reasonably expect to include all peer-reviewed literature, because of the rapid expansion of research in this field shown in Fig. \ref{papers}.  

All the information in this review is either already in the public domain, or appropriate permission has been obtained from the relevant traditional owners. The knowledge in this review is  merely the tip of the iceberg, in that there is much research still to be done, and also because there is a wealth of sacred, and therefore secret, information that cannot be discussed  here. It is also possible that the knowledge in this review is biased to `male knowledge', since the author, like most authors in this field, is male, and is more likely to be told `male stories'. It is possible that a female researcher would write a different and complementary review of the subject.

This review also includes a significant amount of previously undocumented information obtained directly from Indigenous elders and others with traditional knowledge, some of whom are now deceased. For reasons of cultural sensitivity, many of these people cannot be named in this review. This paper therefore differs from conventional astrophysical papers by including a significant number of `private communications',  for many of which the name of the contributor cannot be given. In this review, I handle this unusual situation by citing the private communication in a  similar way to journal references, and referring to them as AC1, AC2, etc. where AC stands for `Aboriginal Contributor'. and list some details (language group, date, etc) in the bibliography, but usually withholding  the name. Further details of these contributors  (name, date, place, and, in many cases, a recording) can be made available in confidence to bona fide researchers. 

The primary goal of most this review and the literature cited is to describe the astronomical knowledge and culture of Aboriginal Australians before the arrival of Europeans in 1788. However, this is not intended to imply that Aboriginal knowledge or culture is frozen or static, and of course Aboriginal cultures continue to develop and evolve, and be infused by other cultures with which they come into contact.

To avoid confusion, I refer to the Aboriginal Australians before 1788 as `traditional Australians', and to their culture and knowledge as `Aboriginal culture' and `Aboriginal knowledge' respectively. 
I also confine this review to the astronomy of `Aboriginal Australians' (i.e. those living in mainland Australia and Tasmania), rather than the broader `Indigenous' grouping, which includes Torres Strait Islanders. 
 The astronomical knowledge of Torres Strait Islanders, a Melanesian people with links to both Aboriginal and Papuan cultures, is the focus of research by \cite{hamacher13c}.
 
To keep the review to a reasonable length, I exclude non-astronomical phenomena such as aurorae, lightning, rainbows, and weather. For brevity, I generally focus on  original research papers, rather than citing later papers that merely report the earlier work, unless they add to it by interpretation or consolidation with other work, or if the earlier work is difficult to access.

\begin{figure}[h]
\begin{center}
\includegraphics[width=10cm]{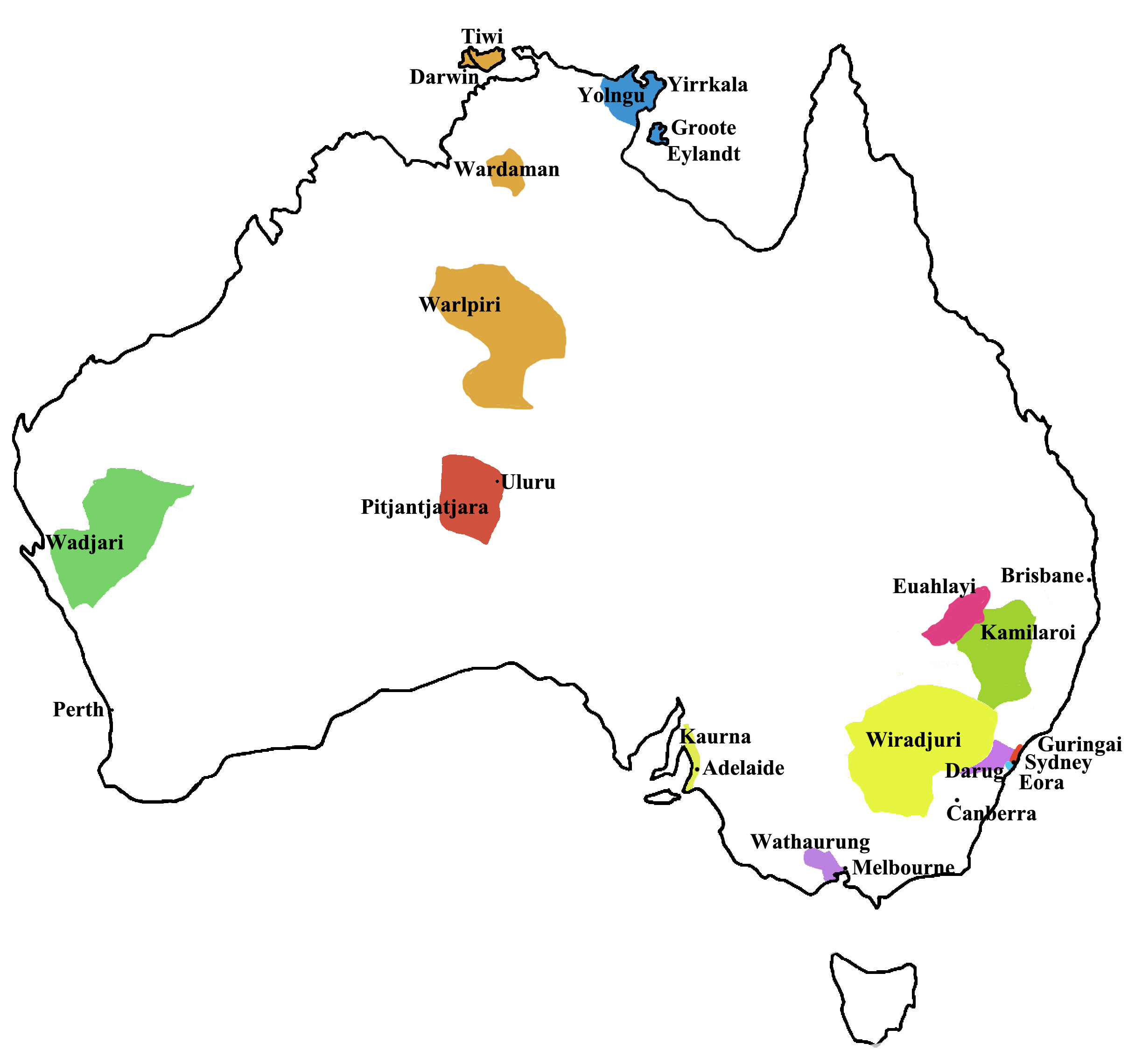} 
\caption{Map showing approximate locations of some of the places and  language groups discussed in the text.}
\label{map}
\end{center}
\end{figure}

\begin{figure}[h]
\begin{center}
\includegraphics[width=8cm]{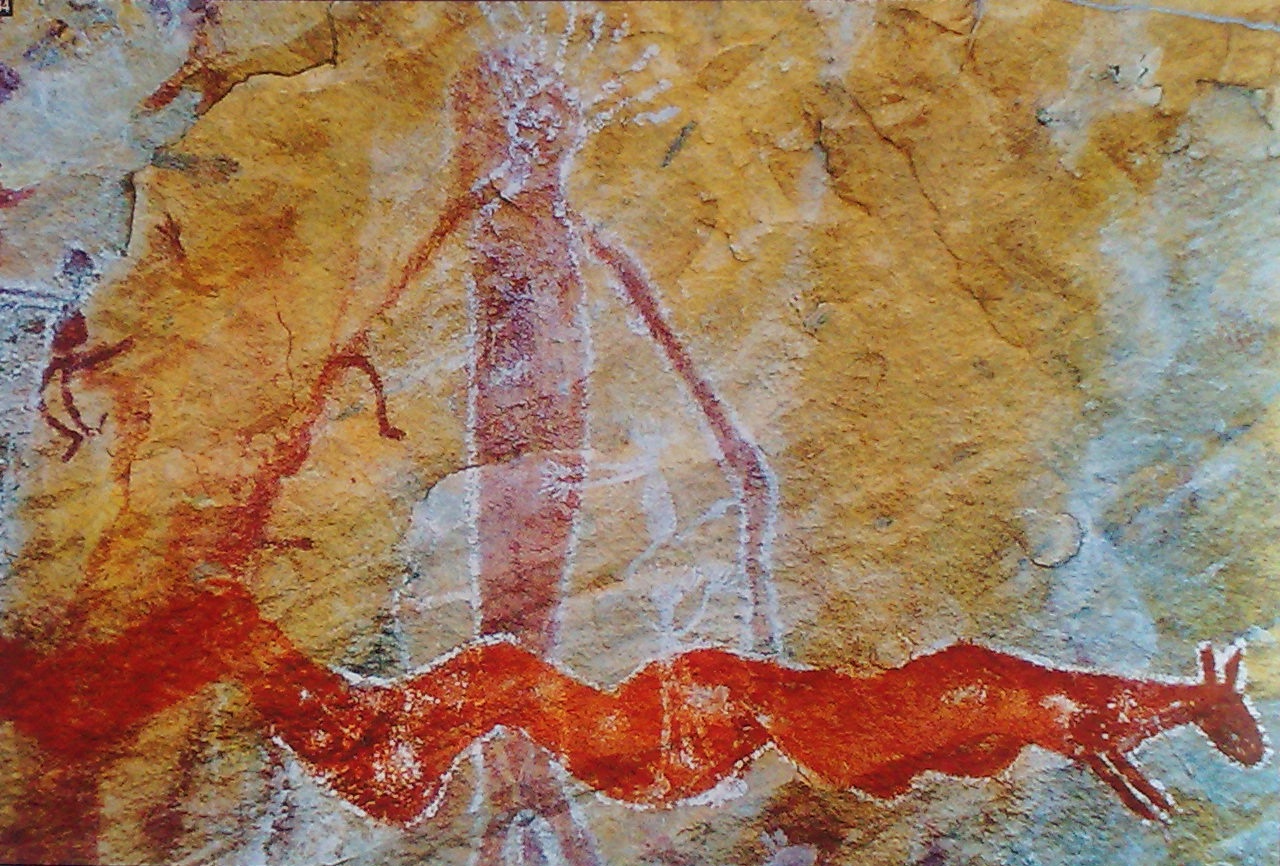} 
\caption{Wardaman rock painting of the Sky Boss and the Rainbow Serpent. The serpent at the bottom represents the Milky Way, and the head of the Sky Boss is associated with the Coalsack nebula, although a researcher could not deduce this astronomical connection without access to the cultural insight of Wardaman elder Bill Yidumduma Harney. (Photo courtesy of Bill Yidumduma Harney)}
\label{skyboss}
\end{center}
\end{figure}

\subsection{Aboriginal Number Systems and \\ Writing}
 \cite{blake81} stated that `no Australian Aboriginal language has a word for a number higher than four,'  despite the existence of well--documented Aboriginal number systems extending to far higher numbers  \citep{altman, harris87, mcroberts90,tindale25,  tully97}.  \cite{harris87} comments: `Statements such as these, which do not even admit five, are not simply misleading; they are false'.

There are many counter-examples to Blake's statement:
\begin{itemize}
\item the many well documented Aboriginal languages with number systems extending to well beyond four  \citep{altman, harris87},  
\item the observation   that traditional Aboriginal people could count in multiples of five  \citep{harris87} or twenty  \citep{tindale25}
\item the Gurindji `no-base' counting up to 50 with no compound words  \citep{altman},
\item the  report  \citep{dawson81} 
 that the Tjpwurong people had a cardinal system extending to 28 - the number of days in a  lunar month - each of which was identified as a place on the body, and as a verbal name which also described that part of the body, 
 \item a series of 28 lines at Moon Dreaming reported as being `moon counting'  \citep{DS},
 \item Harney's  \citep{DS}, report that as a stockman he could count a herd faster than whitefellas, explaining:
`We go five and five all the way and then bunch it up. Go five and five is ten. Then count the number of tens. We call that Yigaga' .
\item the number system of the Gumatj clan of the Yolngu people (who count in base 5) up to over one thousand, although the compound words for large numbers tend to be unwieldy for everyday usage  \citep{altman, harris87},
\item  the account  \citep{krefft65} of two Aboriginal men counting  about 50 bags. Their spoken language, using increasingly complex compound words for numbers greater than 5,  proved difficult for this task, so they used notches in a  stick instead.
\item the report  \citep{gilmore} that `The Aboriginal power to count or compute in his native state was as great as our own. ... I have seen partially trained native stockmen give the exact number of cattle in a group up to four or five hundred almost without a moment's hesitation, yet authorities on the blacks continue to tell us that the Aboriginal only counted to ten or thereabouts'. 
\item everyday observations of Tiwi children counting beyond 50 in their language  \citep{ED}, 
\item the observation that  the Pleiades are called `the Seven Sisters' in many Aboriginal languages,  or, alternatively that there are specifically five, six, or seven sisters  \citep{johnson11}, 
\item the reports that specific numbers of participants are required in some ceremonies  \citep[e.g.][]{berndt43}, 
\end{itemize}

On the other hand,  it has been argued that, while many Aboriginal groups construct `compound numbers' (like the english `twenty-one'), none have words like `one hundred' or `one thousand', although this too is disputed by  \cite{harris87}.
Recently,  \cite{zhou} have argued that there is considerable variation in Aboriginal number systems, and most don't contain higher numbers, but can gain or lose higher  digits over time. On the other hand, Zhou et al. do not discuss the  counter-examples, listed above, to their statement that the upper limit of Aboriginal cardinal numbers is twenty.
 

A further complication is that sign language is an integral part of many Aboriginal languages, or even a self-contained complete language \citep{kendon88, wright80}, and complex ideas can be conveyed silently using fingers. So it is perfectly possible that, in a particular language, Aboriginal people may have been familiar with the idea of `twenty', be able to count to twenty, be able to communicate it by sign language, but perhaps not have a spoken word for the number. For example, Mitchell in 1928 was able to barter food and a tomahawk for 10 days work with Aboriginal guides, using finger-counting  \citep{baker98}. Similarly  \citep{harney59}
reports an old man saying `after that many days - he held up five fingers to give the number as was the custom of the people'.  \cite{morrill64} and William Buckley \citep{morgan52} gave similar accounts that the Mt. Elliott people counted verbally up to 5, then used their fingers, and then the ten fingers of another person, and so on, until they reached a `moon' (presumably 28). The word for `five' was {\it Murgai}, which is quite different from the word for hand ({\it Cabankabun}).

In summary, traditional people were demonstrably able to count to much higher numbers than four, but in some cases using compound words (such as `hand of hands' or `twenty-one') or by using finger-language. Thus Blake's  statement that `no Australian Aboriginal language has a word for a number higher than four' may refer to a linguistic nuance that does not include compound words such as `twenty-one' or  non-verbal languages, but even then, this statement still seems inconsistent with the evidence cited above. Even worse, Blake's statement is often misinterpreted to mean `Aboriginal people can't count beyond four' or `Aboriginal people don't have a concept of numbers greater than four', both of which are obviously incorrect.

It has also been said that Aboriginal people `made no measurements of space and time, nor did they engage in even the most elementary of mathematical calculations'  \citep{haynes00}.
However, as will be shown in this review, there is plenty of evidence that Aboriginal people  made careful measurements of space and time, resulting in elaborate calendars and navigational systems.

\begin{figure}[h]
\begin{center}
\includegraphics[width=6cm]{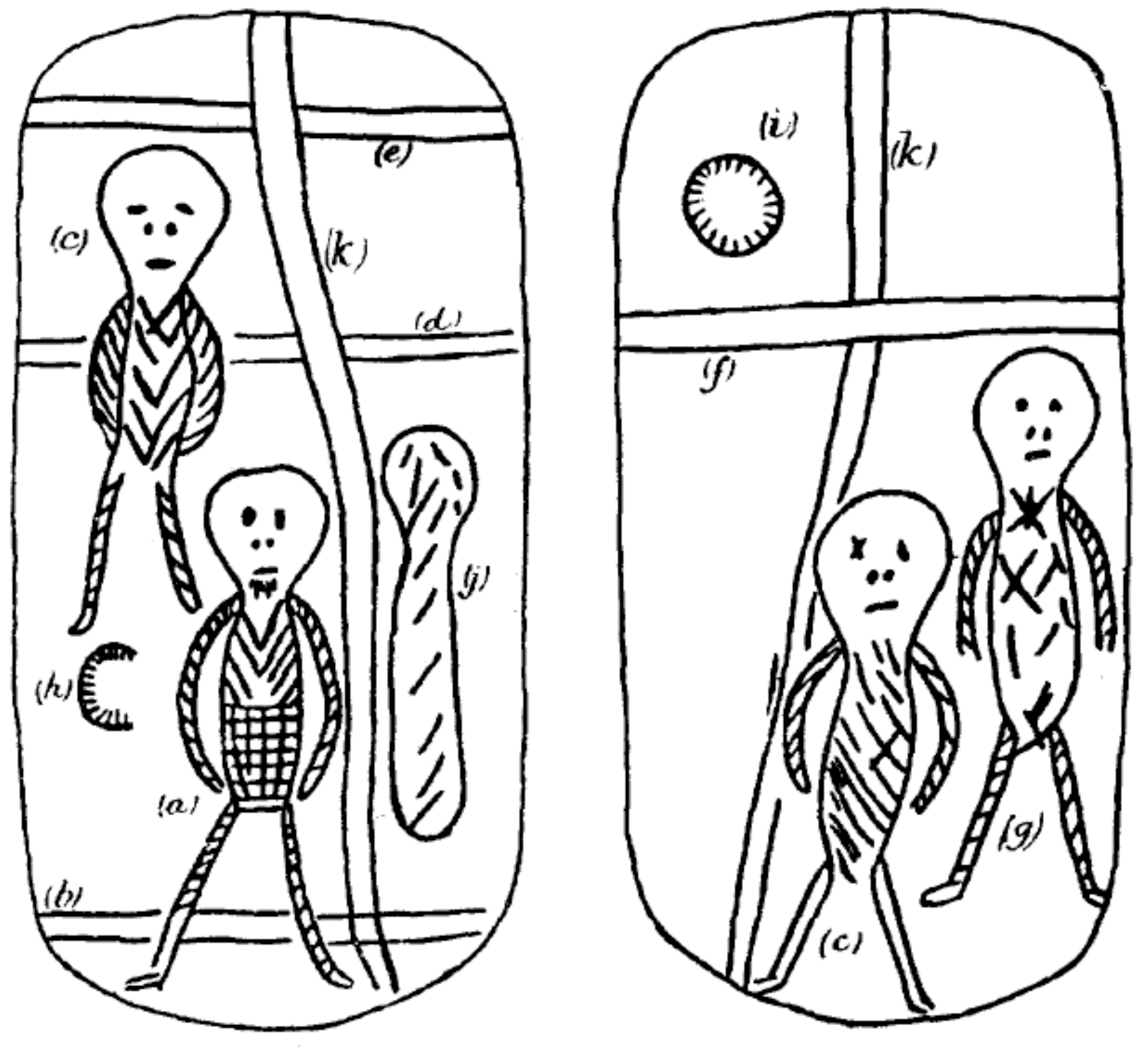} 
 \caption{A message stick, taken from  \cite{n268, mathews97c}, 
 depicting information including time, denoted by the phase of the moon. The message stick states that `Nanee (a) sent the message from the Bokhara river (b), by the hand of Imball (c), via the Birie (d), the Culgoa (e), and Cudnappa (f) rivers, to Belay (g); that the stick was dispatched at new moon (h), and Belay and his tribe are expected to be at Cudnappa river (f) at full moon (i); (j) represents a corroboree ground, and Belay understands from it that Nanee and his tribe are corroboreeing at the Bokhara river, which is their taorai, and, further, that on the meeting of the two tribes at full moon on the Cudnappa river a big corroboree will be held.'  The new moon, which in this context represents a crescent, is depicted in the lower--left of Frame 1, labelled as (h) while the full moon is the full circle depicted in the upper--left of Frame 2, labelled as (i).}
\label{fig:message_stick}
\end{center}
\end{figure}

It is also often stated that Aboriginal people had no written language. While this  appears to be largely true, it is worth noting three possible exceptions. First,  \cite{mathews97c}
explained how the pictograms on a message stick (Fig. \ref{fig:message_stick}) gave detailed information  about the place and time of a future corroboree. Second,   \cite{hahn64} 
discussed how  Aboriginal people in South Australia made notches in their digging sticks to measure their age in lunar months. Third,  \citet{harney09c} identifies the `scratches' in painting and rock-carvings as a variety of symbols, which he says would be understood by other Wardaman people. No doubt further research will uncover more exceptions.

\section{Sun, Moon, Tides, and Eclipses}
\subsection{The Sun}
Many Aboriginal cultures identify the Sun and Moon as a kindly female and a bad male creator sprit respectively  \citep{bates72, fredrick08, haynes00, johnson98,   ED, n230}, although this is not ubiquitous  \citep{berndt93, clarke90, clarke97b, meyer46, morgan52, tindale83}.
For example, the Yolngu people of Arnhem Land in the Northern Territory, identify the sun with Walu, the Sun-woman. In their tradition,  Walu lights a small fire each morning, which we see as the dawn  \citep{wells64}. She then decorates herself with red ochre, the traces of which create the red sunrise. She then sets fire to a torch made from a stringy-bark tree, which she carries across the sky from east to west creating daylight. When she reaches the western horizon, she extinguishes her torch, takes off her red ochre (creating a red sunset) and starts the long journey underground back to the morning camp in the east. An almost identical story is found in the Tiwi Islands  \citep{mountford58}. A similar story is recounted by  \citet{tindale83} from the Tanganekald people in SA
, where the Sun-woman carries fire-sticks instead of trees, mirroring the duty of old women to carry fire sticks when a group is travelling. In some stories  \citep[e.g.][]{mowaljarlai80}, the Sun-woman was originally too hot and burned the land, and so now she hides underground while her daughter, the `little sun', crosses the sky each day, lighting the land.

\begin{figure}[h]
\begin{center}
\includegraphics[width=6cm]{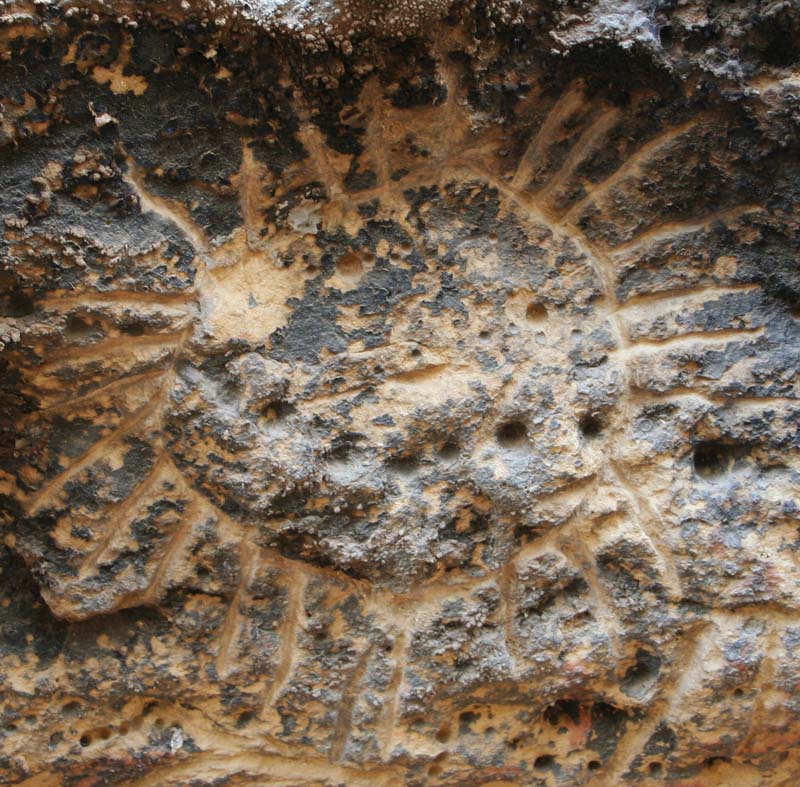} 
 \caption{A representation of the Sun, from the Ngaut-ngaut site in South Australia, described in \S\ref{ngaut}).}
\label{sunfig}
\end{center}
\end{figure}

The annual change in declination of the Sun has a corresponding explanation.  \cite{harney63} 
reports that central desert people said that `the Sun-woman doesn't like the cold, ...  so she rolls away to the northward on a sort of  annual walkabout',  mirroring the northward path in the sky taken by the winter Sun.


 \cite{n268} report accounts describe the sun--woman as an aggressive lover that chases the moon--man, who avoids her advances by zig--zagging across his path in the sky.  This may be a description of the complex motion of the Moon, varying in declination over the course of a month between two extremes, which themselves vary on an 18.6-year cycle. When she eventually catches him, they make love, resulting in an eclipse. \cite{parker05}
report a Euahlayi story that the Sun-woman, Yhi, was a wanton woman who tried to kill or ensnare her enemy, the Moon, but the clever men prevented it, and she continues to chase him across the sky.  \cite{reed65b}
describes how Yhi was in love with the Moon, and 
how she brought life to the world. 

 \cite{n311} also record a Kamilaroi and Euahlayi creation story associated with the Sun. In this story, either the brush turkey or brolga, depending on the version, threw an emu egg to the East, where it hit a pile of logs, which burst into flames, creating the Sun. The Sun then travelled across the sky each day, returning at night, thus creating day and night. A variation says that  a spirit in the sky saw how bright and beautiful the world was, so each night he collected firewood, and then  after sending Venus to warn people, he set fire to it in the morning. The Boorong people also said that the Sun was created by the throwing of an emu egg \citep{stanbridge57}.

\subsection{The Moon}

In most Aboriginal cultures, the Moon is identified with a bad man, sent to the sky for his evil acts. The Yolngu children's story is typical  \citep{wells64, wells73b}: the Moon-man, Ngalindi, was a fat round man, representing the full moon. He was extremely lazy, and refused to bring any food home for his family, but insisted his wives and sons catch all the food. They became very annoyed at this, and started chopping bits off him, making the fat man thinner, like a crescent, which we see as the phases of the Moon.  He managed to escape by climbing a tall tree to follow the Sun, but was mortally wounded, and died (the new Moon).  
(In other versions, he jumps off the top of the tree into the sky but is unable to return.) After remaining dead for three days, he rose again, growing round and fat (the waxing Moon), until, after two weeks his wives attacked him again. The cycle continues to repeat every month. 

In other versions of the story, he's not merely lazy, but commits incest or other taboo crimes. For example, in another Yolngu version  \citep{ac10}  he kills his sons, after which his wives burned him, giving him his white colour with black burn-marks.   \cite{n311} report a Kamilaroi story in which the Moon was punished for having sex with an under-age girl (the youngest of the Pleiades sisters).

In many Aboriginal cultures, the Moon is blamed for bringing death to the world  \citep[e.g.][]{bates72, berndt47,berndt48, berndt89}. For example, in the Yolngu children's version, until the Moon-man first died, everyone on Earth was immortal, but when he first came back to life, he cursed the world and its inhabitants, saying that henceforth he would be the only one who could come back to life after dying. For everyone else, death was to be final. 

A version  \citep{mountford58} of this story from the Tiwi islands is quite different but has significant elements in common. The moon-man, Japara, seduced a married woman, Bima. He persuaded her to leave her young son, Jinani, in the shade of a tree, while Japara and Bima went into the forest together.  One hot day, Bima was careless about leaving Jinani in the shade, and he perished of sun-stroke. When BimaÕs husband, Purrukapali, found out, he assaulted his wife with a throwing stick and banished her into the forest. Although the moon-man, Japara, claimed he could bring the boy back to life in three days, Purrukapali fought and wounded him. He then picked up the body of his son and walked into the sea, cursing the world and saying that henceforth, everything would eventually die and not return to life, thus bringing death to the world. Japara escaped by turning himself into the moon, but didnÕt quite escape from PurrukapaliÕs curse, as he dies for three days each month, but is then reincarnated.  The Tiwi people believe that, when he returns to life, he gorges himself on mangrove crabs, become fat and round, but then becomes sick from over-eating, so he dies again \citep{mountford58}.
The marks on the face of the moon are the wounds that he received in the fight with Purrukapali. Bima became Wayai, the curlew, whose plaintive call expresses her sorrow for the loss of her son and for her foolishness that caused it. A significant difference from the Yolngu version is that the Tiwi Moon-man tries to stop Purrukapali from bringing death to the world.

 \cite{n311} report a Kamilaroi story in which Wahn the crow wanted to bring dead people back to life, but the Moon-man Bahloo refused. In anger, Wahn tricked Bahloo into climbing a tree, which Wahn then made to grow up into the sky, trapping Bahloo in the sky. After he rejected Yhi, the Sun-woman, she enlisted the help of spirits to prevent him from returning to Earth. Similar stories  appear in many other language groups, such as the Wardaman people  \citep{ac11}  who have a story in which the Moon-man is expelled to the sky for breaking a taboo.

 \cite{parker98}
reports a Euahlayi story in which the Moon-man,  Bahloo, asked the first men to carry his dogs across the river. When they refused, the dogs turned into snakes, which bit the men. Bahloo decreed that men would henceforth stay on Earth, and die, which is why men kill snakes whenever they see them. 

The Kaiadilt people, from Bentinck Island (Qld), also call the Moon-man Balu, and  the first appearance of the New Moon in the west triggers a ceremony in which Balu is asked to help with the weather or with food supplies, or to supply an especially low tide  \citep{tindale83}.
Tindale  notes  similarities in ceremony and in the word for the Moon-man between the Kaiadilt and NSW groups such as the Euahlayi, and suggests the name Balu is very old, and was carried to NSW by migrants from the north.

 \cite{n259} discuss exceptions to the widespread masculinity of the Moon. For example, the Ngarrindjeri of South Australia saw the moon as a promiscuous woman  \citep{clarke97a, meyer46}.
Her frequent  sexual liaisons made her thin, represented by the waning moon. Eventually, when she became very thin (crescent moon), the creator-spirit Nurrunderi ordered her to be driven away, causing her absence (new moon). Then she began to eat nourishing roots, causing her to fatten again, and the cycle repeats. The nearby Jaralde people have a similar story, differing in that  the waxing moon represents the moon-woman becoming pregnant  \citep{berndt93}.
  \citet{DS} tell the story of the Moon-man who was poisoned and went to the sky for breaking the law, but then say that the full moon is female.
 
A completely different story  \citep{tindale83}
involves the Moon-man being killed by falling from a burning tree,  leaving only four ribs in the burnt remains of his body. His four ribs were thrown to the four cardinal points (as discussed further in \S\ref{direction}), and three returned like boomerangs, but the fourth sailed over the western horizon and eventually became the moon. The monthly lunar cycle illustrates the burning of his body followed by the appearance of his rib in the west. Other groups  \citep{berndt89} also have stories in which the Moon-man throws boomerangs, explaining how the boomerang-shaped moon moves across the sky.  Escaped convict William Buckley reported that the Wathaurung people believed that the Moon-man sometimes came to Earth, and they sometimes met him, but that the Moon was also a boomerang which was thrown into the air by one group, and then caught and re-launched by another \citep{morgan52}.
 
Potaruwutj people in SA believed the Moon to have been the male ancestor of the quoll  \citep{clarke97a}, and the phases are caused by the fact that he was driven away after trying to steal someone's wife, and is sometimes well fed, and sometimes starving. 

The lunar month is roughly the same length as the menstrual cycle, and many cultures associate the moon with fertility  \citep{fredrick08,n259, johnson98}.
 In some, young women were told not to look at the moon to avoid becoming pregnant  \citep{bates72, haynes97}, and Wardaman people believe that looking at the moon encourages promiscuity in both sexes  \citep{ac9}. Several other Aboriginal groups associate the moon with love, fertility and intercourse, including the Koko-Yalanyu of Queensland  \citep{mcconnel31} and the Lardil people of Mornington Island  \citep{isaacs80, roughsey71}. 
 
  \cite{n311} report that the Kamilaroi moon-man, Bahloo, created female babies.
 \cite{parker98}  noted that the Euahlayi people said that if the Moon was late rising, he'd been making girl babies, and his imminent rising was signalled by the haze that precedes the Moon. 

In cold weather, a halo often surrounds the  Moon, as a result of ice crystals in the upper atmosphere. 
 \cite{parker05} said that the Euahlayi associate this with rain, and that it is a house to keep the Moon-man dry. Others  \citep[e.g.][]{clarke09a} report that a halo around the Moon signifies that the Moon-man is building a shelter to keep out the cold.  \cite{n311} report a Kamilaroi story that the ring is dust kicked up by a rain-dance. The dust creates the clouds, and brings rain.


\subsection{Tides} 
The height of the tides varies with the phase of the Moon, with the highest tide (`spring tide') occurring at the full or new moon.  Many of the Yolngu people, known as the `saltwater people', live on the coast, travel by means of bark canoes, obtain a significant fraction of their food from the sea, and have an intimate knowledge of the motion of the sea and the tides. A traditional Yolngu story  \citep{berndt48, hulley96, mountford56} explains how the Moon causes the tides:  it fills with water from the sea as it  rises through the horizon. When the moon is a crescent or half-full, not much water is transferred, but on a full or new moon, the whole contents of the moon is transferred. The explanation is rather different from the modern scientific explanation, but given the information available to the Yolngu hundreds of years ago, it is a reasonable explanation that is consistent with their model of how the world works. It is  a good example of an evidence-based approach to understanding the world in an appropriate cultural context, resulting in a description with predictive power, in the sense that a Yolngu person, seeing the position and phase of the Moon, can predict the time and height of the next tide. It may be contrasted with Galileo's incorrect explanation of the tides  \citep{galileo} which, unusually for Galileo, was inconsistent with observational data, since it predicted only one tide per day, whose height and timing were independent of the Moon. Unlike the Yolngu model, Galileo's model had no predictive power.

 \cite{warner37} described how the Sun, Moon, and tides had important roles in  Yolngu ceremonies and rituals, and said that the Yolngu `have a most accurate knowledge of the locational, seasonal, and daily variation of the tides. Anyone who has taken a canoe trip with them along the seacoast quickly learns  that this knowledge is immense in detail, well organised, and held by all the men'.

These stories   illustrate how  different explanations of the same phenomenon  (e.g. the phases of the Moon either being a result of the Moon-man being attacked by his wives, or by filling with water) often coexist.  Sometimes different stories are from different language groups or different clans, but often several overlapping stories occur in the same clan and are not regarded as inconsistent  \citep{ED}.

\subsection{Solar Eclipses}

A total solar eclipse is visible from only a small area on Earth, and so, from any one location, a total eclipse is typically seen only once every ten generations. Partial eclipses are much more common, but are not normally noticed unless observers have been pre-warned  \citep{n259}. It is therefore surprising that explanations of  solar eclipses are very common in Aboriginal mythology. These explanations imply a remarkable continuity of learning. 
For such a myth to have been created, someone must have seen an eclipse, associated it with a story passed to them from many generations previously, and then strengthened and retold so that it was eventually passed on to someone many generations later.  \citet{tindale83} gives an example of a specific solar eclipse in 1793 still being the core of a story over one century later.

The first attempt to compile Aboriginal stories about solar eclipses  \citep{n230} concluded that, while solar eclipses were widely viewed as a bad omen, there was also good evidence that at least three Aboriginal groups (the Euahlayi, Yolngu, and Warl\-piri people, each from a  different state of Australia) recognised that an eclipse was caused by a conjunction between the Sun and Moon. A  more detailed study of eclipses   \citep{n259} found that solar eclipses were widely regarded as an omen of impending disaster, or a sign that someone was working black magic. Consequently, solar eclipses were feared by many Aboriginal people, although some were unafraid of them   \citep{peggs03, tindale83}.

Many Aboriginal communities across Australia would try to ward off this evil by chanting a spell  \citep{parker05} or by throwing sacred objects at the sun  \citep{rose57,spencer99}.
In Wardaman culture, an eclipse can be prevented by Djinboon, the head of the sun-clan  \citep{harney68}.

The explanation of the eclipse varies widely from group to group. Several groups recognised that it was caused by a  conjunction between the sun and the moon,  often attributing it to the sun-woman and the moon-man making love. For example, Daisy Bates recounted  \citep{bates44} how, during the solar eclipse of 1922, the Wirangu people told her that the eclipse was caused when the Sun and Moon became `{\it guri-arra }: husband and wife together'.  \citet{n217}  speculate that a rock engraving in NSW (shown in Figure \ref{basinpic}) may be an illustration of the conjunction of the Sun-woman and the Moon-man causing an eclipse.
Some other groups believed that a solar eclipse was caused by the sun being covered by something, but not necessarily the moon  \citep{n259, morgan52}, 
and a few attributed it to a completely different cause, such as the Pitjantjatjara belief that bad spirits made the sun ÒdirtyÓ  \citep{rose57}
and the Wardaman belief that an evil spirit swallowed the sun   \citep{harney68}. \cite{morrill64} reported that the Mt. Elliott people attributed an eclipse to one of their group covering the Sun.
According to  \cite{tindale63}, 
the Ngadjuri may have associated  an eclipse with a story in which the Sun went dark after an evil old woman was killed, but was restored by throwing  a ritual boomerang  to the East.

 \cite{n311} report Euahlayi stories that the Sun-woman tried to ensnare or kill her enemy, the Moon-man, and she continues to chase him across the sky.  \cite{fredrick08}
suggested that this story concludes by the Sun eventually catching the Moon, causing an eclipse. Similarly, the Yolngu have a tradition that on rare occasions, the sun--woman ({\it Walu}) manages to capture the moon--man ({\it Ngalindi}) and consummates their relationship before he manages to escape, explaining a solar eclipse  \citep{warner37}.

\subsection{Lunar Eclipses}
Unlike solar eclipses, lunar eclipses are visible from anywhere on the side of the Earth facing the Moon, and so are a relatively common sight from any location. Like solar eclipses, lunar eclipses are often viewed as a bad omen by Aboriginal groups, and often regarded as a sign that someone has died  \citep{n259}. On the other hand,  the Aboriginal people near Ooldea, South Australia saw three lunar eclipses, which they called `pira korari', in a year, but  paid little attention to them  \citep{tindale34}.

I was staying at the remote Yolngu community of Dhalinbuy during a total lunar eclipse on 28 August 2007. Apart from showing mild interest in the phenomenon, with which they were clearly familiar, the Yolngu people  showed no particular reaction to it, and certainly showed no signs of fear or distress.

Like solar eclipses, lunar eclipses were  widely interpreted as something covering the Moon, but in only one case was it attributed to the relative positions of the Sun and Moon  \citep{n259}.  \citet{n230} have pointed out that, in this case,  to reason that it is precisely this alignment which has caused the eclipse is an impressive intellectual feat, given that the Sun and Moon are diametrically opposed in the sky.

In other cases, the Moon-man is thought to be covering his face or covered by a blanket  \citep[e.g.][]{DS},
 or the Moon is covered by the shadow of  a Man who is walking in the Milky Way. The red colour of the Moon during a lunar eclipses was widely attributed to the Moon-man having blood on his face, although the Kurnai  \citep{massola68}
and the Murrawarri  believed a red moon signified that the Moon-man was a murderer  \citep{mathews94}.

\subsection{Relationship between the Earth and the Sky}
\label{relationship}
It is interesting to speculate on whether the Sun's journey back to the East at night implies that traditional Aboriginal people  regarded the Earth as being a finite body which the Sun might pass underneath. In about 1928 the anthropologist Warner  \citep{warner37}
recorded the Djunkgao myth from Yolngu people who had had almost no previous exposure to  western civilization.  A Yolngu man told him `Those two women ... followed the sun clear around the world'. Warner said that the informant illustrated this by putting his hand over a box and under it and around again, suggesting that the Yolngu people at that time had reasoned that the apparent motion of the Sun around the Earth showed that the Earth was not flat and infinite, but finite. We shall probably never know whether this worldview extended to the idea that the Earth is a ball in space. Similarly, 
 \cite{smith30} 
reported that some Aboriginal elders who `study the stars and their positions in the heavens at night' said at intervals during the night that that `the earth has already turned' and each `day' meant `the earth has turned itself about', suggesting that they knew the Earth rotated, although this may have  been influenced by cultural contamination from the West.

Apart from these two isolated accounts, most other accounts  \citep[e.g.][]{ash03,clarke15a, n311,n318} reflect a widely-held belief that the Earth and the Sky are two parallel worlds which mirror each other, and the sky (sometimes called the `skyworld'  \citep{clarke03b}) is a reflection of the terrestrial landscape, with plants and animals living in both places. Often, each astronomical object  is thought to have a counterpart on Earth, and vice-versa.  \citet{mowaljarlai93} said `Everything has two witnesses, one on Earth and one in the sky. This tells you where you came from and where you belong'. 

\cite{leaman16a} have shown that, at least around Ooldea, SA, animals in the skyworld were associated with particular stars or celestial objects, often chosen such that the seasonal appearance of the star is linked to the breeding cycle of the animal.

 \citet{teichelmann41} said Aboriginal people in Adelaide believed that: `all the celestial bodies were formerly living on Earth, partly as animals, partly as men, and that they left this lower region to exchange for the higher one. Therefore all the names that apply to the beings on Earth they apply to the celestial bodies'.  \cite{n318} reported that one of the participants said that `everything up in the sky was once down on Earth because thatÕs the way it started out, and the sky and the Earth reversed'. This closeness between the land and the sky, and the symmetry between them, perhaps explains why sky knowledge is so important in ceremony. 
 
Clever men are said to be able to move between the land-world and the sky-world, often by climbing a tall tree, being pulled up by a rope, or walking to the top of certain high hills  \citep{clarke90, clarke97a, clarke15a, howitt04, kaberry35}. In south east Australia it was widely believed that the sky was held up by four enormous wooden props or trees  \citep{haynes00}, and there was a call around the time of British colonisation for supplies to be sent to the old man who maintained  the eastern prop which was in danger of rotting and collapsing  \citep{howitt04, massola68, morgan52}. 

The sky is often regarded as being relatively close to the Earth, perhaps the distance of a long spear throw \citep{clarke15a}, and several reports involve  spirits moving between the land and the sky.  \citet{tindale83} notes that ancestral beings started on Earth, but moved to the sky, sometimes merely by walking to the horizon where the Earth meets the sky. Some groups believed that the connection to the underworld was through a  cave  \citep{clarke09a}. 

Many groups believed that all the celestial bodies were formerly living upon Earth, partly as animals, partly as men, and that they moved from the Earth to the sky  \citep[e.g.][]{akerman14, hamacher15a, teichelmann41}. Many groups believed that stars and  planets are the ancestors of living people, who  therefore have  kinship links to particular stars  \citep[e.g.][]{clarke97a}. The Karadjeri people believed the sky consisted of a dome made of hard shell or rock, and the stars are the spirits of dead people \citep{piddington32}.

The regions of the sky itself may correspond to the different groups on Earth. For example,  \cite{maegraith32} reports that Arrernte and Luritja people divide the sky into two great `camps'. Stars to the east of the river of the Milky Way are said to be Arrernte camps, and  stars to the west are said to be Luritja. 

 \cite{mountford58} 
reports that the Tiwi people believe that the Universe has four flat levels: (a) a subterranean world, Ilara, (b) the Earth on which humans live, (c) the sky-world Juwuku, and (d) an upper world, Tuniruna. 
 The Earth itself is flat and has an edge, over which the Sun-woman can be seen to descend in the evening. She then travels along a valley in Ilara to reach her camp-site in the East. Aboriginal people in SA also believed in an Underworld, which was also known as the `Land to the West'  \citep{clarke97a}, through which the sun-woman travelled back to the east each night.
 \citet{clarke03a} reports a range of views, from different parts of Australia, of how the Sun and Moon return each day to the east, either underground, or by a road just beyond the northern or southern horizon. 
 \citet{mountford58} reports that the Tiwi say that the Moon-man used to travel by a path beyond the southern horizon, but now returns by a route beyond the northern horizon. 
Some Yolngu say that the sun-woman turns into a king-fish and swims under the earth, implying that the earth is a floating island  \citep{wells64}.

\section{Stars and Constellations}

\begin{figure}[h]
\begin{center}
\includegraphics[width=8cm]{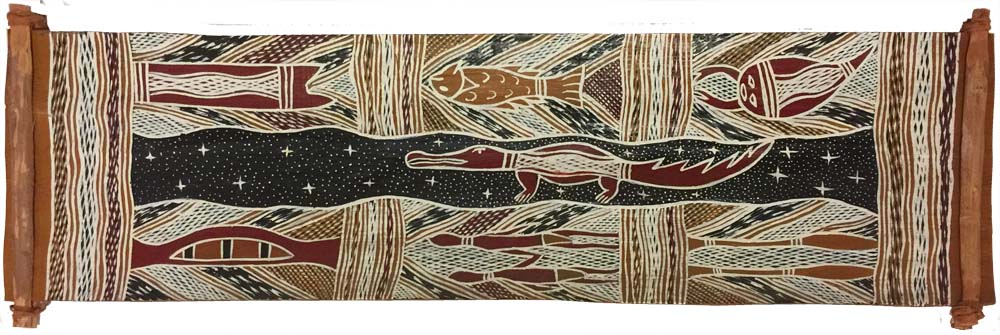} 
\caption{A Yolngu view of the Milky Way: a traditional bark painting by Yarawu. The crocodile represents Scorpius. This particular painting is modern (~2000 AD) but follows a traditional design.}
\label{yarawa}
\end{center}
\end{figure}

Stars are a central part of Aboriginal sky-lore, and are often associated with creator spirits.  Harney  \citep{n315} recounts how the twinkling of the stars represents the stars talking to each other. There are many recorded stories, too numerous to list here, associated with individual stars, throughout the different Aboriginal cultures. \cite{fredrick08} lists a number of them. In many cases, individual stars are associated with ancestors or creator spirits  \citep[e.g.][]{berndt43}.
As pointed out by the anonymous referee of this Dawes review,  this focus on stars is not widely found in other ethnoastronomical studies, and the reason for this different is unclear. It is unlikely to result from the interests of the researchers, since stars are also widely portrayed in traditional Aboriginal art, such as Figures \ref{art},\ref{yarawa}, and\ref{pole}. A comparative study to answer this question would be interesting. 

 \citet{clarke09a} has noted that individual stars occur more often in Aboriginal folklore than `constellations', but that stars are more likely to be noted if they are part of a recognisable constellation.   Nevertheless, as in traditional European and other cultures, many Aboriginal groups identify particular groups of stars in the sky as `constellations'. In many cases, these are  remarkably similar to the European constellations. In addition, many Aboriginal cultures also identify objects in the sky formed from the dark spaces between the stars, a concept that is completely absent from European astronomical lore, perhaps because the dark clouds in the Milky Way are less conspicuous in the Northern Hemisphere. For convenience, I term such structures `constellations', although I acknowledge that the word `constellation' is not strictly appropriate  because these structures are not formed from stars.

Here I discuss a few notable examples, but many more Aboriginal constellations,  some of which  are listed in Table 1, are also well-known and documented,

\subsection{Orion}

The European constellation of Orion takes its name from Greek  mythology, in which Orion was a hunter. He first appears in literature in HomerÕs Iliad  \citep{iliad}, where he is said to be the son of Poseidon, the King of the sea, and a Cretan princess. 
However, the story was probably already ancient in HomerÕs time, and was used in  ancient Greek calendars  \citep[e.g.][]{planeaux06}. 

\begin{figure}[h]
\begin{center}
\includegraphics[width=6cm]{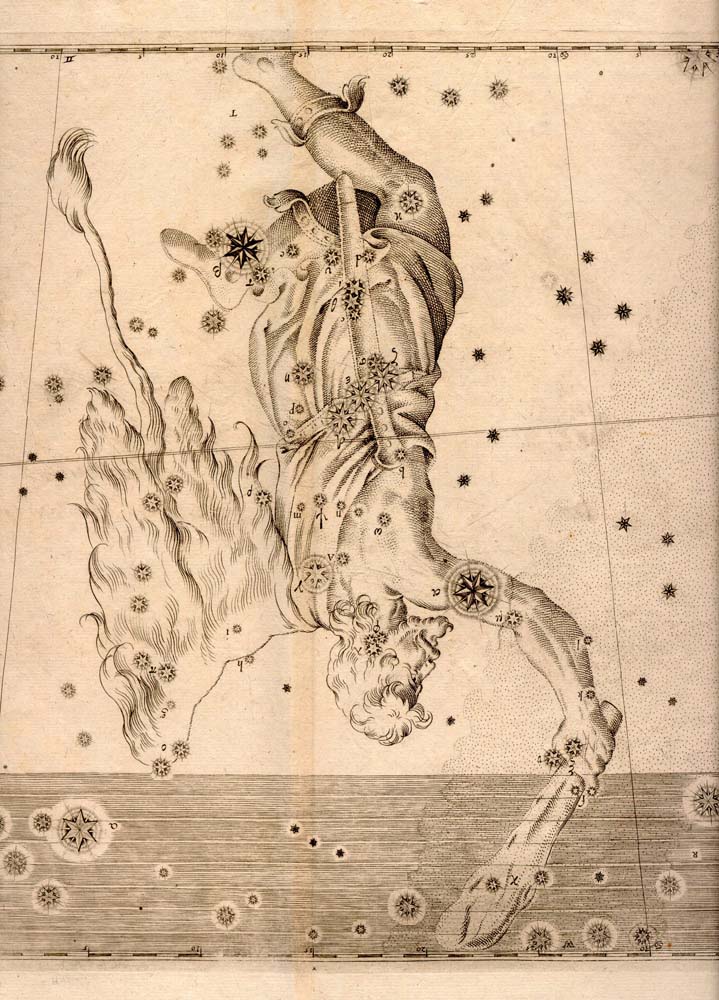} 
\caption{An engraving of Orion from Johann Bayer's Uranometria, 1603, inverted to show the view from Australia. Creative Commons Attribution-Share Alike 3.0}
\label{orion}
\end{center}
\end{figure}

In this western tradition, the stars of the Orion  constellation symbolize the Greek Hunter, shown in Fig. \ref{orion}. The three prominent stars are his belt, and the  Orion nebula is his sword.  There are also suggestions (Hugh Cairns, private communication) that the Orion Nebula was originally his penis, and that the story was changed to a sword to protect human sensibilities.  

\begin{figure}[h]
\begin{center}
\includegraphics[width=6cm]{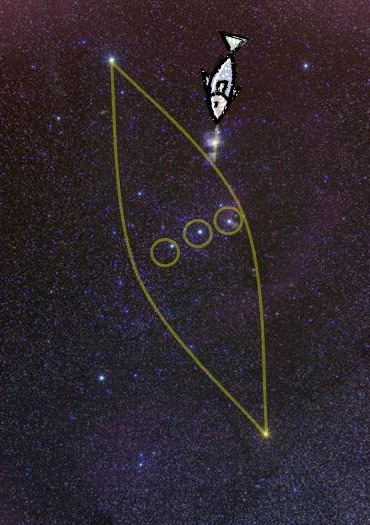} 
\caption{A Yolngu interpretation of Orion. Drawing by the author based on Yolngu oral and written accounts.}
\label{djulpan}
\end{center}
\end{figure}

Curiously, most Aboriginal cultures also associate Orion with a hunter, or a young man, or a group of young men, or a male ceremony.   For example, in Yolngu culture, Orion is called  Djulpan. A typical Yolngu story is that the three stars of Orion's belt are three brothers  in a canoe, with Betelgeuse marking the front of the canoe, and Rigel the back of the canoe. The brothers were blown into the sky by the Sun-woman  after one of them ate his totem animal, a king-fish, whose consumption was strictly prohibited by Yolngu law  \citep{davis89, davis97, wells73c}. 
The Orion nebula is the fish, and the stars of Orion's sword are the line still attached to the fish (Fig.\,\ref{djulpan}).  Many similar stories  in other Aboriginal cultures  associate Orion with young men, particularly those who are hunting or fishing  \citep[e.g.][]{massola68, mountford39, mountford76}. For example,  the Kaurna people of SA  \citep{clarke90, gell42, teichelmann40} see Orion as a group of boys who hunt kangaroo and emu on the celestial plain.  \cite{mathews94}
report that the Murrawarri said that Orion wore a belt, carried a shield and stone tomahawk, and their name for the constellation (Jadi Jadi) means either `strong man' or `cyclone'.  \citet{bates25} says that people over a great area of central Australia  regarded him as a `hunter of women', and specifically of the women in the Pleiades, and that the male initiation ceremony includes an enactment of Orion chasing and raping women. The ceremony may only take place when Orion is {\em not} in the sky, which is consistent with the report  \citep{n311} that, in Kamilaroi culture, Orion's setting in June is associated with the male initiation ceremony.

\subsection{Pleiades}
\label{pleiades_section}

The young open stellar cluster known as the Pleiades, or Seven Sisters, is a cluster of hot, blue, young stars which have only recently 
emerged from their dusty womb, and they are still surrounded by hot blue strands of glowing gas. They take their name from the Pleiades of Greek mythology, in which they were the daughters of Atlas and Pleone. 
After Atlas was forced to hold up the sky, he could no longer protect them from the lusty Orion, who tried to rape them. 
To save them from this fate, the almighty Zeus transformed them into stars. However, the rotation of the sky makes it appear that the constellation Orion still continues to chase the Pleiades.
 \citet{krupp94} 
notes that they appear in the mythology of many cultures, often as seven women, young women, or `daughters'. 

In nearly all Australian cultures, the Pleiades are female, and are associated with sacred women's ceremonies and stories. The fact that these secret stories cannot be told to men was the basis for a famous legal battle in which female elders tried unsuccessfully to stop the Hindmarsh Island Bridge, SA, being built on sacred ground. The elders lost the case partly because they were unable to reveal the stories to the court  \citep{bhathal06, bhathal11c}.

\begin{figure}[h]
\begin{center}
\includegraphics[width=6cm]{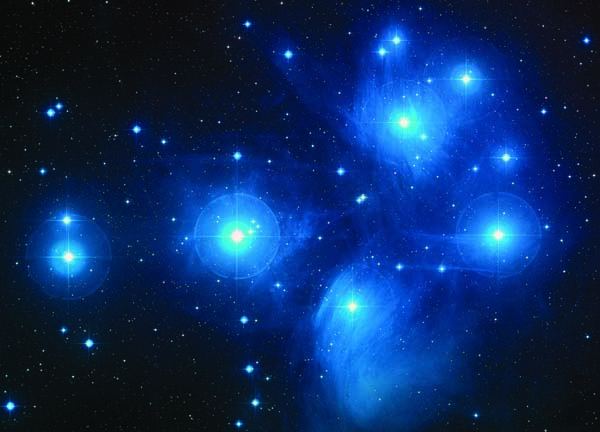} 
\caption{A colour-composite image of the Pleiades from the Digitized Sky Survey
Credit: NASA/ESA/AURA/Caltech}
\label{pleiades}
\end{center}
\end{figure}

Many Aboriginal cultures associate the Pleiades with a group of young girls, or sisters  \citep[e.g.][]{andrews05,  clarke90, clarke97a, clarke09a, n311, harney59,   leaman, massola68, n230, n255}.  \citet{johnson11} argues that Aboriginal Pleiades stories can be divided into four main geographical areas: mainland Australia, Arnhem Land, Torres Strait, and Tasmania. Each area has  different versions of the Pleiades story, except that there is no known Pleiades story from Tasmania.  Most mainland Aboriginal stories portray the Pleiades as girls chased by the young men in Orion  \citep[e.g.][]{johnson11}, which is very similar to the traditional European myth about these constellations, while  most Arnhem Land stories portray the Pleiades as partners of the men.

Curiously, in many Aboriginal cultures, the Pleiades are  called the seven sisters, just as in the Greek tradition. This is odd because no combination of seeing, quality of eyesight, or sky brightness would cause a human to count seven stars. Instead, humans can typically perceive four, six, or  eight stars. Even after taking proper motion into account  \citep{n316} they would not have appeared as seven stars at any time in the last 100,000 years, unless one of the  brightest stars has long-term variability which has so far eluded detection.

There are also several accounts  \citep[e.g.][]{n311, kyselka93} 
in which one (or occasionally two) of the sisters has died, is hiding, is too young, or has been abducted, so only six (or five) are visible.  \cite{n311} reports a Kamilaroi story in which the Pleiades are five sisters searching for their two lost sisters, and their mother (Venus) is also searching for her lost daughters. These stories appear to be an attempt to reconcile the seven sisters in the stories with the five or six stars which are actually visible in the sky.

In two versions of the story from NSW and Victoria, Orion consists of boys who dance at night  to music made by the girls in the  Pleiades  \citep{parker05, smyth78}.
The Kamilaroi people also believe the Pleiades to be a nest of honeybees  \citep{tindale83}, and the Tiwi people believe they are  a group of kangaroos being chased by the dingoes in Orion  \citep{mountford76}.

The Pleiades are also important as an element of Aboriginal calendars. Hunter-gatherers, such as the Yolngu people,  have a well-defined seasonal cycle of movement  throughout their land. It is important that they move camp at the right time, to catch the Barramundi, or harvest berries before the birds get them, or harvest the raika roots while they are still fresh and tender.  For example,  \cite{ac5} said of the Pleiades:
`Seven sisters Écome back with turtle, fish, freshwater snakes and also bush foods like yams and berriesÉ.The stars come in season when the food and berries come out, É They give Yolngu bush tucker, they multiply the foods in the sea -  that's why Yolngu are happy to see them'.


 Many traditional Aboriginal stories refer to the sisters as pursued by the young men in Orion \citep{harney59, leaman,tindale83}, 
or occasionally by Aldebaran  \citep{n311}, which is curiously similar to the traditional European myth about these constellations. In a Central Desert version \citep{harney59},
  the girls are being chased towards Uluru by the young Orion men from the North, and escape by fleeing into the sky.  \cite{harney59} reports the story from the Pongi-pongi (Kandjerramalh) people at length concluding with `Can't stop following them ... that Manbuk [Orion] always chase them seven dog-sisters along that big sky road'.
 
Similarly, in Kamilaroi culture, Orion is known as the young men who loved, and pursued, the Pleiades
 \citep{greenway78, mountford76, parker05,  ridley75}.

Versions of the story from south-west Australia often feature the girls being protected by their dingoes, but this detail tends to be absent in versions from south-east Australia, from which  \citet{tindale83} deduced that the story may predate the arrival of dingoes in Australia in about 5000 BC. The association of the Pleiades with dingoes may also stem from  the harvesting of dingoes by the Kuwema (or Kandjerramalh) people as a food source at the helical rising of the Pleiades  \citep{harney63, n214,  tindale63}.


It is remarkable that the various versions of the Orion and Pleiades stories are very similar to the Greek mythological culture, in the following respects:
\begin{itemize}
\item Orion is male and the Pleiades are female
\item Orion is chasing the Pleiades
\item The Pleiades are seven sisters, even though the cluster is not visible as seven stars under any combination of seeing and eyesight.
\end{itemize}

One putative explanation for this similarity is that it is a result of European cultural influence. However, the widespread distribution of these stories, and their reporting soon after first contact, suggest that they are far older than European contact. Similarly, it is unlikely that they can be attributed to the annual pre-1907 cultural contact with Macassans (see \S\ref{scorpius}) in the North of Australia.

Another putative explanation is that Aboriginal people independently devised the stories in a sort of cultural convergent evolution, perhaps reflecting the fact that the `masculine' bright stars of Orion follow the beautiful little cluster of the Pleiades as the sky rotates, although this fails to explain other similarities such as the insistance on {\it seven} sisters.

An alternative hypothesis  \citep{n316} is that this story dates back to about 100,000 BC, before our ancestors left Africa, and was carried by the people  who left  Africa to become Aboriginal Australians, Europeans, and Asians. It is planned to test this hypothesis by sequencing the evolution of the story around the world.

\subsection{The Milky Way}

The Milky Way is widely recognised by traditional Australians across Australia, although interpretations vary greatly. 

Many Aboriginal groups see the Milky Way as a celestial river along which the sky people travel in their canoes and gather food  \citep[e.g.][]{mountford56,mountford58} 
alongside which reeds are growing  \citep{clarke90}, or sometimes it is seen as a canoe itself  \citep{clarke09a}.
In the Yolngu interpretation \citep{wells64}, the nebulae represent campfires of their ancestors, and some Kamilaroi people interpret the stars as fires,  lit by spirits of the dead  \citep{parker05}, whose smoke causes the diffuse light.
In SA the smoke of the Milky Way is to guide spirits of the dead to the eternal camp fires  \citep{clarke97a}.  \cite{n311} report a Kamilaroi story that the two bright patches either side of the Milky Way in Sagittarius (the Galactic bulge) are the spirits of the sons, but their bodies are the large rocks which can be seen today on either side of the stone fish traps on the Barwon River at Brewarrina. The dark patches in the Milky Way are said by several groups to be water holes inhabited by monsters  \citep[e.g.][]{clarke97a}.

The Nukunu people of SA viewed the Milky Way as a large ceremonial pole which became a large box gum tree  \citep{clarke09a}. 

In northern parts of Australia, the Milky Way is often interpreted as the Rainbow Serpent  \citep{ED}, or, in the Kimberleys, as one of the two supreme (and serpent-like)  Creator-beings  \citep{akerman14}.  \citet{berndt89} report a story from the Ngulugwongga people in which the Milky Way is a rope thrown up to help two sisters avoid the incestuous advances of their father, the crocodile-man, who is often associated with Scorpius in that part of the world.

\subsection{Crux: the Southern Cross}

\begin{figure}[h]
\begin{center}
\includegraphics[width=6cm]{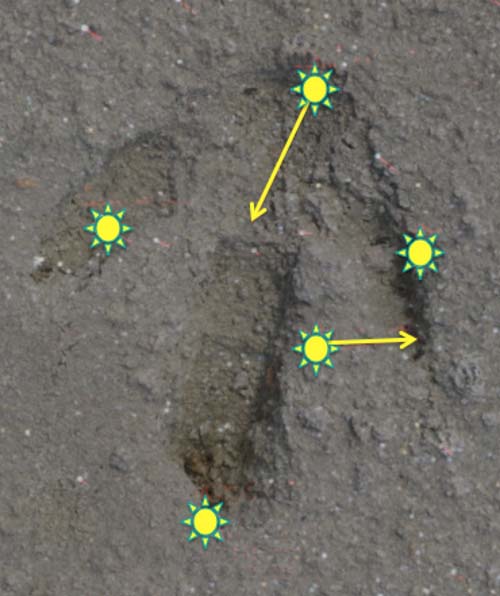} 
\caption{The stars of crux (the Southern Cross, showing their measured proper motions, so that, more than 10,000 years ago, they did not resemble an Emu's foot. Adapted from  \cite{hamacher11a}}
\label{crux}
\end{center}
\end{figure}

The Southern Cross (Crux) is one of the best-known constellations of the southern sky. It is curious that, while Orion and the Pleiades have similar stories throughout Australia, the interpretation of crux seems to vary markedly from group to group. 

For example, Kimberley people say that Crux is an eaglehawk \citep{kaberry39}, while  \cite{maegraith32} reports that the central desert people (Arrernte and Luritja) include the stars of Centaurus in the eaglehawk of crux.  \cite{mountford76} 
reports that the central desert people say that it's an Eagle, the Coalsack is the EagleÕs nest and the two pointers ($\alpha $ Cen and $\beta$ Cen) are the Eagle's throwing sticks. Yolngu people say that the Coalsack is a rock-cod  \citep{wells64}, and the Boorong people  say that the Southern Cross is a tree, and $\gamma$ Crux (the star at the top of the cross) is a possum, who has just been chased up the tree by an emu, represented by the Coalsack, whilst the emu in turn is being chased by two brothers who are the pointers  \citep{stanbridge57}. Kamilaroi people describe Crux as a gum-tree  in which the first man to die was laid. The tree was then lifted into the sky by a spirit, and the eyes of the spirit and the dead man became the stars of Crux, followed by two white cockatoos (the pointers) chasing their tree into the sky  \citep{parker05, parker14, sveiby06}.

The Murrawarri tell a story in which a rival group tried to steal the sacred fires, resulting in people from both groups, together with the fire, being lifted into the sky, where they became Crux, and the Pointers are two guards  \citep{n311, mathews94}.
Crux is particularly important to the Wardaman people  \citep{DS}, and its orientation defines their calendar and their annual cycle of dreaming stories.

 \citet{clarke15a} report that the Kamilaroi interpretation of crux is a large river red gum tree, typical of those that grow along the inland creek systems  \citep{ash03}, while in the Anmatyerr language of Central Australia, the Southern Cross is called {\it Irrety Ingka}, meaning `eaglehawkÕs foot'  \citep{green10} or {\it Iritjinga}  \citep{maegraith32}.

It was noted above that Orion and the Pleiades have very similar stories across Australia, and around the world. So why should the equally conspicuous constellation of the Southern Cross seem to have a different story in nearly every Aboriginal language group?
 If the hypothesis  \citep{n316} regarding the extreme antiquity of the Pleiades and Orion stories is correct, this provides a natural explanation. Crux is very much further south than Orion and the Pleiades. To reach Australia from Africa, the humans first travelled  north to the Middle East, and then down south through Asia before reaching Australia. Orion and the Pleiades would have been visible throughout this journey. However, crux would have been below the Southern Horizon during the time they were in northern latitudes. This part of the journey probably took  tens of thousands of years  \citep{hudjashov}, and any stories associated with crux may have been forgotten during this time, and so a diversity of new stories were created.

\subsection{Scorpius}
\label{scorpius}
Scorpius is one of the most distinctively shaped constellations in the sky,  with a long curved tail extending from a `head', and many cultures, both in Australia and elsewhere, identify it as a scorpion.
 \cite{n311} report a kamilaroi story in which the appearance of Scorpius heralds the  the dry season, when scorpions are most prevalent. 

Many Australian cultures identify Scorpius  as a crocodile, as shown in Fig. \ref{art}. For example, the constellation is known in Arnhem Land, where crocodiles are plentiful, as Baru, which is the Yolngu word for crocodile. Other groups identify Scorpius as an Eagle  \citep{mathews04} or a brush turkey  \citep{n311}.  \cite{n311} also report a Kamilaroi tale that meteors come from Scorpius, perhaps reflecting the radiant point of the $\alpha$ Scorpiids and the $\omega$ Scorpiids.  \cite{maegraith32} tells a complex story of Scorpius, involving several actors, including Antares as the `red ochre woman'.  \cite{mountford76} tells a central desert story of a young couple fleeing after breaking traditional law, in which the stars in the tail of Scorpius are the tracks of the men pursuing them. 

Like many constellations, the heliacal rising of Scorpius is used in several cultures to mark the transition from one season to another 
 or to mark the time when a food source becomes available. In Arnhem Land, the heliacal rising of Scorpius is also said to mark the arrival of the Macassans, who from before 1640 to 1907 used to sail from  Indonesia to catch Trepang (sea cucumber) and also traded with the Yolngu people  \citep{orchiston16}.

Other Kamilaroi stories  \citep[e.g.][]{n311} suggest the dark clouds close to Scorpius are more important than the stars of Scorpius, 
and can represent demons, tree-carvings, initiation scars, a whirlwind, or the tail and claws of Scorpius.   \cite{parker05} report a Kamilaroi story in which a travelling route has to pass two dark spots in Scorpius, which are devils to catch spirits of the dead. 

Curiously, Scorpius is said to be a crocodile amongst the Kamilaroi people of NSW  \citep{parker05, sveiby06, n311}, even through crocodiles have not been known in NSW for 40,000 years  \citep{gillespie01}.  A Kamilaroi tale  \citep{n311, sveiby06} explains how Narran and Coocoran Lakes in NSW were formed when Baiame killed a crocodile which had eaten his wives. 
So why do Kamilaroi identify Scorpius as a crocodile, and have stories about crocodiles?
Possible explanations are (a) it represents a distant memory of the time  when crocodiles were common in NSW, or (b) the story and identification have been brought  from further north by Aboriginal travellers, or (c) it is a story imported in modern times. (c) seems unlikely, as there are several  traditional Kamilaroi stories embodying the crocodile. (b) is particularly interesting given the argument mentioned above by  \cite{tindale83} that the Kamilaroi are linguistically close to people from the north of Queensland, and may have migrated to NSW from there. Perhaps the crocodile stories are a distant memory from when Kamilaroi people lived in northern Queensland.

\subsection{The Emu in the Sky}
\label{emu}
\begin{figure}[h]
\begin{center}
\includegraphics[width=6cm]{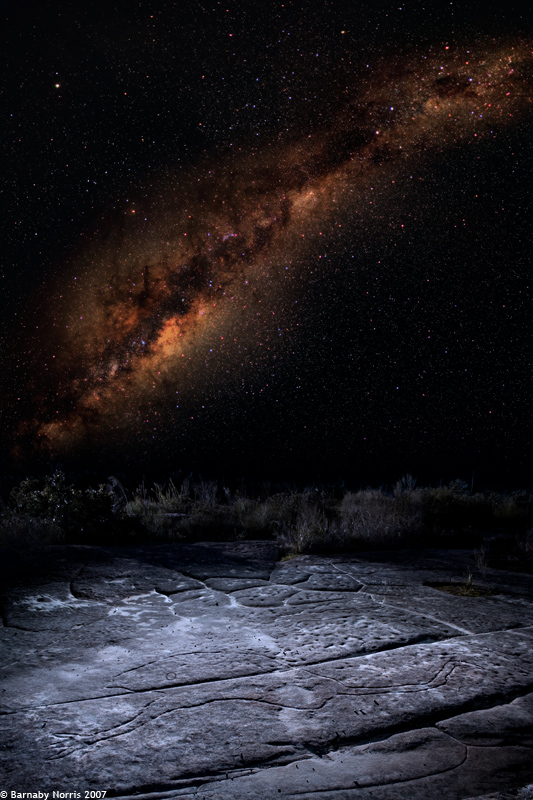} 
 \caption{An image of the `Emu in the Sky' constellation, taken above the Emu engraving in Kuring-gai-Chase National Park, NSW, as it would appear in an August evening before the European occupation of Australia. This image, commissioned by the author for the cover of the book `Emu Dreaming', required several months of work by Barnaby Norris, as described by  \citet{a111}, and won a `Eureka' award. The engraving below is thought to be a representation of the Emu in the Sky, as discussed further in \S\ref{rockart}.}
\label{emupic}
\end{center}
\end{figure}

\begin{figure}[h]
\begin{center}
\includegraphics[width=6cm]{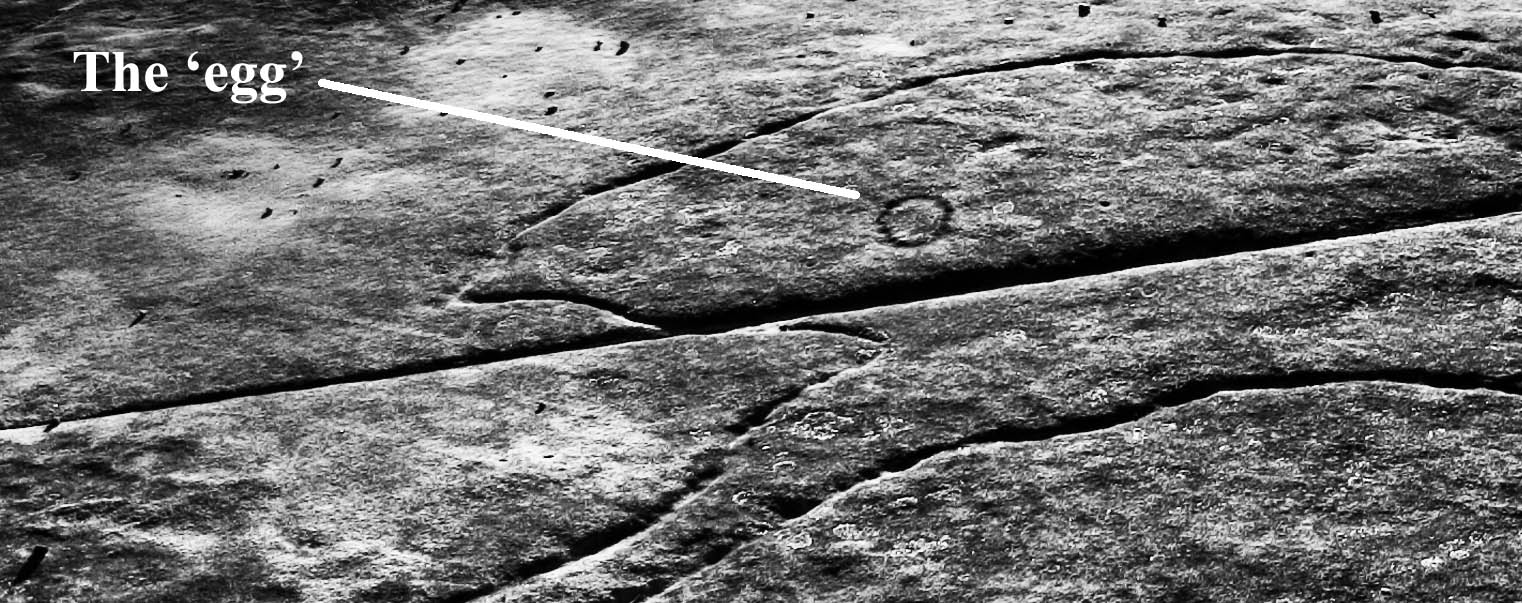} 
 \caption{A detail from Figure \ref{emupic}, showing the `Emu Egg' which was first recorded during the making of Figure \ref{emupic}, despite the emu engraving having previously been recorded many times by archaeologists. The egg became obvious as a result of using powerful low-level flash lamps for the photography. It has since been `re-grooved' by National Parkes and Wildlife Service, and is therefore now obvious to the casual visitor.}
\label{emuegg}
\end{center}
\end{figure}


The emu is a large flightless bird that is found throughout most of Australia. It is unusual in that, once the female has laid the eggs, the male takes over all care and feeding of the eggs and then the chicks, while the female wanders off and never sees her chicks hatch. Many Aboriginal groups draw an analogy with human male babies, in that, while females produce babies, the men turn the male babies into men, through the process of education and initiation. Thus the emu is often associated with initiation ceremonies. The relationship between the emu and initiation ceremonies is discussed further in \S\ref{bora}.

The emu's representation in the sky, the `Emu in the Sky',  is perhaps the best known Aboriginal constellation, and is known by many different Aboriginal cultures across Australia.  It   consists not of stars but of the dark clouds within the Milky Way.There are several variants, but Figure \ref{emupic} shows one of the most common interpretations
 \citep[e.g.][]{basedow25, DS, n318, love87, massola63, ED}, in which the dark cloud known as the Coalsack forms the head, and the body extends along the body of the Milky Way, through the constellations of Scorpius and Sagittarius. As viewed from the Earth, the Emu subtends an angle of $\sim$ 90$\deg$, which is far larger than any European constellation, and is a magnificent sight when viewed in a  dark sky.

In this interpretation, the Emu's legs stream out behind it, which is curious, as the legs of real-life emus   are always underneath the body, even when the emus are running. This discrepancy is explained in stories which are sacred and sadly cannot be discussed here. In one public version  \citep{reed78}, 
the legs stream out because the emu is flying, because emus used to be `equipped with powerful wings and spent their entire lives disporting themselves above and through the clouds ... Not one of them had ever landed on the earth'. They lost the ability to fly when a female emu landed on earth and was tricked by a brolga into taking off her wings, after which her descendants were never able to put them back on.

Another interpretation from the Boorong people  \citep{stanbridge57} and Kamilaroi people  \citep{ridley73} describes a smaller emu which is identified as the Coal Sack.  Both  \cite{DS} and  \cite{n318} describe how the precise interpretation of the Emu in the Sky can vary depending on the time of year.
There are also many other sacred interpretations that cannot be discussed here.

Interestingly, Indigenous Brazilians identify  the same dark clouds identified above as a large `Emu in the Sky' (i.e. Crux through to Scorpius) as a `Rhea in the Sky'  \citep{afonso09}.

\section{Planets}

Aboriginal elders had an intimate knowledge of the sky, which was passed down from generation to generation. 
To anyone with such a detailed knowledge of the sky, the difference between the fixed stars and the wandering planets is blatantly obvious, and planets therefore naturally receive special attention in Aboriginal cultures. The ecliptic, called Yondorrin in the Wardaman language, also has special significance: `The planets come straight across like you and I doing walk, pad up and down, walking backwards, forwards, make a little track there, a pad... Pad is straight across country. Yondorrin shows how!'  \citep{DS}.

Five planets (Mercury, Venus, Mars, Jupiter, and Saturn) are visible to the naked eye, and all were recognised by Aboriginal groups \citep{mountford58, mountford76, DS}. The Tiwi  believed that  Mercury, Venus, Mars, and Jupiter were the wives of the Moon-man. In many Aboriginal cultures, Venus appears to be far more significant  than any other planet.

\subsection {Mercury}
Few stories about Mercury are known, but the Tiwi see it as one of the wives of the Moon-man   \citep{mountford58} and the Wardaman see it as a little girl who is wronged by the Moon-man  \citep{DS}. 

\subsection {Venus - the morning star}
\label{venus}

\begin{figure}[h]
\begin{center}
\includegraphics[width=2cm]{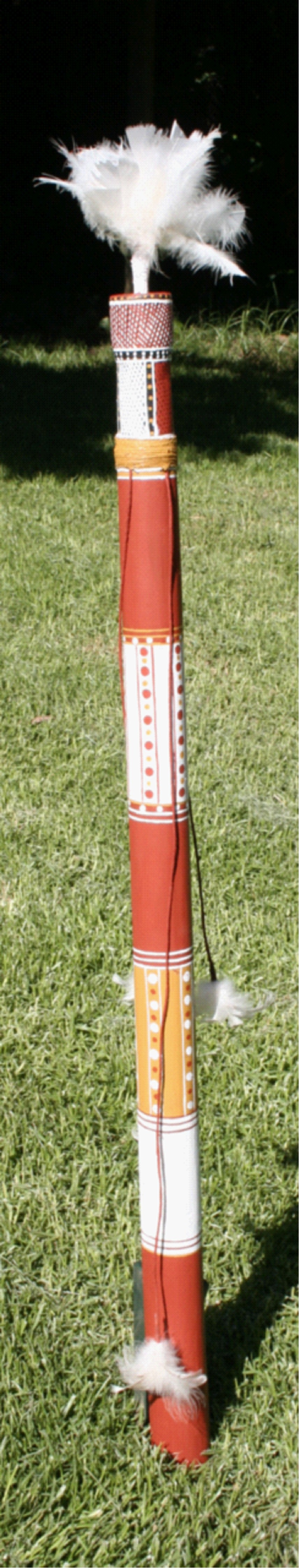} 
 \caption{A `Morning Star Pole', which is used by Yolngu people in a morning star ceremony, in which they communicate with Barnumbirr, or Venus. The tuft of magpie-goose feather on top represents Venus, and the other tufts, on pandanus strings, represent other stars close to Venus. This pole was made by Yolngu artist Richard Garrawurra, from Elcho Island.}
\label{pole}
\end{center}
\end{figure}

Because the orbit of Venus in inside that of the Earth, Venus is always seen close to the Sun. Because the sky is too bright to see Venus when it is high in the sky, Venus is only ever visible either as a morning star just before sunrise, or as an evening star just before sunset. In most Aboriginal cultures, Venus as an evening star is regarded as being different from Venus as a Morning star, although they are recognised as being related, or complementary  \citep[e.g.][]{berndt48}.

Venus as a `Morning Star' is particularly important in several Aboriginal cultures, and is known as Barnumbirr in Yolngu culture. She is often associated with death, and is said to guide the spirits of the dead to their spirit-world  \citep{allen75, tindale83}. Barnumbirr was a creator spirit who left her island of Baralku to lead the first humans to Australia  \citep{ED, wells71}. After crossing the coat of Australia, she continued flying across the land, describing the land below her in great detail, naming   and creating the animals and places. As she flew  westwards across the land, she named waterholes, rivers, and mountains   in considerable detail, including defining the territory of  clans, and the areas where people had fishing rights. Her song therefore not only forms a basis of Yolngu law, but describes a navigable route across the land. The path that she followed is now known as  a `songline', or navigational route, across the Top End of Australia, so that her song is effectively an oral map. 

The Yolngu explanation \citep{allen75,ED,wells71} for the fact that  Venus is never seen  high in the sky is that  Barnumbirr's sisters tied a rope to her  to prevent her flying too high and perhaps getting lost. This rope appears  \citep{ED} to be the faint conical glow of the zodiacal light, caused by extraterrestrial dust in the solar system, which is readily visible in the clear dark skies and low latitude of northern Australia. 

\begin{figure}[h]
\begin{center}
\includegraphics[width=7cm]{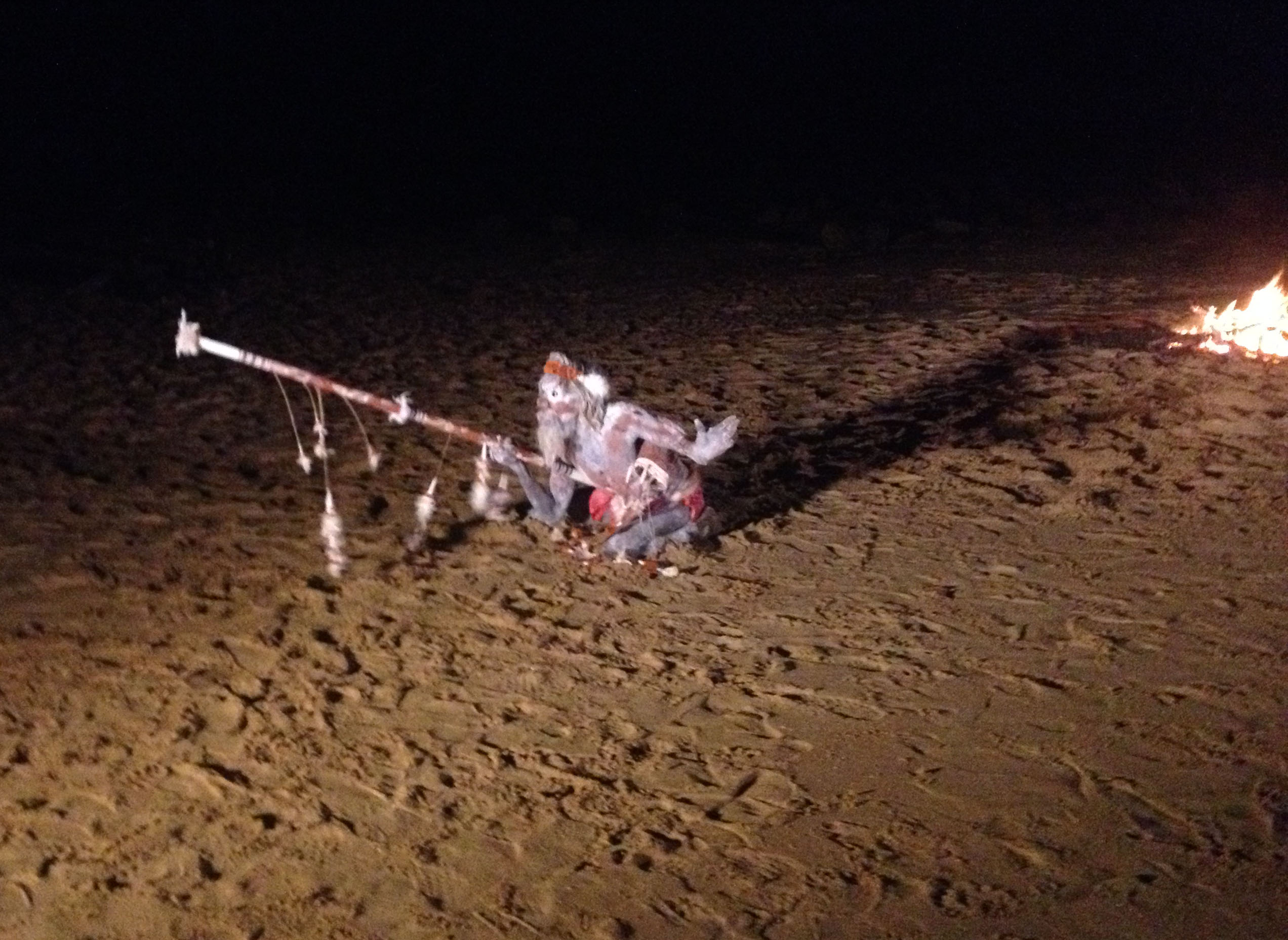} 
\caption{A Yongu elder, Gali, performing the Morning Star Ceremony on Elcho Island.}
\label{gali}
\end{center}
\end{figure}

Some Yolngu clans still hold the ÒMorning Star CeremonyÓ (see Fig, \ref{gali}) as part of the funeral process. This ceremony starts at dusk and continues to a climax as Banumbirr (Venus) rises before dawn. The ceremony features a `Morning Star pole' (see Figure \ref{pole}), which is a pole about two metres long, decorated at the top with Magpie Goose feathers, representing Venus and also the Lotus flower, and other tufts of feathers representing other stars near Venus. During the ceremony, the `Morning Star Pole' is used to help the participants communicate with their ancestors, with Banumbirr's help. The communication is said to be via the morning star pole to Venus, and then down the rope to Baralku, where the ancestors live.

In summary, there are several important features of the Yolngu Morning Star ceremony:
\begin{itemize}
\item Barnumbirr's song is effectively an oral map, enabling navigation along the `song-line'
\item Yolngu tradition includes the knowledge that Venus never moves far from the Sun, which is explained by a rope (the zodiacal light) binding Venus to the land. 
\item since the Morning-Star ceremony needs to be planned, and Venus rises before dawn only at certain times (roughly every 1.5 years), Yolngu people also keep track of the path of Venus well enough to predict when to hold the Morning Star Ceremony.
\end{itemize}

 \cite{n311} report a Kamilaroi story of  an eagle hawk  who lived in a giant yarran tree near the Barwon River, and hunted people for food, until some young men  set his home on fire, and he died, becoming the Morning Star. Other Kamilaroi stories associate Mars and Venus as the (red and blue) eyes of the creator-spirit Baiame.

\subsection{Venus - the evening star}
\label{venus}
Venus spends about 45\% of its time as `the Evening Star' and about 45\% as `the Morning Star', with the remaining 10\% of its time too close to the Sun to be observed with the naked eye. 
The question is sometimes asked whether traditional Australians recognised that the Evening Star and the Morning Star were the same object. The question immediately becomes a semantic and cultural one, as while it was widely recognised by traditional Australians that these two objects are closely related and in some sense complementary  \citep[e.g.][]{berndt48, ED}, calling them `the same object' only makes sense if they are recognised to be a body in orbit around the Sun. From a purely observational point of view, they are obviously {\bf not} the same object - one appears in the morning and the other appears in the evening!

 \cite{n311} report the names of Venus used in NSW, and describe how Venus is known as the laughing star, a rude old man, possibly referring to the twinkling of Venus when it is close to the horizon.

In Yolngu, Venus as an evening star is called `Djurrpun' but is recognised as being closely related, or perhaps complementary, to Barnumbirr, the morning star. The morning star is prevented from rising too high by a rope attached to the mythical eastern island of Baralku, whilst the evening star is prevented from rising too high by a rope held by the `spirits in the west'  \citep{berndt48, n287}.

The interpretation is  complicated by reports  \citep{wells64, wells73c}  that the evening star indicates the time when {\it rakai} (also {\it raika} or {\it rakia} or water chestnut or {\it eleocharis dulcis})
should be harvested. Even the meaning of {\it rakai} is unclear, with both spike rush  ({\it eleocharis palustris} \citep{haynes92, mountford56} and lotus (also called a `lily' in Arnhem Land, or {\it Nelumbo nucifera})  \citep{ED} being suggested as translations. The confusion arises partly because the lotus bulb (called {\it dirpu} in Yolngu)  is ready for digging just after the {\it rakai} is ready  \citep{wells64}.
But in any case,  Venus may appear as an evening star at any time in the year and therefore cannot indicate the time of year for harvesting a plant. While investigating this, I was told  \citep{n230, n219, ED}  by the traditional custodian of this story  \citep{ac4} that the Evening Star in this context is not Venus, but the star Spica, which does indeed set just after the Sun at the time of year when {\it rakai} nuts are ready for harvesting. 

On the other hand, I was also shown a  `Evening Star rope', known as {\it Laka}  \citep{n219} which was made of pandanus twine, twisted together with possum fur and orange lorikeet feathers. Woven in to the rope were the yellow-white marbles of the {\it rakai} nuts. Laka was said to be closely related, and complementary, to the `Morning star pole', and yet was also associated with the harvesting of {\it rakai}. It was said to be used in a  funeral ceremony to send someone off to the evening star. In the ceremony, the mourners hold  the rope on their heads, joining their spirits to that of their deceased as they say farewell.

The identity of the `Evening Star' therefore appears to be ambiguous, and may in fact refer to different astronomical objects depending on context.

 \cite{n311}  report that, amongst the Kamilaroi, the appearance of the Evening Star is a sign to light the sacred fire, which is relit every evening until Venus becomes a Morning Star. Like the Yolngu ceremony, the Kamilaroi Morning Star ceremony also involves a `Morning Star pole', but the Kamilaroi pole it is used horizontally, and represents a bridge between the  moieties. 

 \citet{tindale83} does not distinguish between Venus as a morning star and as an evening star, but reports that in the Central desert, several groups associate Venus with bringing life-supporting rain, and gives a detailed description of a ceremony to encourage Venus to bring the rains.

\subsection{Mars}
 \citet{tindale83} reports a story from SA that Mars was a younger brother of Nepele, a major creator being.  Shortly after being initiated, and thus covered in  red ochre, he stole Nepele's wives and escaped with them into the sky, where all three now wander. So the red colour of Mars comes from the red ochre, and the two wives may be Jupiter and Saturn. 

Most groups viewed Mars as male  \citep{fredrick08}, although Mars was female to the Gundidjrnara people  in Victoria  \citep{dawson81}. 

\subsection{Jupiter}
 \citet{tindale83} notes that Jupiter is generally male and often associated with the Moon-man and with fire-making, and recounts  a story from SA in which Keibara, a bustard-being, tries to take fire from humans and eventually ends up as the planet Jupiter.
A story found across NSW  \citep{fraser88, fraser01,n311,tindale83}, 
 says that Jupiter is a young boy roaming the heavens. His mother, the Sun, sent men to spear him at a time while he was  low  in the western sky. Another Kamilaroi story  \citep{mathews04, mathews05} 
says that Jupiter had a reddish colour because  he ate  roasted yams, although there is some evidence that this may have referred to Mars  \citep{n311, johnson98}. 

\subsection{Saturn}
There are few recorded stories about Saturn. \cite{mountford76} reports that the Central Desert people regarded Venus and Saturn as two brothers, and Jupiter was their dog, and Saturn and Jupiter spend much of their time finding food for Venus. \cite{ridley73} reports that the Kamilaroi regard Saturn as a small bird called Wuzgul.

\section{Comets and Meteors}
Some Aboriginal languages are reported not to distinguish between meteors and comets, although this may represent confusion in the mind of the recording anthropologist rather than confusion by the people who speak the language.  For example,  \citet{tindale83} does not distinguish between comets and fireballs, and a story about comets reported by  \citet{spencer99} appears to be identical to a story about meteors reported by  \citet{piddington32}. Detailed studies  \citep{n242,n244,n230} report  many different views of comets and meteors, although both are often  associated with bad omens and death. In some cases  \citep{hamacher11b} they are manifestations of spells intended to cause death.

In general, comets and meteors are viewed in  many traditions, both in Australia and in other countries, as unexpected events that disrupt the orderly flow of the heavens and are therefore probably bad things  \citep{clarke09a, hamacher14b, howitt04}.
 \cite{collins98} says of the Eora people in Sydney `To the shooting of a star they attach a degree of importance; and I once, on an occasion of this kind, saw the girl Boo-roong greatly agitated, and prophesying much evil to befall all the white men and their habitations'.

\subsection{Meteors}
 \citet{n242} report many stories and beliefs associated with meteors, many of which are associated with death, although some have positive connotations and some negative. Several groups associate the appearance of a meteor to signify someone's death  \citep[e.g.][]{n311, kaberry35,piddington32}  with the direction of the meteor often indicating where they have died.
 \cite{harney09c}
tells how, after someone dies, and their body is placed in a forked stick to decay, a meteor will shoot from their body back to his country, so people there know that he has died. Meteors can also {\em cause} death:  \cite{harney59}
 tells how a meteor is used to punish by death the breaking of law, and   \cite{peck25} reports a Kamilaroi story in which a `great bright light, burning blue, travelling at enormous rate' killed a group of people.   \cite{allen75} report a Yolngu story of a lonely fire spirit who came to earth as a meteorite to bring fire to the people, but accidentally caused massive fires and destruction to the people when he accidentally touched the Earth. Even now, to touch or smell the hot ashes from a recently-fallen meteorite will cause death  \citep{wells71}. 
The Tangani people (SA) believed that a devastating smallpox epidemic had been brought by a meteor-man from  {\it crux} \citep{clarke03a, tindale31}.

Meteors can also signify the creation of a new life. In Kimberley and Kamilaroi cultures, the same meteor that signifies death can also bring a new baby  \citep{akerman14,n311}. In the Wardaman reincarnation cycle \citep{harney09b}, dead spirits go up to the Milky Way through the star Vega, which functions as a gateway, and then go through a series of ceremonies in Sagittarius, led by the star Altair (identified by the Wardaman as a wedge-tailed eagle), in which they are mentored by the star Arcturus (identified as a rock cod). After completing the ceremonies, they fall to the Earth as a meteorite, and make their way to a creek, where they are fed algae by the rock cod while they wait for their mother to pass by. When she does, they leap up into her, and she becomes pregnant.  \cite{n311} report a similar story from the Kamilaroi culture: the spirit of a dead person flies up to the sky, and after a stay there, returns to Earth as a meteorite. The spirit then waits behind a yarran tree in a birthing area.  \citet{basedow25} notes that Aboriginal people on the `north coast' believe that a spirit returns as a meteorite, and then searches for its old skeleton.

 \cite{mountford58} 
reports that the Tiwi people believe that meteors are the  Papinjuwaris, a group of ghoulish meteor-men who live in the far west, where the sky meets the Earth. The Papinjuwaris carry corpses to their land where they eat them, and they drink the blood of sick people, causing them to die. This story has a striking resemblance to the Luritja `debil-debils' reported by  \cite{hamacher14a} from people living close to the Henbury meteorite crater. Similarly, a Kuninjku tale from Arnhem Land says that `The malevolent Namorrorddo (shooting star being) has huge clawed hands to grasp the souls of humans and rush away with them across the sky.'  \citep{agnsw04}.

Annual meteor showers have not been reported in the literature of Aboriginal astronomy, with two possible exceptions. First is the  Kamilaroi story noted in \S\ref{scorpius} that meteors come from Scorpius, perhaps reflecting the radiant point of the Scorpiid meteor showers. The second is the speculation by  \cite{morieson07}, noting the Boorong association between the Mallee fowl and the constellation Lyra  \citep{stanbridge57},  that the Lyrid meteor shower in April may be associated with the way the male Mallee fowl kicks showers of pebbles from the egg-mound to prepare the mound for egg-laying.


\subsection{Meteorite Craters and Impacts}
Many Aboriginal stories  \citep[e.g.][]{berndt89, jones89,peck25} describe a meteor falling to earth, causing a crater. For example,  Tnorala (Gosse's Bluff), a 142-million-year old crater, is said to be  caused by the falling baby of the Morning Star (the father) and the Evening Star (the mother) \citep{n243}. Similarly, the Wolfe Creek crater has a local Indigenous tradition that it was formed by a falling rock, which is tempting to regard as a folk-memory of a historical event, except that the crater was formed 300,000 years ago, before any humans came to Australia  \citep{hamachergoldsmith13, hamacher14b}.  Anecdotal evidence from other known meteorite impact sites is also frequently reported  \cite[e.g.][]{hamacher14a} although these have to be compared with similar stories from areas {\em without} any evidence for a meteorite impact to assess their significance.

\begin{figure}[h]
\begin{center}
\includegraphics[width=7cm]{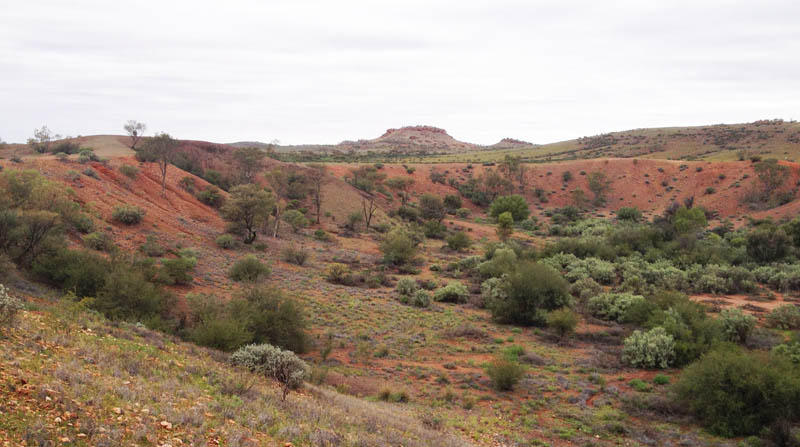} 
\caption{The largest of the meteorite craters at Henbury, NT. The impact, 4700 years ago, was probably witnessed by local Aboriginal inhabitants.}
\label{henbury}
\end{center}
\end{figure}


Nevertheless, there are four meteorite impact craters that were formed since humans came to Australia, and  \cite{n243} explored whether any memories of their formation remain in oral tradition. Although they found some stories associated with local meteorite craters, they also found many such stories in areas where no meteorite impacts had been recorded, and concluded that no more stories were found near known impact craters than anywhere else, suggesting that the stories were randomly distributed and not associated with particular craters. For example, they did not find any significant local stories associated with the best-known meteorite crater to be formed within human habitation, at Henbury, NT, which was created about 4700 years ago.  \citet{alderman31} states that `None of [the local Aboriginal people] had any ideas as to the origin of the craters ... [and] showed no interest in the craters' and  \cite{harney63}
 was unable to find any stories associated with the crater, although he reported that  `the area was under some sort of a taboo that had been handed down from ``long time ago'''.
However,  \cite{hamachergoldsmith13} and  \cite{hamacher14a} disagree with the earlier studies, having found stories that were not previously available. It appears that people living near the Henbury crater do indeed have an oral tradition that recalls the devastating impact which no doubt killed a significant number of their population. 
Aboriginal oral traditions have even helped to identify a previously unknown meteorite fall \citep{hamacher14a}.

\subsection{Comets}
Comets were generally regarded as bad omens  \citep{n244}. For example, Comet C/1843 D1, visible in SA in March 1843, was thought to have been created by northern sorcerers (possibly the Pitjantjatjara people) to kill European settlers who had imprisoned an Aboriginal man  \citep{clarke90, clarke97a, clarke03a, eyre45}.  \cite{kunari86} said that the Pitjantjatjara believed comets to be a manifestation of a large dangerous man called Wurluru, who eats flesh raw, including human flesh.  \citet{n244} have compiled an extensive catalogue of Aboriginal perceptions of comets, such as that of the Euahlayi who regarded comets as bringers of drought  \citep{parker05}. Several groups associated comets with smoke, spears, or the Rainbow Serpent.  \citet{trezise93} even suggested that the Rainbow Serpent  may have originated with the 76-year cyclic appearances of HalleyÕs Comet.  The Wathaurung were said to regard comets as spirits of those who have been killed away from home, and are making their way back to their own country \citep{morgan52}. 

\subsection{Representation of Meteors and Comets in Art}
\begin{figure}[h]
\begin{center}
\includegraphics[width=7cm]{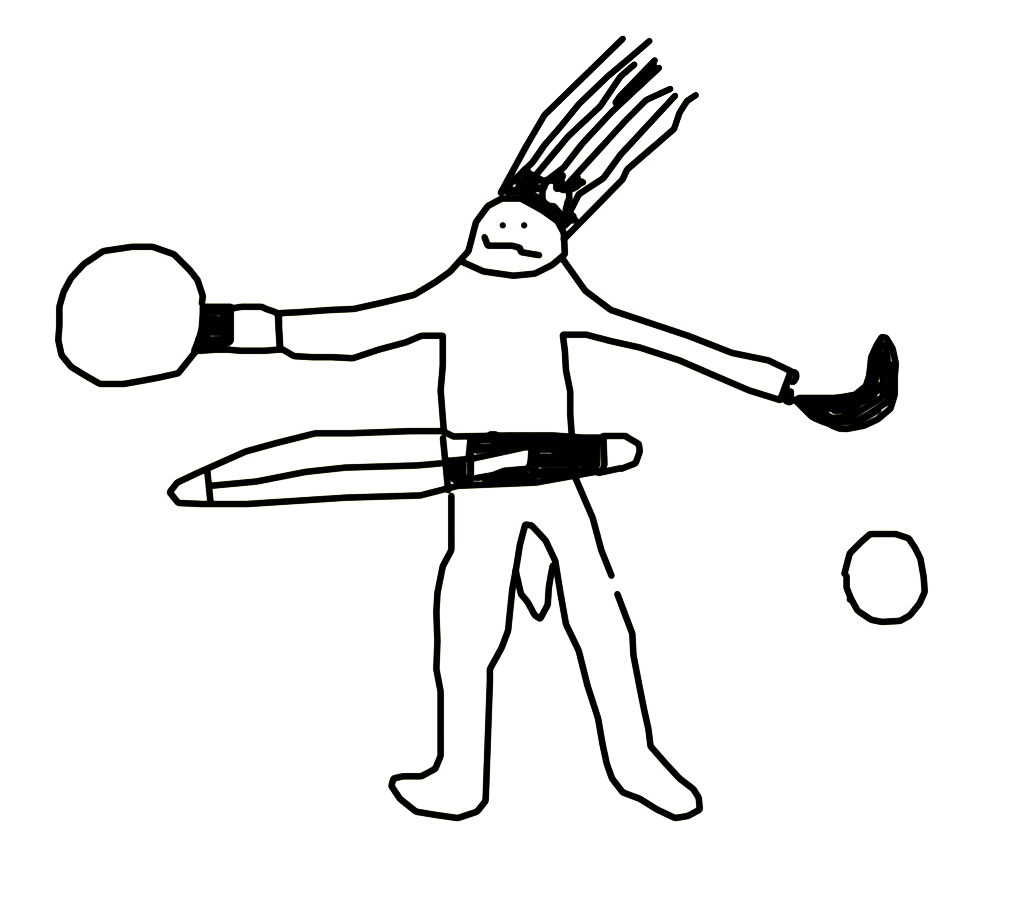} 
\caption{A possible representation of a comet in Sydney rock engravings: Bulgandry, near Woy-woy, from  \cite{n220}}
\label{bulgandry}
\end{center}
\end{figure}

There are at least 150 separate traditional stories of meteors, and at least 25 of comets  \citep[e.g.][and references therein]{n255}, but there are few examples of depictions in  art  \citep[e.g][]{jones89} although Figure \ref{art} shows one possible example.
Amongst Sydney Rock Engravings are several representations (e.g. Fig.\ref{bulgandry}) of culture heroes with distinctive hair arrangements that resemble comets  \citep{n220, elkin49}, in which case a comet may have been viewed as an apparition of the creator-spirit. Other engravings elsewhere in Australia also resemble comets  \citep[e.g.][]{clark86}.

\section{Other Galaxies}

\subsection{The Magellanic Clouds}
Several stories  \citep[e.g.][]{haynes96,mountford56,ED}  report  that the Magellanic Clouds are the campfires of an old couple, living far from the young people,  who live along the banks of the Milky Way and whose campfires are signified by nebulae. The younger people hunt fish and gather food from the Milky Way, and bring them to their old people to cook on their fire. Other stories   \citep{mountford56} identify them as two sisters, or a mother and daughter, or two men, or two spirits.

 \cite{n311} report several Kamilaroi stories that the Magellanic Clouds were a place where dead people went.  \cite{howitt04} reported a  Kamilaroi story that a dead man's spirit goes to a dark patch (called endless water or river) in the Magellanic Clouds.  \cite{n311} report a Kamilaroi women's story that the Magellanic Clouds were openings to heaven, and  \citet{kaberry39} reports a Wolmeri story that they are the camp of the Moon-man, and a Lunga story that they are poisoned fish.

Other stories of the Magellanic Clouds identify them as a pair of brolgas    \citep{massola68,ridley73, stanbridge57} while others identify them as two brothers who collect the spirits of dead people  \citep{bates04,mountford76}, or two culture heroes \citep{piddington30}. Adelaide Aboriginal people believed that the Magellanic clouds were the white ashes of two rainbow lorikeets which had been treacherously killed  \citep{clarke90}, and  Kimberleys people say they are lily-roots that exploded while being cooked by a creator-being  \citep{akerman14}. The Wardaman people believe they are smoking hot ashes, and are  associated with rain, fog, and hail  \citep{DS} 
 and are called on in rain-making ceremonies  \citep{ac9}.

\subsection{M31}


M31, the Andromeda Nebula, is easily visible to the naked eye as far south as Sydney, where it appears for a few hours as a faint smudge above the northern horizon. In northern Australia it is  visible, but faint, well up in the sky. It would therefore be surprising if there were no references to it in Aboriginal stories.

 \citet{wells64} 
describes the Larrpan, a canoe said by Yolngu people used to carry the dead to the mythical island of Baralku: 
`To those whose eyes can look more deeply than most into the night sky, it is given sometimes to behold a tiny canoe of light. It is not a star that twinkles as do other stars, but is a soft light. As a new canoe shines when it is far away on the water and the sun is low .. so this canoe shines far away in the sky'. Wells does not mention where in the sky this canoe may be found,  but her description strongly suggests M31, although this is yet to be confirmed. This  should not be confused with the Larrpan identified as a meteor, which signifies the canoe returning to Earth.

\section{Stone arrangements}

\begin{figure}[hbt]
\includegraphics[width=8cm]{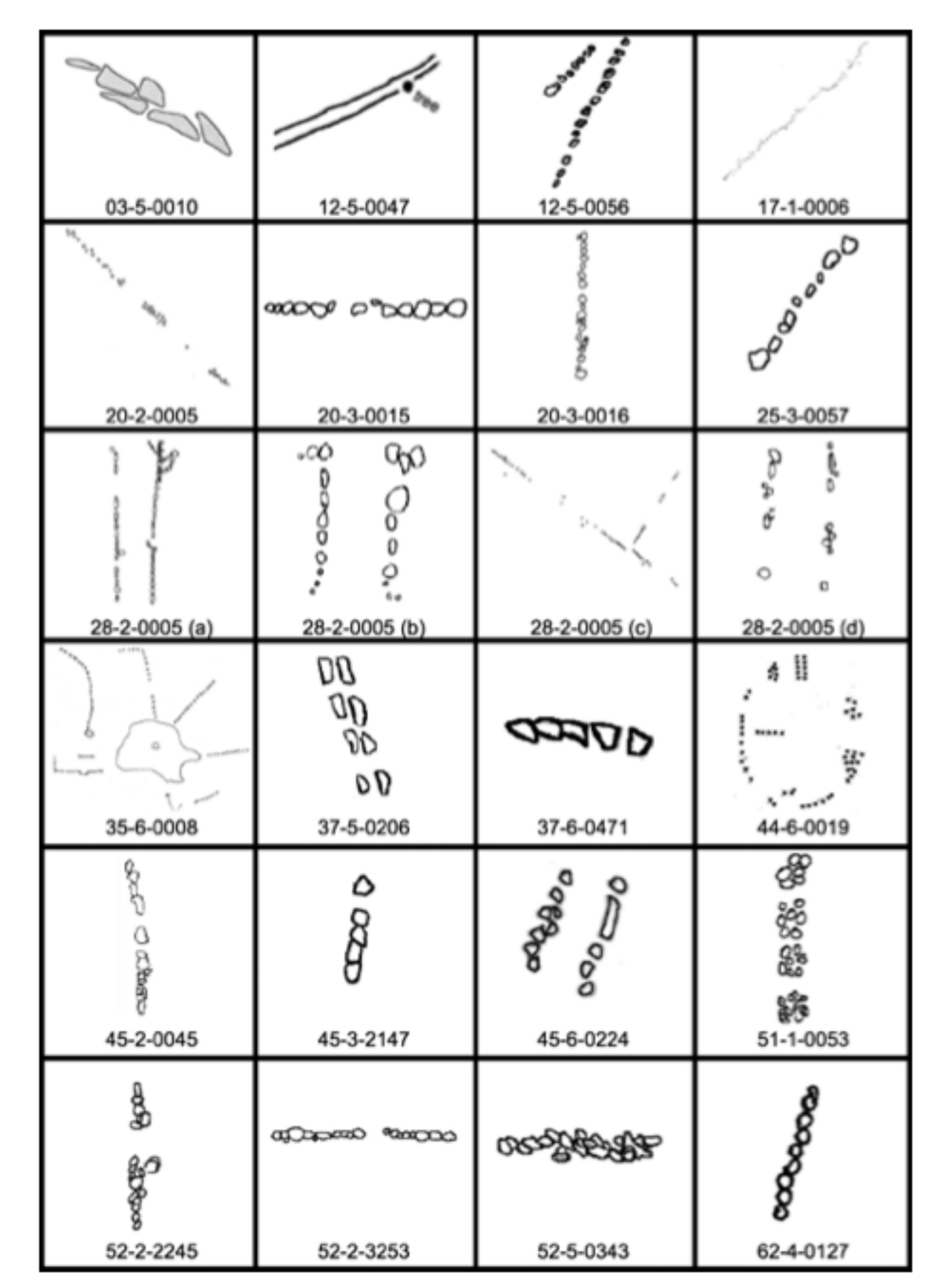}
\caption{Plans of the stone rows analysed by  \cite{n278}, and reproduced by permission of Australian Archaeology journal.  } 
\label{rows}
\end{figure}

Throughout Australia are arrangements of stones, ranging from simple lines of stones, each a few cm across, to large arrangements of large stones weighting several tons and carefully embedded in the ground, and which are  similar to the bronze-age megalithic structures found in north-west Europe. While some are clearly functional, such as fish-traps  \citep{mcbryde73}, others appear to be ceremonial  \citep[e.g.][]{mccarthy40} and may indicate the direction of a landmark, or reflect the surrounding landscape. or mimic a land feature. Many are situated in elevated positions commanding a panoramic view  \citep[e.g.][]{flood97}.
 There is widespread circumstantial evidence reported in the literature that many are aligned roughly north-south, although this had not been systematically investigated  until recently \citep[][and references therein]{n278}.

\subsection{Linear Stone Arrangements in New South Wales}
Many stone rows are known in NSW. Their use is uncertain but they are presumed to have a ceremonial function 
 \citep[e.g.][]{ attenbrow02, bowdler83,mcbryde74}.  \cite{n278} have compiled a sample of 32  of  these stone arrangements, selected to have an unambiguous direction, and some of which are shown in Fig. \ref{rows}. 
They concluded that the stone rows were unambiguously aligned preferentially north-south or east-west, but, curiously, the north-south peak seemed closer to magnetic north than to true north. This may be partly because the original surveyors did not always indicate whether their directions were relative to true or magnetic north, thus introducing an uncertainty of  $\sim$12$\deg$, corresponding to the  magnetic declination (the difference in bearing between true north and magnetic north).

\begin{figure}[hbt]
\includegraphics[width=8cm]{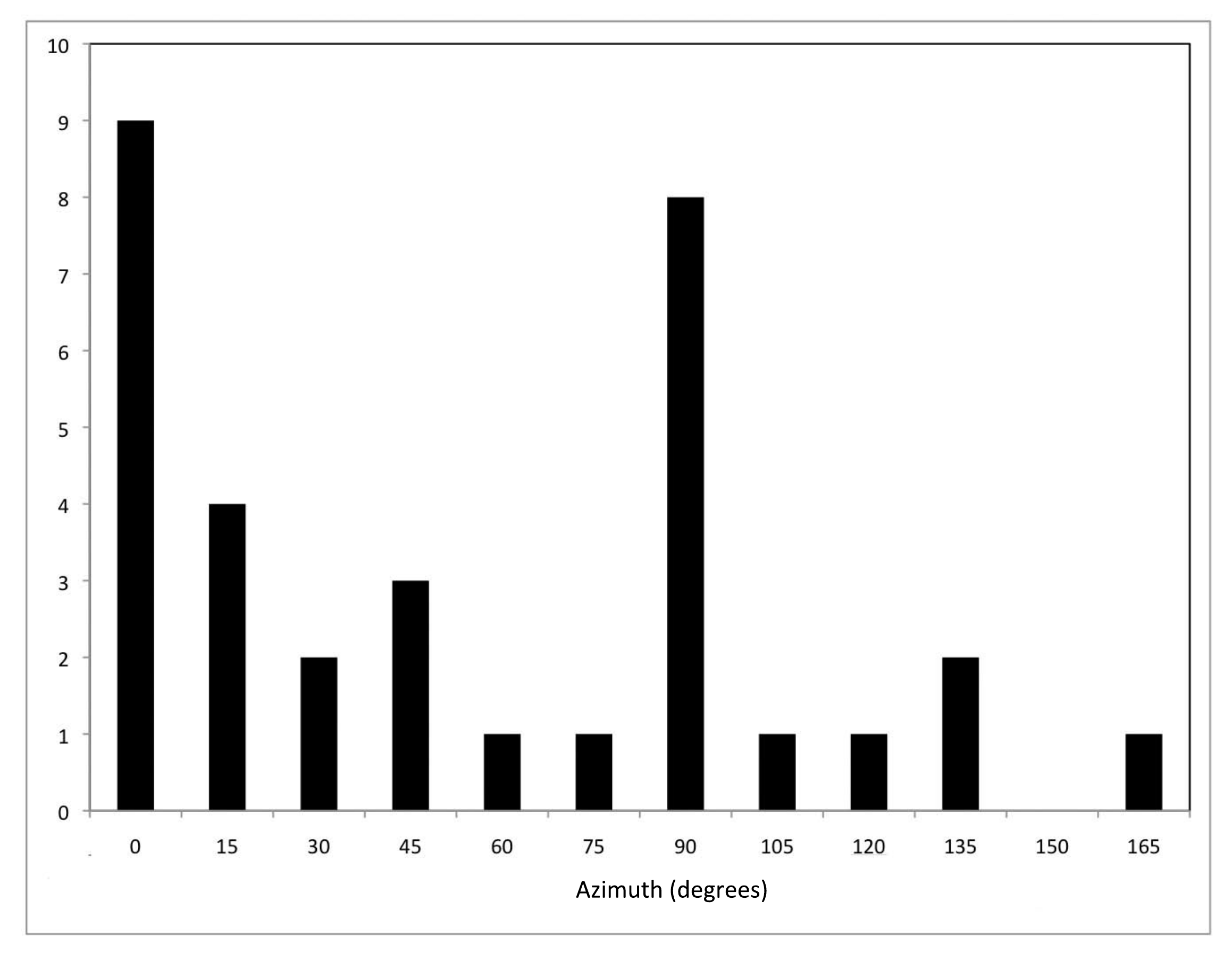}
\caption{Histogram of the orientation of the stone rows, adapted from  \cite{n278}. 0$\deg$ corresponds to north-south, and 90$\deg$ corresponds to east-west.  } 
\label{rowplot}
\end{figure}

Here I have accommodated this ambiguity by re-binning the data used by  \cite{n278} into 15-degree bins, which is larger than the $\sim$12$\deg$ magnetic declination, thus minimising the effect of the ambiguity. The results are shown in Figure \ref{rowplot}. Over half the stone rows are oriented either north-south or east-west. 

To determine whether this result could be caused by pure chance, a `Monte Carlo' test was run, in which a large number of stone rows were simulated with randomly varying azimuths. Of the one million tests conducted, less than 1000 showed peaks as high as those shown in Fig. \ref{rowplot},  so  the probability of getting the result shown in  Fig. \ref{rowplot} by chance is less than 0.1\%. It is therefore clear that, unless there is an error in either the surveys or in the site selection, the builders of the rows deliberately aligned them either north-south or east-west, with an accuracy of a few degrees. However the result by  \cite{n278} should be regarded as preliminary, and it is important to repeat this work with a larger sample and carefully surveyed data.

The significance of this result is discussed further in \S\ref{direction}

\subsection{Bora sites}
\label{bora}
Bora sites are sacred places in south-east Australia in which young Aboriginal males were initiated. They generally consist of two circles of stone or earth of different diameter connected by a pathway. The larger circle (usually to the north of the smaller one) is regarded as a public space, while the smaller circle  is restricted to the initiates and elders.  Although we have eyewitness accounts of the initiation ceremony at bora sites  \citep[e.g.][]{mathews97a}, initiation ceremonies in south-east Australia largely ceased once the traditional lands were occupied by European settlers, and many of the details of the ceremonies are lost.

There is a strong association throughout Australia  \citep[e.g.][]{berndt74, love87,love88}  between the emu and  initiation ceremonies, partly because emu eggs, once laid by the mother, are then hatched and reared by the male emus. This is seen as analogous to the initiation process, by which initiated men turn  boys into men. Similarly, there is an association between the `Emu in the Sky' constellation (see \S \ref{emu}) and the Bora sites  \citep{n318}. The Wardaman people  \citep{DS} also associate the Emu with initiation,  with initiation ceremonies being timed for when the `Emu comes down to drink' (i.e. the head of the emu, in crux, approaches the horizon).

\begin{figure}[hbt]
\includegraphics[width=8cm]{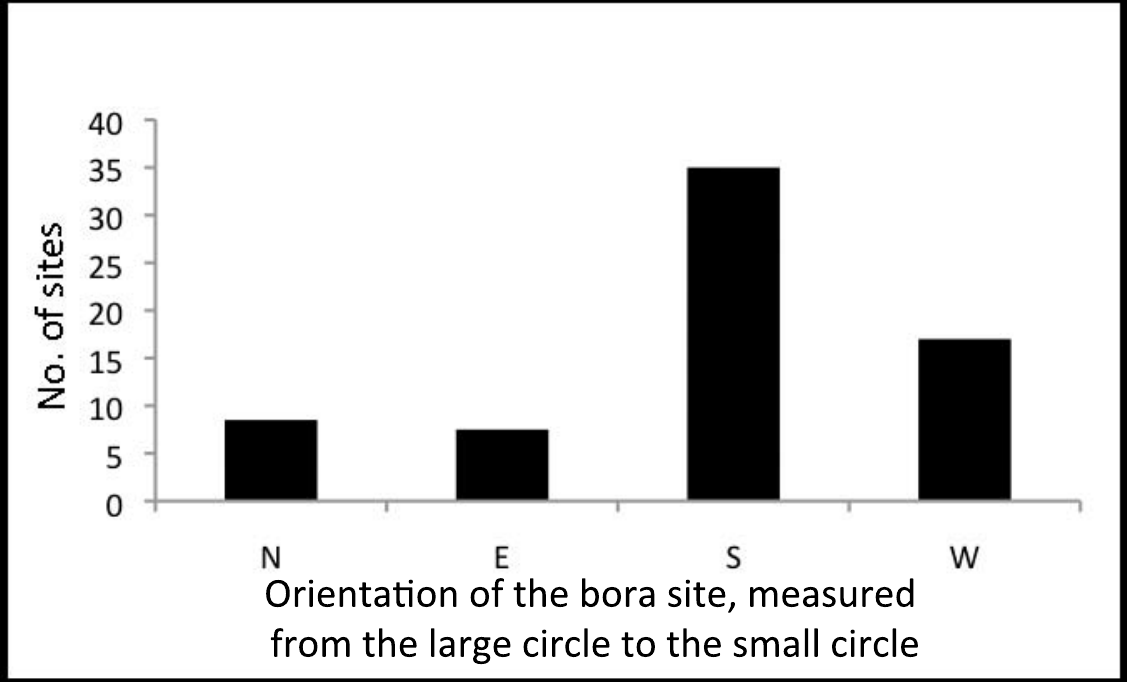}
\caption{Histogram of the orientation of the bora sites, taken from  \cite{n282}. } 
\label{borapic}
\end{figure}

For example, Fig. \ref{emupic} shows an Emu engraving at the Elvina Track initiation site at Kuringai Chase National Park, NSW.  \cite{love88} argued that  initiation ceremonies were timed to coincide with a  vertical orientation of the emu in the sky, and that the bora sites, consisting of a northern large circle and a southern small circle, was mirrored in the sky by two dark clouds, possibly the Coal Sack and the larger circle of dark clouds near Scorpius which makes up the `body' of the emu. This  was supported by the result by  \cite{n282} that bora sites are preferentially aligned to S-SW, (see Fig. \ref{borapic}) which is consistent with them being aligned on the vertical emu at azimuth 231$\deg$. Recent ethnological studies have shown this hypothesis to be consistent with Kamilaroi tradition. \citep{n318}.

\begin{figure}[hbt]
\includegraphics[width=8cm]{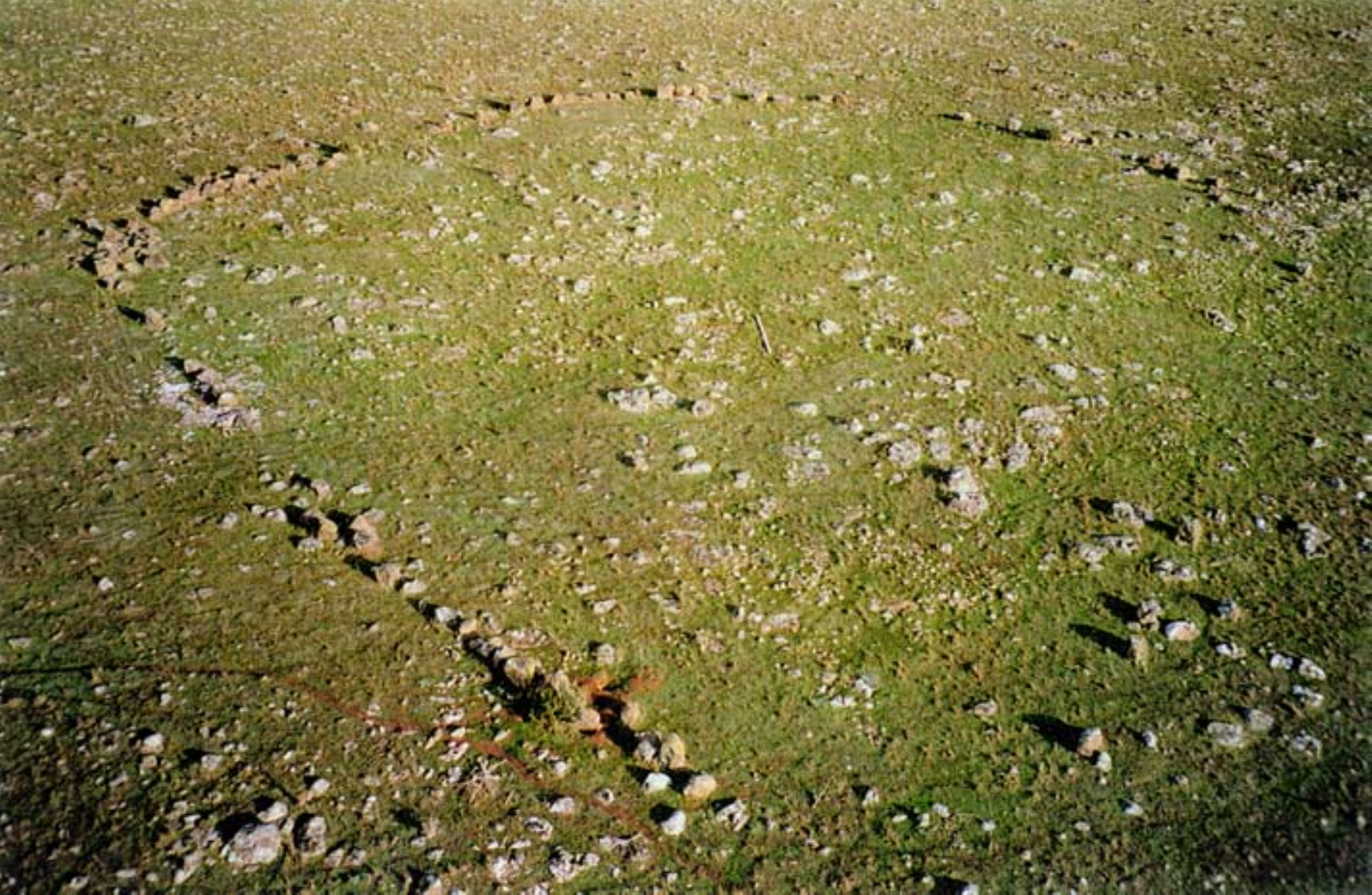}
\caption{Aerial view of the Wurdi Youang site, reproduced with permission from  \cite{marshall99}, looking west.  } 
\label{wurdi1}
\end{figure}

\begin{figure}[hbt]
\includegraphics[width=8cm]{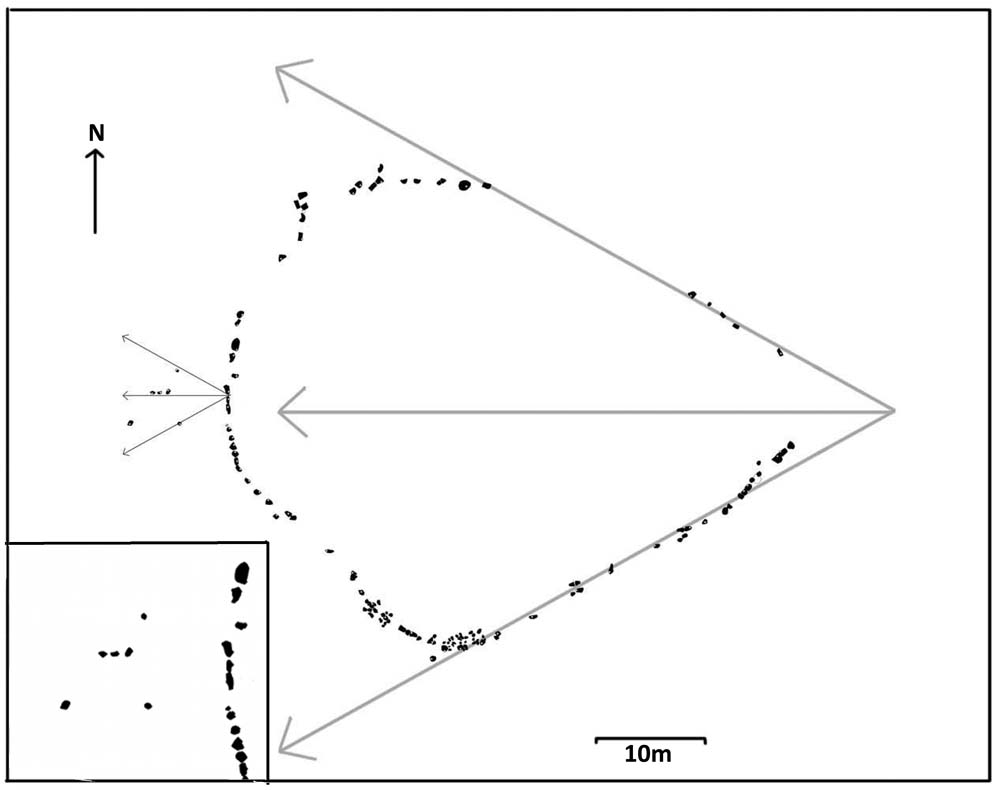}
\vspace{2mm}
\caption{A plan of the Wurdi Youang stone arrangement, adapted from  \cite{n258}. The arrows indicate the directions to the equinoxes and solstices, and are superimposed on the the outliers, left, and the ring (right). Note that these directions are \emph{not} adjusted to fit the ring, but are defined astronomically. Thus while the straight sections of the ring are not well defined, and not exactly straight, this diagram shows that they are well aligned to the same astronomical directions as the alignments over the outliers. The inset at the bottom left shows the outliers.} 
\label{wurdi2}
\end{figure}

\subsection{Stone arrangements in Victoria}
\subsubsection{Wurdi Young}
\label{WY}
The Wurdi Youang stone arrangement, near Little River in Victoria, also known as the Mount Rothwell Archaeological Site,  is an egg-shaped ring of stones, shown in Fig. \ref{wurdi1}, built by the Wathaurung people.  It measures about 50 metres along its major axis, which is oriented almost exactly East-West, and its constituent basalt stones range from small rocks about 0.2 m in diameter to standing stones up to 0.8 m high, some of which appear to be supported with trigger stones. At its highest point, the western apex, is a prominent group  of three stones  about 0.8 m high, with smaller stones, referred to as outliers, nearby.   \cite{morieson03} suggested that sight lines from a gap between the two largest stones over the outliers  mark the position on the horizon where the Sun sets on the solstices and equinoxes.
A survey by  \cite{n258}  confirmed the Morieson hypothesis, and also  found that the straight sides of the arrangement  indicate these same directions to the solstices.  \cite{n258} also found that  the three prominent stones at the western apex of the arrangement, as viewed from the eastern apex, mark the point where the sun sets at equinox. The astronomical alignments at this site, shown in Fig. \ref{wurdi2}, are therefore indicated by  two independent sets of indicators, effectively ruling out a hoax or chance alignment. Norris et al.  also use a Monte Carlo analysis to show that these alignments are unlikely to have arisen by chance, and conclude that the builders of this stone arrangement appear to have deliberately aligned the site on the astronomically significant positions of the setting sun at the solstices and equinoxes. 

Wurdi Youang is the only Aboriginal site known to indicate significant astronomical positions on the horizon other than the cardinal points,  suggesting that other such sites may be discovered in the future.

Since the publication of  \cite{n258}, a possible connection to Orion has been pointed out independently by Reg Abrahams (private communication) and by Mark Mania (private communication). First, as noted in  \cite{n258}, the three large stones appear to mimic three hills on the horizon (locally known as the `Three Sisters Hills'). The line from the three stones to these hills is currently about 10$\deg$ away in azimuth from Orion's belt, but in about 2000 BC, Orion's belt would have set over the three sisters hills as viewed from Wurdi Youang. Second, the arrangement of the outliers (shown in the inset in Figure \ref{wurdi2}) shows some resemblance to the brightest stars in the constellation of Orion. With the three stones in the middle pointing to the setting place of Orion's belt over the three sisters hills in about 2000BC. Is this a genuine astronomical association or mere coincidence? Without more data, it is difficult to tell.

\begin{figure}[hbt]
\includegraphics[width=8cm]{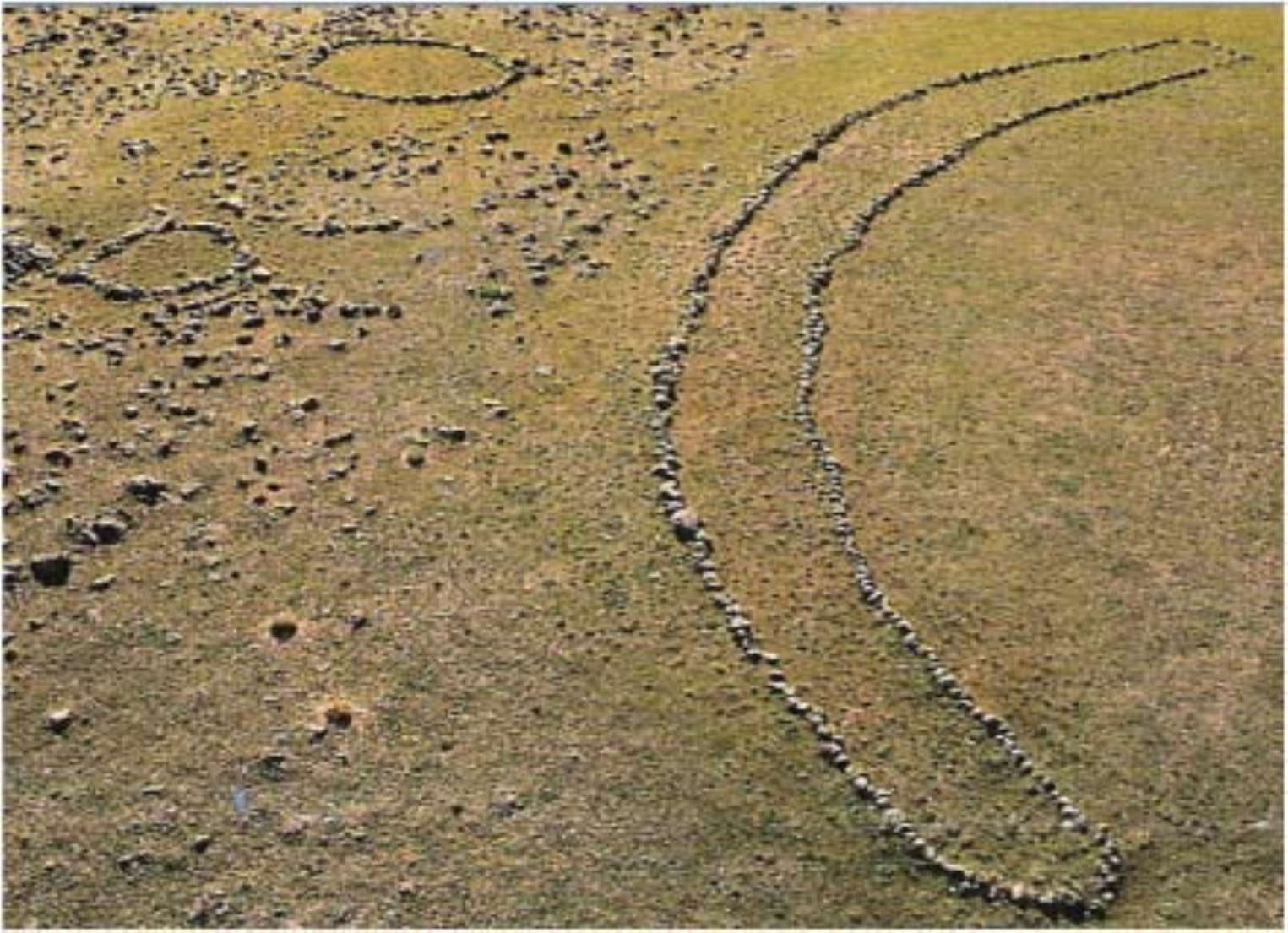}
\caption{Aerial view of the Carisbrook stone arrangement, reprinted with permission of  \cite{VAS86}. } 
\label{carisbrook1}
\end{figure}

\begin{figure}[hbt]
\includegraphics[width=8cm]{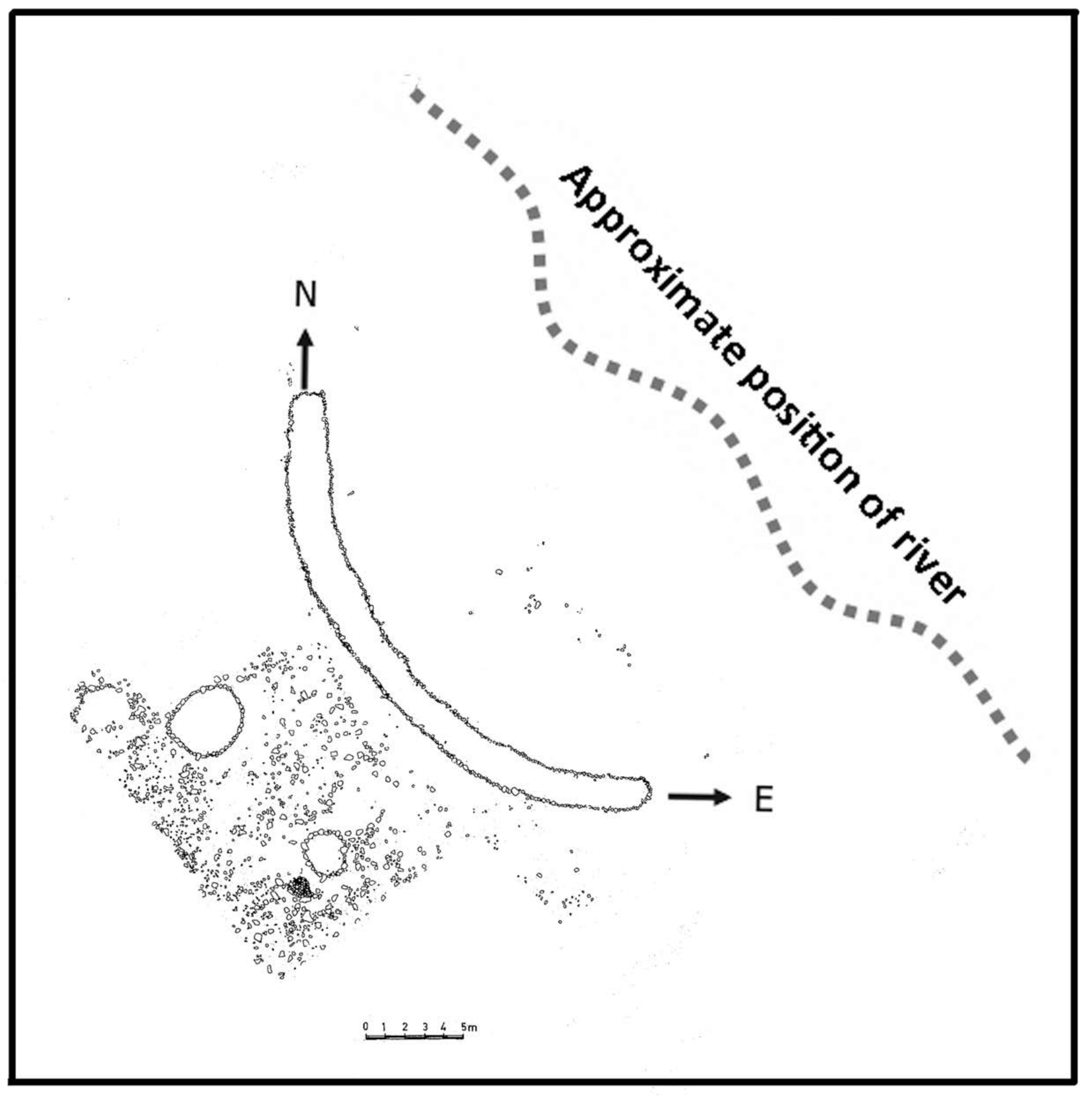}
\caption{Plan of the Carisbrook stone arrangements, adapted from  \cite{coutts77}. Azimuths are measured relative to true north.} 
\label{carisbrook2}
\end{figure}

\subsubsection{Carisbrook}
 \cite{massola63} reported the rediscovery of a remarkable stone arrangement near Carisbrook, Vic., shown in Figures \ref{carisbrook1} and \ref{carisbrook2}. It is described by Massola as a boomerang shape. While its purpose is unknown, it is likely to have been a ceremonial site, and it is possible that the large and small rings represent Bora rings similar to those found in NSW. As first noted by  \cite{morieson94}, the position of the rings relative to the large boomerang arrangements also resembles the position of the  large and small Magellanic Clouds relative to the Milky Way, although we know of no additional evidence to support this hypothesis.

Here I point out that the arrangement is remarkable in that the `boomerang' turns through a full right angle along its length, and is therefore unlike conventional boomerangs used in south-east Australia, which typically turn through a smaller angle. Furthermore, we point out that one end of the arrangement is oriented east-west, while the other is oriented north-south, each with an accuracy of a few degrees.  As in the case of the linear stone arrangements described above, such precision can only be achieved by making astronomical observations, implying an astronomical connection to this site.

\subsubsection{Other Victorian Stone Arrangements}
\begin{figure}[hbt]
\includegraphics[width=8cm]{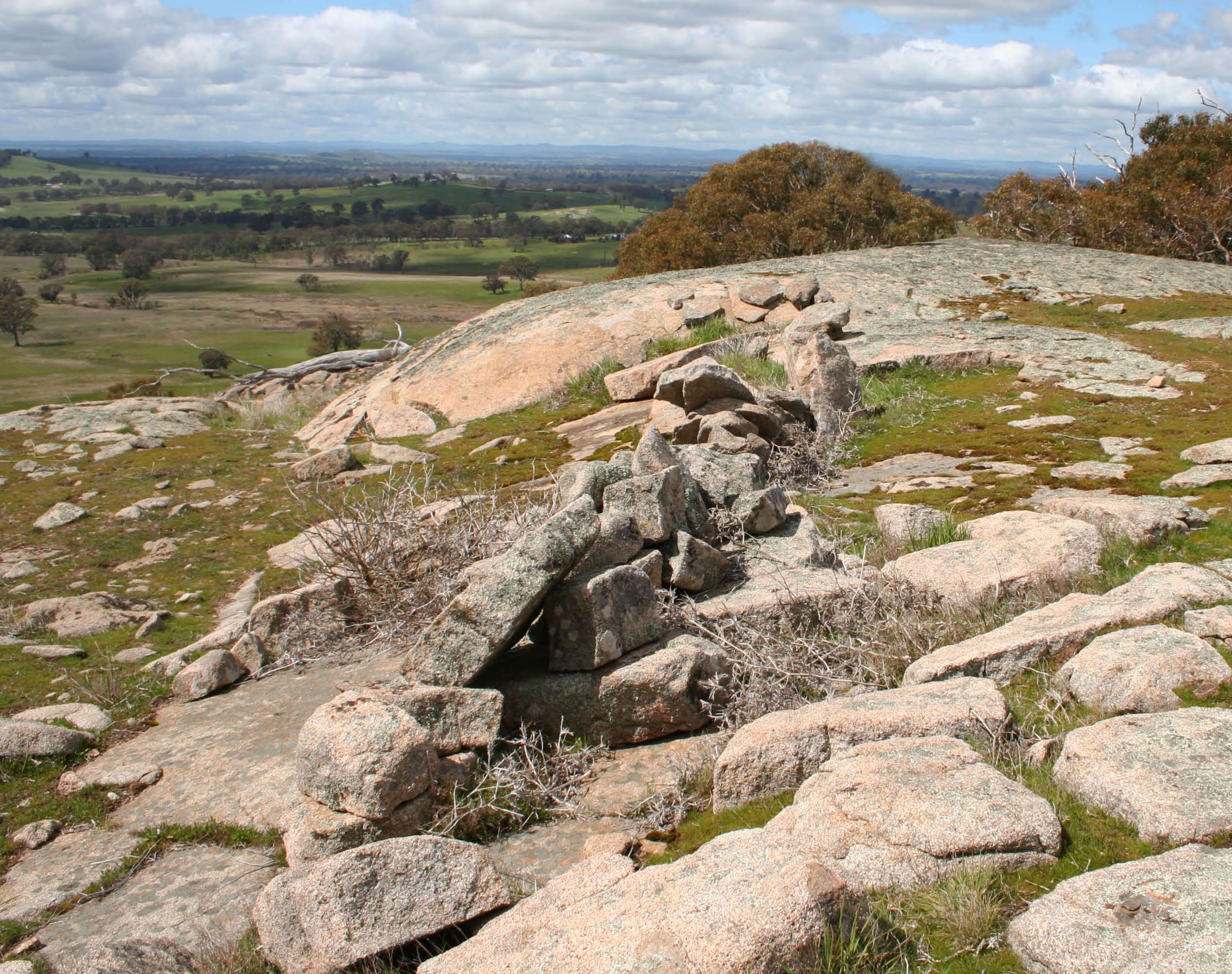}
\caption{View of a putative stone arrangement on Mt. Barker, Vic. The arrangement appears similar to those studied in NSW by  \cite{n278}, and has an orientation roughly north-south, but it has not yet been classified in the literature as an Aboriginal stone arrangement.} 
\label{barker}
\end{figure}

These studies of stone arrangements are hindered by the sparsity of data. For example, only four major stone arrangements in Victoria have been reported in the literature, and so the evidence of knowledge of the position of the Sun at the solstices relies on just one site - Wurdi Youang. However, several other putative stone arrangements are known, such as that shown in Figure \ref{barker}, and there may well be others, but they lack the archaeological or ethnographic evidence that would enable them to be classified as Aboriginal stone arrangements. 

\section{Astronomical Dating of Aboriginal Astronomy}

There are several ways in which astronomical phenomena could in principle be used to date a site, although they have had limited success in Australia, as discussed by \cite{hamacher11a}.

The precession of the equinoxes is caused by the Earth's axis tracing a circle of radius 23.5$\deg$, relative to the stars, over a period of about 26000 years. This results in a shift in the apparent positions of stars by $\sim$1$\deg$ every 72 years. Thus the position on the horizon at which a star sets 2000 years ago could differ from that seen today by nearly 30$\deg$. If an archaeological site is known from ethnographic evidence to point to a particular star, this precession can in principle be used to date the site  \citep[e.g.][]{thom67}. However, no such sites are known in Australia.

The declinations of the Sun and Moon, and hence their rising and setting positions, are unaffected by this precession. However, the apparent declination of the Sun is affected by a much smaller effect, the nutation in the obliquity of the EarthÕs rotational axis, which varies by about 2.4$\deg$ over a period of 41000 years. However, no Aboriginal site has yet been found showing sufficient accuracy for this to be useful for dating the site.

A further effect, stellar proper motion, causes the stars to shift their apparent position relative to each other over time. Although this is a small effect,  \cite{hamacher11a} has shown that the shape of {\it crux} (the Southern Cross) would have looked significantly different 10000 years ago (see Fig. \ref{crux}), so that stories in which {\it crux}  is identified with an emu's foot cannot be more than 10000 years old.

\begin{figure}[hbt]
\includegraphics[width=8cm]{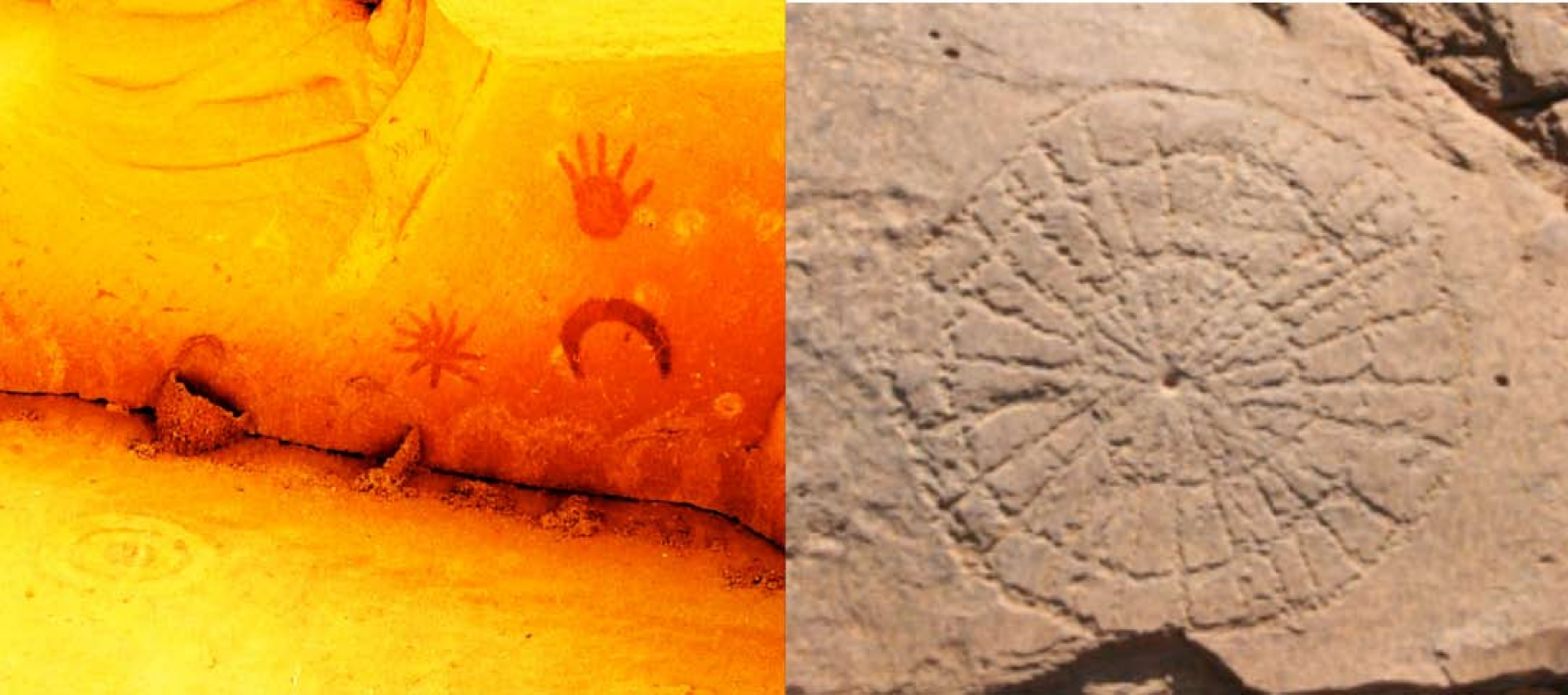}
\caption{Left: the `Supernova' pictograph in Chaco Canyon, New Mexico, which may represent the 1054 `Crab' supernova witnessed by Chinese astronomers. Right: The `bicycle wheel' engraving at Sturts Meadows, NSW, which has been claimed to represent a supernova.
} 
\label{sn}
\end{figure}

Supernovae can in principle be used to date sites. For example, the famous painting, shown in Figure ~ \ref{sn} in Chaco Canyon, New Mexico, is said  \citep[e.g.][]{eddy79} to represent the supernova of 1054AD, resulting in the Crab Nebula, although this claim has  been disputed \citep[e.g.][]{williamson87}  and \cite{krupp10} places this image in the context of other crescent/star images found in the American south-west, not all of which  can be images of the supernova.  \cite{murdin81} suggested that a rock engraving (Figure \ref{sn} ) at Sturts Meadows, NSW was also a supernova, although without any supporting evidence this seems speculative.  Other speculations include my own  that that the `Emu's egg'  (see Figure \ref{emuegg} ) may represent a once-obvious but now vanished object in the sky, such as a supernova, while \cite{hamacher14c} concluded that there was no evidence that the Yolngu `Fisherman Story'  \citep{wells73c}, or any other Aboriginal story,  refers to a supernova.

Stellar variability could also, in principle, be used to date a site or story.  \cite{hamacherfrew10} argue that the Boorong reference to {\it Collowgullouric War} noted by  \cite{stanbridge58} is the hyper--giant variable star Eta Carina during an eruptive period in the 1840s, when it became the second brightest star in the sky. While not relevant to dating a site, it has also been noted  \citep{ bates25, fredrick08, leaman, n268} that the Pitjinjatjara  story that Betelgeuse was a hunter whose burning club becomes faint and then brightens, may be a record of  the variability of Betelgeuse.

\section{Aboriginal Time-keeping and Calendars}
\subsection{Calendars and Seasons}
\label{solar}
Aboriginal hunter/gatherers typically moved throughout different camps on their land in an annual cycle  \citep[e.g.][]{thomson39}. For example, the Yolngu moved up to the Arnhemland escarpment to harvest the nuts and berries that grew there, and down to the rivers to catch barramundi and harvest rakai nuts. The timing of the moves was critical, otherwise  animals would eat the crop before the humans arrived  \citep[e.g.][]{wells64} and so most Aboriginal cultures had well-defined seasons. Over much of the southern half of Australia, Aboriginal groups had four seasons  \citep{berndt93, stanbridge58} while in  north Australia they often had five  \citep{baker99, kaberry39}, six  \citep{thomson83}, or even as many as 13  \citep{nova15, tipiloura14}.  

The seasonal changes were defined by many factors such as  changes in the weather, the arrival of migratory birds, the flowering or fruiting of plants, altered behaviour of animals, and the sky, particularly the heliacal rising of stars   \citep[e.g][]{clarke09a, clarke03a, clarke15a, davis97, hamacher15a, haynes92, n214, n287}.  `Big Bill' Neidjie, the last of the Gagadju/Kakadu male elders described the process: `I look at star, I know just about time for wet season, maybe time for dry season, I know from star'  \citep{neidjie85}.  These seasons typically have names which are descriptive of the season, as in northern Australia where even now the year is divided into the `wet' and the `dry'. Excellent reviews and descriptions of Aboriginal seasons are given by  \cite{BoM, clarke09b, csiro15, johnson98}.
 
For example, the heliacal rising of the Mallee-fowl constellation (Lyra)  in March signifies to the Boorong people of Victoria that the Mallee-fowl are about to build their nests, and when Lyra disappears in October, the eggs are laid and are ready to be collected  \citep{stanbridge57}.  \cite{n311} describe how the Kamilaroi people use different parts of the Emu in the Sky as markers: when the head appears in February, people start moving from their summer camp. When the legs appear in April, people go home to their winter camp. When the neck and legs disappear around August-September, the belly is still there, and represents the egg, which is now developing into a chick.
A sample of astronomical markers is listed in Table\ref{seasons}.

In some cases, the stars are thought to be the causes of the change of season, rather than mere markers. For example,  \cite{mountford39} reports that the Anyamatana	 believe that the women of the Pleiades cause the first frosts of winter, by scattering frost from their bodies.  \cite{mathews04} and \cite{clarke15a} report that women in the central deserts should not look at the Pleiades during winter nights, because this would increase the severity of frosts.

\begin{table*}[h]
\begin{center}
\caption{Examples of the significance of the heliacal rising of  stars in Aboriginal traditions,  
adapted and extended from  \cite{n268}.}
\centering                          
\begin{tabular}{llllll}            
\hline
Group			&  	Location			&	Object					& 	Meaning					&	Reference\\
\hline
Anyamatana		&	South Australia		&	Pleiades				& 	First frosts of winter 	&  \cite{mountford39}\\
Boorong			& 	NE Victoria		&	Vega					&	Mallee Fowls Build Nests		&	 \cite{stanbridge58}\\
Kaurna			& 	Adelaide Region	&	Parna (Fomalhaut?)					& 	Start of Autumn Rains		&	 \cite{gell42, hamacher15a}\\
Kuwema			& 	Daly River, NT &	Orion					& 	harvest dingo puppies		&	 \cite{harney59, tindale83}\\
Pitjantjatjara		& 	Central Desert		&	Pleiades					& 	Dingo Breeding Season		&	 \cite{tindale83}\\
Pitjantjatjara		& 	Central Desert		&	Pleiades					& 	Start of winter		&	 \cite{clarke03a}\\
Wardaman	& 	Northern Territory		&	Leo	& 	Start of ceremonial year			&	 \cite{DS}\\
Warnindilyakwa	& 	Groote Eylandt		&	$\upsilon$,$\lambda$ Scor	& 	Start of Dry Season			&	 \cite{mountford56}\\
Yolngu			& 	Arnhem Land		&	Scorpius					& 	Arrival of Macassans		& 	 \cite{mountford56}\\
Yolngu			& 	Arnhem Land		&	Pleiades					& 	start of winter	& 	 \cite{mountford56}\\
Yolngu			& 	Yirrkala, Groote Eylandt		&	Arcturus					& 	time to harvest rakai \& dirpu 	&  \cite{mountford56,  haynes92}, \S\ref{venus}\\

\hline                            
\end{tabular}
\label{seasons}
\end{center}
\end{table*}

\subsection{Lunar Markers of time}
While a seasonal calendar was used to mark the seasons, shorter periods, including the timing of ceremonies, were often based on lunar phases  \citep[e.g.][]{kaberry39}. For example, Harney  \citep{DS} 
says: `Time, how long we stay? Go by the moon. Moon counting. Three moons. One moon is four week. Three moons is twelve weeks.'
 \cite{berndt43} report how a ceremony is held `during the first new moon of the new season' and   \citet{morphy99} 
reports how a ceremony  is held over many days during a season when there is plenty of food, ending with a full moon.  \citet{harney09c} says `This one, when the new moon come, thatÕs the time for initiation, Gandawag-ya nunyanga Gandawag means Òthe new moon.' \cite{morrill64} said that the Mt. Elliott people measured `time by moons and wet and dry seasons',  where the wet and dry seasons refer to the seasonal calendar as described in \S\ref{solar}.

 \cite{hahn64} discussed how  Aboriginal people in South Australia made notches in their digging sticks to measure their age in lunar months, and   \cite{flood97}
  reports an Aboriginal interpretation of a series of parallel  Panaramitee crescents as representing an elapsed time in half-moons, and 
suggests that a diprotodon tooth, over 20,000 years old, engraved with 28 grooves, may represent a lunar calendar.

A full moon was also the optimum time for travelling at night (see e.g. Figure \ref{fig:message_stick}).

 However, while lunar phases were used to mark periods of time, they differ from a  true lunar calendar in that there are no named lunar `months' that might be arranged in a sequence, or be given names. Instead, the seasons (described in \S\ref{solar}, and which in some cultures are little more than a month) are used for this purpose.  \cite{n268} give further examples of how lunar phases were  widely used by Aboriginal groups. 

\subsection{Timekeeping}
Other than the commonplace practice of using the position of the Sun during the day as guide to the time of day, and the orientation of the stars to mark time at night  \citep[e.g.][]{DS, harney09b}, there are few recorded instances of using the sky to measure time. One exception is the Yaraldi of South Australia, who divided the day into seven sectors: before dawn, dawn with rising sun, morning, noon, afternoon, sun going down, including twilight, and night  \citep{berndt93}. 

\section{Direction, Songlines and Navigation}
Like the Pacific Islanders, Torres Strait Islanders were skilful sea navigators  \citep[e.g.][]{hamacher13c} but are outside the scope of this review.   Coastal Aboriginal groups (often called the `saltwater' people) were adept at building, operating, and navigating canoes, but their craft did not permit ocean-going travel.  Yolngu people navigated around the coast using stars as guides  \citep{wells73c}, and the constellation Djulpan (corresponding to Orion, the Hyades, and Pleiades) was particularly associated with navigation during the wet season. Elsewhere in Australia, most Aboriginal navigational skills were focussed on land-travel, which is the focus of the rest of this section.

\subsection{Aboriginal trade routes}

\begin{figure}[hbt]
\includegraphics[width=8cm]{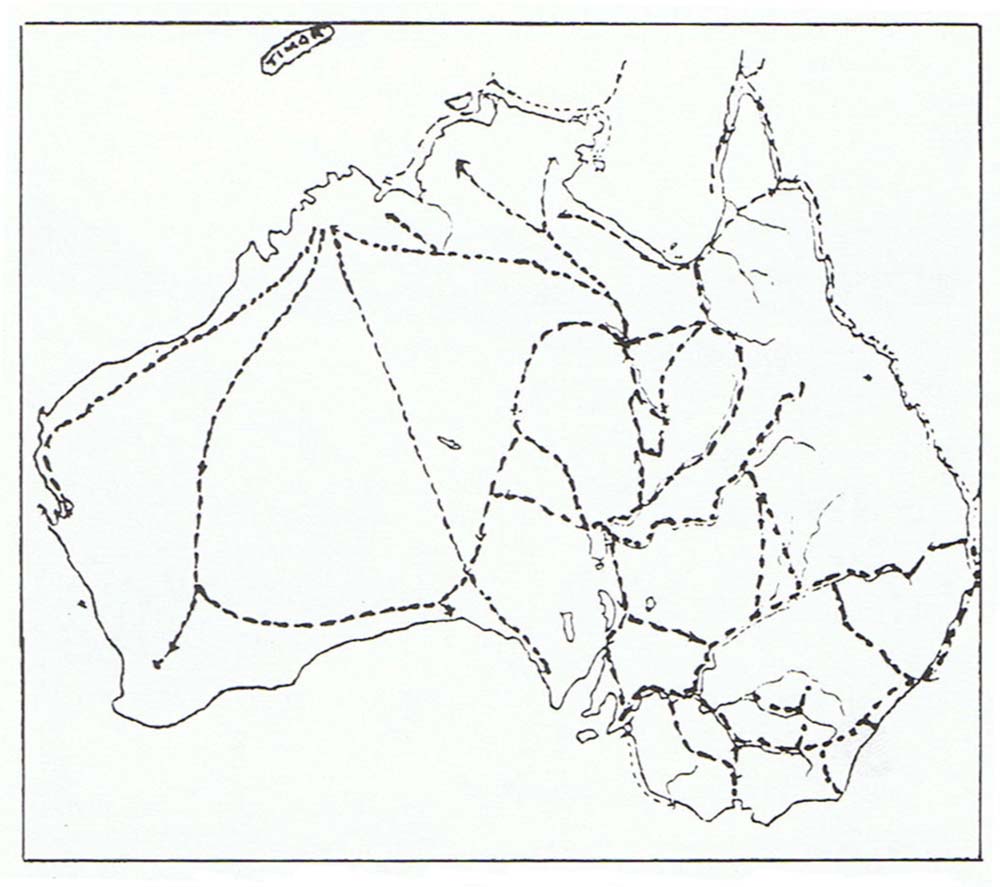}
\caption{The Aboriginal trade routes of Australia, taken from  \cite{mccarthy39}.
} 
\label{traderoute}
\end{figure}

As far back as 1939,  \cite{mccarthy39} showed that Australia was criss-crossed by Aboriginal trade-routes 
(see Figure \ref{traderoute}), and these have been described in detail by  \cite{flood83,mcbryde00, mulvaney99}.  \citet{mowaljarlai93} describes how the trade routes connect all Aboriginal people: `The lines are the way the history stories travelled along these trade routes. They are all interconnected. It's the pattern of the sharing system.'
These routes were used to trade commodities such as bunya nuts, pituri (a nicotine-based narcotic), stone axes, ochre, tools, and even stories  \citep{n322,reynolds81}. 

 \citet{sutton98} describes Aboriginal maps drawn on paper, mainly obtained by  Tindale, Mountford, and the Berndts, from various Aboriginal groups. Although the use of paper is obviously not traditional, the readiness with which these maps are drawn suggests great familiarity with the concept, presumably based on traditional `mud-maps'. However, the paper maps include much greater detail, such as names, watercourses, and special places, and are sometimes marked  in terms of a days walk from water to water. 

 \cite{kerwin10} has shown unambiguously that traditional Australians navigated long distances both for trading and ceremonial purposes, but it is curious that this aspect of Aboriginal culture was largely unexplored by anthropologists in the last century, who frequently described  a route  followed by creator-spirits on earth  or in the sky, but did not  connect them to navigation \citep{n315}. For example,  \citet{elkin38}
appears to have heard at least one songline without noting its significance.
 \cite{tindale74} and \cite{mountford76}
were aware of ancient Aboriginal tracks across large parts of Australia, and commented on the difficulty of navigating large distances, but did not explore navigational techniques. Mountford even discusses a songline: ` the route of Orion and the Pleiades extends from the Warburton range in Western Australia through the Rawlison, Petermann, Mann, and Musgrave ranges, reaching Glen Helen, in the country of the Western Arrernte.' but did not explore whether it was used for navigation. 

There are several accounts that Aboriginal people had no astronomical navigation, summarised by  \cite{haynes00} who say `the Aborigines seem not to have used the stars for purposes of navigation'.   It is unclear why such accounts are so discrepant from the evidence presented here. Certainly there are large differences between different groups, and perhaps earlier studies happened to be confined to  groups who did not use astronomical navigation. This discrepancy illustrates the danger of extrapolating a result from one group to include all Aboriginal language groups. Alternatively, perhaps that information was regarded as secret at the time. This latter hypothesis is a plausible explanation of the puzzle  \citep{maegraith32} that the Arrernte/Luritja people could accurately locate stars and constellations, and knew that crux was circumpolar, but apparently didn't use this knowledge for navigation. 

\subsection{Direction}
\label{direction}
Many groups of traditional Australians were familiar with the concept of cardinal directions  \citep{n278, levinson, lewis76, kaberry39}. For example, central to the culture of the Warlpiri people in central Australia are the four cardinal directions that correspond closely to the four cardinal points (north, south, east, west) of modern western culture  \citep{wanta08, n287}. \cite{DS} recount how the cardinal directions were given to the Wardaman people by the blue-tongue lizard, who threw boomerangs to the four cardinal points, and while these closely correspond to the modern European directions, it is interesting that Harney refers to `west' as `down' whereas Europeans would refer to `south' as `down'. 
Boomerangs were also thrown towards the four cardinal points in a Kokoimudji (Qld) story  \citep{tindale83}. \cite{hunter93} also reported that the Darug people in Sydney called `the four principal winds by the following names: The North, Boo-roo-way; The South, Bain-marree; The West, Bow-wan; The East, Gonie-mah'.

 \cite{levinson} reports that the Guugu Yimithirr language has direction so deeply embedded in it (and presumably in the psychology of those who speak it) that cardinal directions are used rather than concepts of left or right or behind. For example, a Guugu Yimithirr speaker will say `the chair to the west of you' rather than `the chair to your left'.   \cite{levinson} also reports that  Guugu Yimithirr speakers have a remarkable sense of direction, and are able to indicate cardinal directions to an accuracy of about 13$\deg$. A similar number is reported by  \cite{lewis76} for Western Desert peoples. But the most remarkable aspect of Levinson's paper is that, to remember these absolute (as opposed to relative) directions, people's memories of an event must include the directions as well as the appearance. Levinson concludes that not only do Guugu Yimithirr speakers speak a language that  requires storage and computation of orientation and absolute directions, but they also think in this way when {\em not} speaking the language. 

However, some other language groups have cardinal directions which can be quite different from the European concept.  For example, some groups have very loose definitions of the cardinal points  which may vary significantly, and apparently arbitrarily, from place to place  \citep{Breen93, n278}. Even in these cases,  east and west are usually associated with the rising and setting positions of the sun, and the words for east and west are often based on the word for the sun. Burials were often aligned facing east  \citep{collins98, dunbar43, mathews04}, 
while initiation sites were often aligned roughly north-south (see \S\ref{bora}).  \citet{n278} have shown that a sample of linear stone arrangements  are oriented north-south with an accuracy of a few degrees (see \S\ref{rows}). So how did traditional Australians measure cardinal directions so accurately?

There is no evidence that magnetic compasses were known to traditional Australians, although this cannot be ruled out definitively, given circumstantial evidence that they regarded iron meteorites as having special properties. The familiar mnemonic of finding south using a simple geometric construction using the stars of Crux and Centaurus  only works if someone else has already worked it out.  \cite{n287} identified   three techniques available to traditional Australians  to determine cardinal directions:
\begin{itemize}
\item In the Northern Territory, so-called `magnetic' termite mounds are aligned north-south with an accuracy of about ten degrees, to minimise solar heating of the mound  \citep{grigg77}. 
\item The direction of the setting or rising position of the sun or moon can be marked each day, and the mid-point between these ends indicates due west or east respectively, with an accuracy of a few degrees,  This may have happened  at Wurdi Young, discussed in \S{\ref{WY}}.
\item  Similarly, the direction of a circumpolar star such as the Southern Cross could be marked, and the mid-point of these would indicate south. 
\end{itemize}
 Except in the Northern Territory, directions to a precision of a few degrees as seen by  \cite{n278} can only be determined by careful astronomical observation.

\subsection{Land Navigation by the stars}
\label{navigation}
Wardaman people in northern Australia travelled mainly at night, when the air was cool and the stars visible as guides \citep{ac9}.  \cite{DS} describe a Dreaming Track in the sky which helps navigation on the ground, and  \citet{harney09c} 
describes how night travellers make a mud dish containing burning sticks, which they then wear on their heads to provide light. He also describes  \citep{DS} 
how you can use your shadow, cast by the Moon, as a compass.

It was described in \S{\ref{intro}} how initiated men knew the sky intimately, and could name most  stars in the sky visible to the naked eye.  \cite{n315} describes how Wardaman senior elder Harney has an intuitive mental map of the moving sky, and its changing  relationship to the land, which can be used directly for navigation: `the walking and dark on foot all around the country in the long grass, spearing, hunting, ... but each night where we were going to travel back to the camp otherwise you donÕt get lost and all the only tell was about a star. How to travel? Follow the star along.'

 \cite{kerwin10}
describes how other language groups are also able to navigate using stars, and the explorer Mitchell describes how his Aboriginal guides were able to navigate by night using an `instinctive knowledge of the ground and a recollection ... as true as a compass, of the direction to any spot to which they chose to go'  \citep{baker98}.   \cite{maegraith32} noted that Aboriginal men had a good understanding of the  apparent motions of the stars through the night, and the slow annual rotation of the constellations, and were  able to describe the positions of a constellation at different times of the year. 

Thus at any time in the night, they had a strong sense of direction, and could couple that to their knowledge of topography to navigate. For frequent journeys, directions would be coupled directly to the constellations (e.g. `you go on the right hand side of the emu'   \cite{DS}).
The ecliptic had special significance  \citep{DS}:
`The planets come straight across like you and I doing walk, pad up and down, walking backwards, forwards, make a little track there, a pad' and
`Travelling pathway joins to all different areas, to base place, to camping place, to ceremony place, where the trade routes come in; all this sort of things. The Dreaming Track in the sky, the planets come straight across'.

On the other hand, many language groups were reported to be  unwilling to navigate by night, apparently because of the greater danger from spirits, and chose instead to travel by day  \citep{n322,lewis76, johnson98}, including groups who knew the sky intimately  \citep{maegraith32}. 

\subsection{Songlines and Dreaming Tracks}
 Section \ref{venus} described the songline followed by Venus, or Barnumbirr, in Yolngu culture. Many such songlines are known, and are used for long-distance navigation. Songlines, or `dreaming tracks', are typically the path said to have been followed by a  creator spirit or ancestor across the land. Sometimes they are also associated with a path in the sky. 

While star maps do exist in Aboriginal paintings and possibly in rock engravings, no Aboriginal star maps intended for navigation have been recorded. Instead, navigational knowledge is committed to memory in the form of songlines, which may therefore be regarded as `oral maps'. In a culture with no written language, but with a strong tradition of memorising oral knowledge, this is probably the optimum way of recording and transmitting navigational information  \citep{DS}.

The English word `songline' was coined by  \cite{chatwin} in his eponymous novel, but the concept is deeply embedded in traditional Aboriginal cultures, in which they are   often referred to as `Dreaming Tracks'.  \cite{clarke03a}  calls them `strings' because they connect different people and sacred sites. 

It is now well-established that traditional Australians traded goods the length and breadth of Australia  \citep{gammage11, kerwin10} , and it appears that these trading routes also follow the song-lines   \citep{gammage11, mulvaney99}. Many such songlines are  documented  \citep[e.g.][]{n322, kerwin10}, and were used  for trading materials such as ochre, trading intellectual property such as songs and dances, and for attending ceremonies  \citep{lee09}.

 \cite{wositsky99} say that:
`Songlines are epic creation songs passed to present generations by a line of singers continuous since the dreamtime. These songs, or song-cycles, have various names according to which language group they belong to, and tell the story of the creation of the land, provide maps for the country, and hand down law as decreed by the creation heroes of the dreamtime. Some songlines describe a path crossing the entire Australian continent.' Descriptions of the creations of songlines by creator spirits may be found in  \citet{harney09c}.

For example, as described in \S\ref{venus}, a songline associated with the planet Venus starts at Yirrkala in Arnhem Land, where the Yolngu believe Barnumbirr (Venus) crossed the coast as she brought the first humans to Australia from the east  \citep{allen75, ED}. Her song, contained within the Yolngu Morning Star ceremony, describes her path across the land, including the location of mountains, waterholes, landmarks, and boundaries. The song therefore constitutes an oral map, enabling the traveller to navigate across the land while finding food and water. It is said by Yolngu elders at Yirrkala   \cite{ac2} that the same song is recognisable in a number of different languages along the path from east to west, crossing the entire Òtop endÓ of Australia. The song changes along the route, being longer and more `sing-song' in the east, and shorter, and broken into short sharp segments, in the west  \citep{ac2}. The same songline is said to  cross the coast again near Geraldton  \citep{ac3}.

Many modern highways in Australia are also said to follow songlines, presumably because the first western explorers  sought advice from Indigenous guides. For example, Wardaman elder Bill Yidumduma Harney witnessed  \citep{wositsky99, ac7} his grandfather leading Europeans along part of a songline that extended from Arnhem Land, through Katherine, to Western Australia, to form what is now the Victoria Highway. The Europeans would blaze (cut marks with knives) the trees, and then buffalos would drag a tree-trunk along the blazed trail to clear the track.

The Great Western Highway and the Bells Line of Road, from Sydney to the Blue Mountains, are also said by Darug elders  \citep{ac6} to be songlines. Supporting evidence includes the Darug rock engravings found close to the path of the Great Western Highway through the Blue Mountains.

Songlines on the ground were sometimes mirrored by songlines in the sky  \citep{DS}.  \cite{ac1} described how the songlines on Earth were mirrored by the songlines in the sky, formed when the Creator Spirits moved to the sky,
so that knowledge of the sky formed a mnemonic for tracing a route on Earth. 

\begin{figure}[hbt]
\includegraphics[width=8cm]{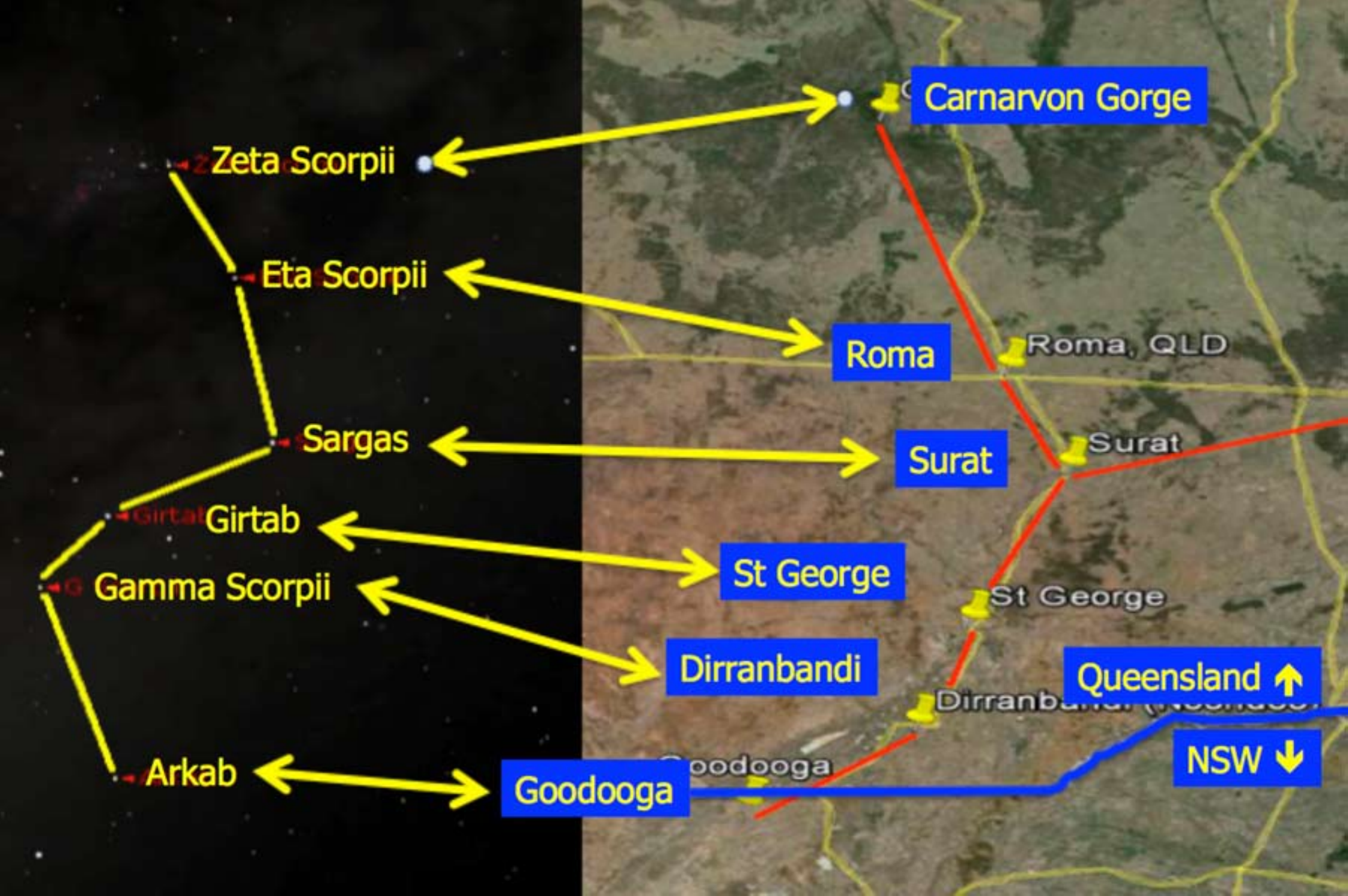}
\caption{A Euahlayi songline  \citep{n322} extending from Queensland into NSW. The song describes the path on the ground, which is mirrored by the stars of Scorpius. The stars provide an aide-de-memoire for the song}.
\label{songline}
\end{figure}

This mirroring is also reported amongst  the Euahlayi people  \citep{n322,fuller16}, including the eaglehawk songline (Achernar to Canopus to Sirius) that extends from Heavitree Gap at Alice Springs to Byron Bay on the East Coast, and the Milky Way/Black Snake/Bogong Moth songline connecting Normanton on the Gulf of Carpentaria with the Snowy Mountains near Canberra. Figure \ref{songline} shows a songline extending from Queensland into NSW.  In such songlines, the stars themselves do not seem to be used to navigate, and the path in the sky does not necessarily exactly mirror that on Earth. Instead, the stars are used to illustrate the song, and are used as a memory aid to help recall the words of the songline.

\section{Rock art}
\subsection{Sydney Rock Engravings}
\label{rockart}

Four weeks after the First Fleet arrived in Sydney Harbour in 1788, Governor Phillip took a long-boat out from Sydney, up the coast, to what is now known as West Head, where he met the local Guringai people, and commented on their friendliness \citep{phillip89, tench93}. 
Four years later he returned to the place to find that the Guringai people had gone, killed by smallpox or driven away by the loss of their land and water.  However, their beautiful rock engravings remain as their legacy throughout Kuring-gai Chase National Park.

Thousands more Aboriginal rock engravings can be found  across the Sydney basin, and thousands more have been  destroyed by roads and buildings. The locations of most of the remaining ones are kept secret to avoid further damage, but a few are public  \citep{attenbrow02, norriswww, stanbury}. Some of the finest accessible sites are located in Kuring-gai Chase National Park, NSW, once home to the Guringai people. Some engravings may represent astronomical objects. For example,  \cite{n311} report a Kamilaroi story that some rock engravings represent  brush turkeys, and there is an engraving of the turkey with a clever man pointing to it in the sky, telling a story which is a prophecy. Here I describe several other examples. None have definitive evidence linking them to the sky, but some have  strong circumstantial evidence.
 
In Kuring-gai Chase National Park, close to the Elvina Track, is a finely engraved emu (see Fig. \ref{emupic}). The artist has taken great care to make the engraving anatomically correct, with the distinct shape of the emu back, and the bulge of the gizzard. However,
its legs trail behind it, in a position that would be unnatural for a real emu. However,   \cite{cairns96} pointed out that the position of the legs is similar to that of the Emu in the Sky. This hypothesis is supported by the observation  \citep{ED} that, (a) the orientation of the emu on the ground mirrors that of the emu in the sky above it, and (b) this orientation occurs at the time of year when Emus are laying their eggs. This assembly of circumstantial evidence seems to support Cairns' hypothesis that this engraving is a picture of the Emu in the Sky rather than a real emu. Other supporting circumstantial evidence is that the Elvina site is said to be an initiation site, and that emus are closely connected with the initiation process (see \S\ref{emu}).

If the emu engraving is a representation of an astronomical object, then we might  expect to find other rock engravings of other astronomical objects such as the Sun and Moon, and certainly examples of solar images exist elsewhere, such as those at Ngaut Ngaut (Fig. \ref{ngautpic}). 
Star-like motifs, typically consisting of a small circle with lines radiating outward, are found in several Sydney rock engravings  \cite[e.g.][]{sim66}, although their meaning is unknown. 

Crescent shapes are also common, and may represent the moon, although they have also been attributed to boomerangs  \citep[e.g.][]{mccarthy83}. However, boomerangs from the Sydney region typically have straight sides and rounded ends, whereas the engravings generally have a curved shape and pointed ends, which more closely resembles a lunar crescent  than a boomerang.  \cite{ED} interpret this difference as support for the hypothesis that the Sydney engravings contain a significant astronomical component. This hypothesis is supported by other obvious engravings of boomerangs which clearly {\bf do} resemble boomerangs rather than the moon.

\begin{figure}[hbt]
\includegraphics[width=8cm]{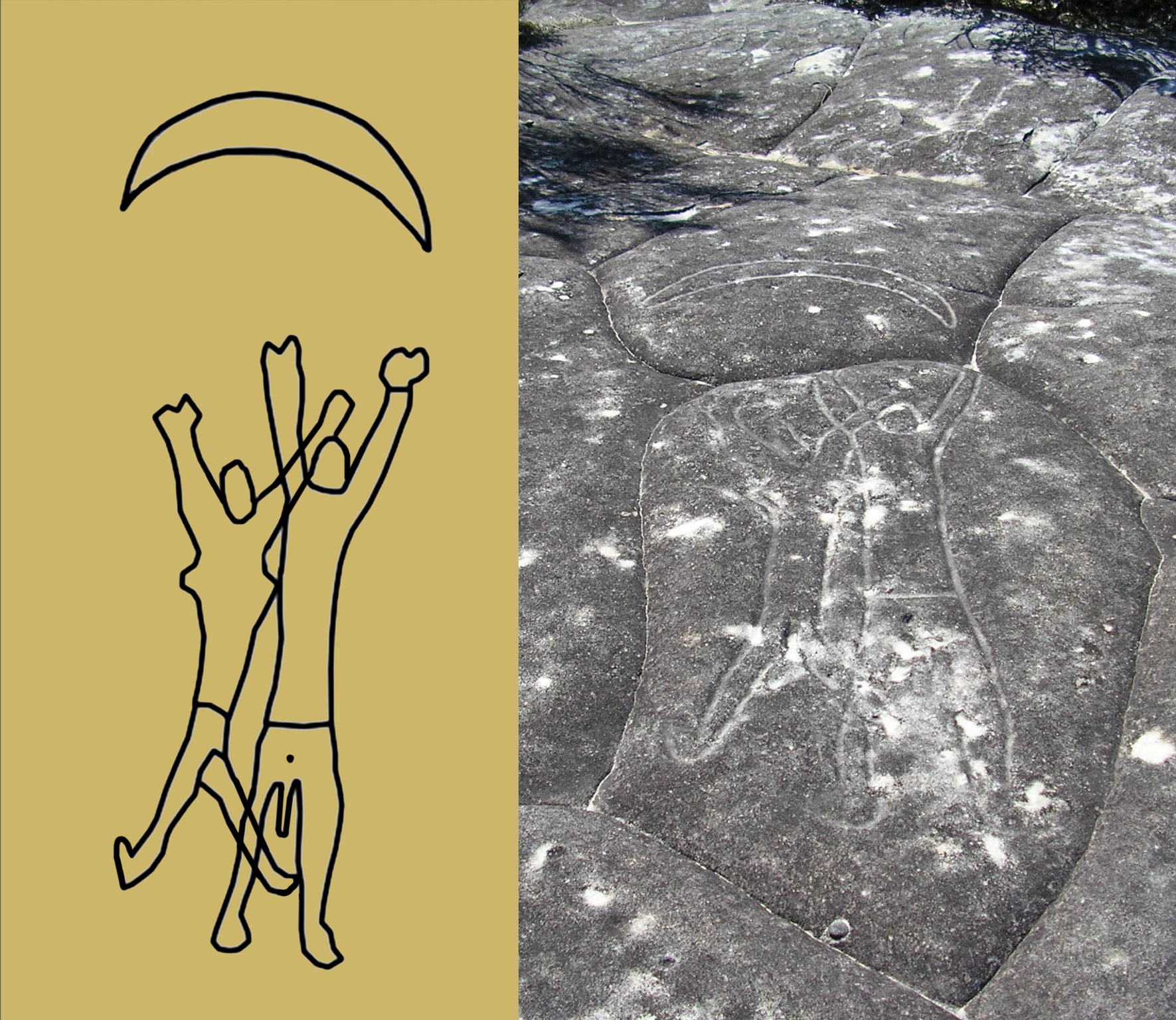}
\caption{A rock engraving from the Basin Track, Kuring-Gai Chase National Park, NSW, showing a crescent which has been variously described as a boomerang, a crescent moon, and an eclipse. From  \cite{n217}}
\label{basinpic}
\end{figure}

Amongst the crescent motifs found in the Sydney engravings are several  that depict a man and woman under or near crescent shapes, such as that at the Basin Track engraving (Figure \ref{basinpic}).  These depict the crescent  with the  horns pointing down, a configuration never seen in the moon. However, eclipses can occur as a crescent with the horns pointing down, and, for example,  one such eclipse took place on the morning of 8 August 1831, in the direction indicated by the engraving  \citep{n255}. However, the Guringai people had been largely destroyed by this date, and so the engraving was presumably made after an earlier eclipse.  Under the crescent, the man and woman overlap, and  \cite{ED} suggested this may be an image of the Moon-man and Sun-woman partially covering each other during the eclipse. The late John Clegg (private communication) suggested that a hermaphrodite figure near this engraving may represent the Moon-man and Sun-woman fully superimposed during a total eclipse. An alternative interpretation involving childbirth has been suggested by \cite{bhathal11c}.

Near Woy-Woy, NSW, is a
rock engraving of a creator-spirit
named Bulgandry (Fig.\ref{bulgandry}), who may be a manifestation of Baiame \citep{ED}. He holds a disc in one hand and a crescent in the other, and his hair resembles the tail of a comet. Whilst he could be holding a shield and a boomerang, the crescent is not shaped like a boomerang, and the local Darkinjung people had oval shields, not circular, and are generally shown in engravings with markings on them. So it is tempting to speculate that he is holding the Sun and Moon, and this engraving may refer to a Dreaming story about the Sun and Moon. T. Leaman (private communication) has suggested that Bulgandry may be Orion, and his outstretched arms the ecliptic.

\begin{figure}[hbt]
\includegraphics[width=8cm]{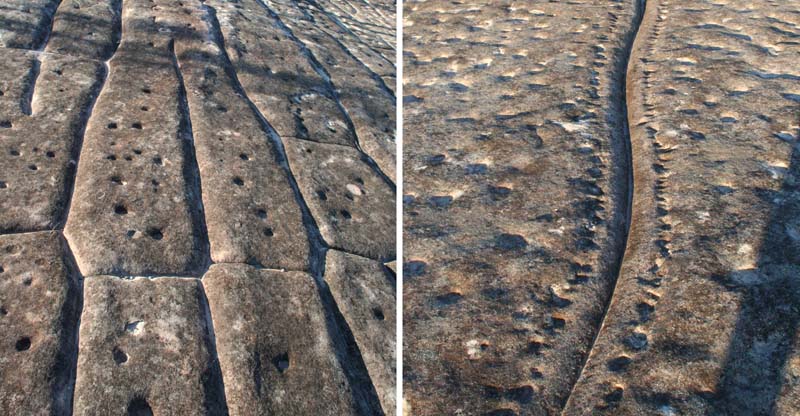}
\caption{(left) Cupules at the Elvina site which resemble constellations. It is unknown whether these are intentional, as argued by  \cite{cairns05} or simply chance resemblances. (right) A line of cupules from the Elvina site, demonstrating that at least some of the cupules are man-made.}
\label{elvina}
\end{figure}

Even more speculative is the hypothesis \citep{cairns92, cairns96, cairns05} that the thousands of small cupules, each a few cm deep, decorating the rock surface near the Emu engraving at the Elvina site may represent constellations. Certainly some of them are similar to the shapes of constellations (see Figure \ref{elvina}, left) but there are so many  cupules that some constellation-shapes are likely to happen by chance. It is not even clear whether the cupules are man-made or natural. Despite  arguments that they are natural  \citep{bednarik08}, there is strong evidence that at least some are man-made  \citep{branagan93b}. For example, some are on a steep slope where water cannot collect, and some  are clearly arranged in man-made lines (see Figure \ref{elvina}, right). Unfortunately, to test Cairns' hypothesis, perhaps by surveying the cupules and testing whether there are more recognisable constellations than expected by chance, would be an enormous undertaking.

\subsection{Other Engraving Sites}
\label{ngaut}

\begin{figure}[hbt]
\includegraphics[width=8cm]{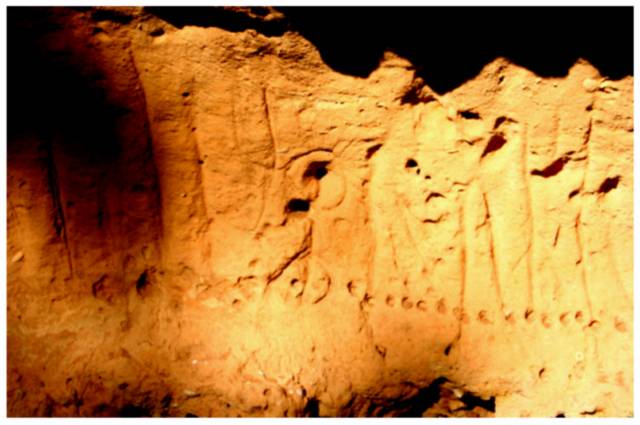}
\caption{The carvings at Ngaut-Ngaut, SA, said to represent lunar cycles  \citep{a111}.}.
\label{ngautpic}
\end{figure}

At a site called `Ngaut Ngaut', close to the Murray River, north of Adelaide, the Nganguraku people engraved images of the Sun (Fig.\ref{ngautpic}) and Moon \citep {curnow06}. Next to these images are a series of dots and lines carved in the rock, which according to traditional knowledge  \citep{n255,tindale86, ac8}, represent the `cycles of the Moon' (Fig. \ref{ngautpic}). The author has carefully recorded these, and searched for evidence of any cyclic pattern that may suggest lunar cycles, but none are apparent. Unfortunately, the traditional knowledge of the Nganguraku was suppressed by Christian missionaries, and without supporting ethnographic evidence, it is very difficult to `decode' such engravings.

Throughout Australia are many other engraving sites, such as the thousands of Panaramittee engravings in central Australia \citep{flood97} (including the claimed supernova engravings shown above in Figure \ref{sn}.) and thousands more in Western Australia  \citep[e.g.][]{bednarik77}. Amongst them are many crescent shapes, star-shapes, and other potential astronomical symbolism \citep{flood97},
but so far none have been carefully studied for potential astronomical significance. The Panaramittee engravings in Sturt Meadows and Mootwingee, NSW, contain many examples that may represent the sun, moon, and stars  \citep[e.g.][]{clarke09a}.

\begin{figure}[hbt]
\includegraphics[width=8cm]{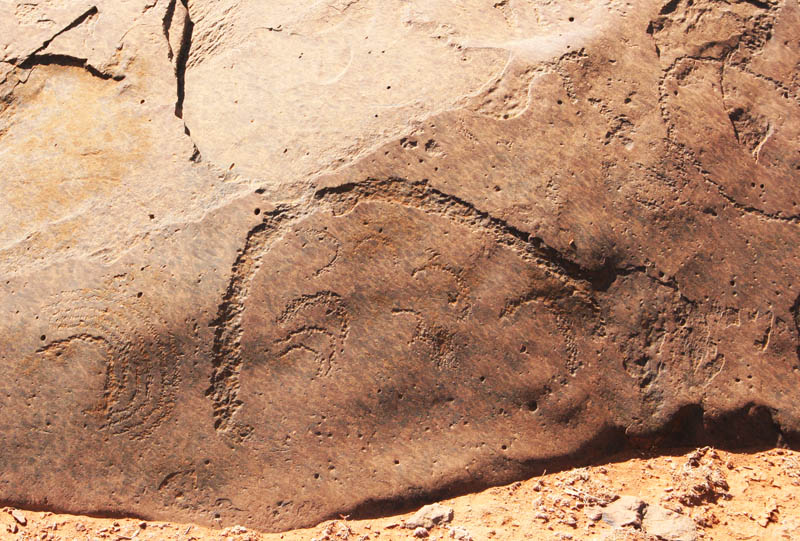}
\caption{Panaramittee engravings in Sturt Meadows, NSW, which resemble astronomical figures, but for which there is no supporting ethnographic evidence.}.
\label{sturt}
\end{figure}

\subsection{Rock paintings}

Fig. \ref{art} is obviously astronomical to the untutored eye, even if we don't understand its full significance. On the other hand, the astronomical rock painting from the Wardaman people shown in Fig.\ref{skyboss} is not obviously astronomical. Wardaman country has many such images  \citep[e.g.][]{lee09}, 
many of which have deep astronomical significance, although space precludes discussing them in detail here. It is instructive that, without the traditional knowledge of Bill Yidumduma Harney, it is very unlikely that any western researcher would understand that  paintings such as Fig.\ref{skyboss}  represent astronomical constellations.  Australia abounds with such paintings, only a few of which can be interpreted by non-Aboriginal people, and so it is important to realise that our studies represent only the tip of an iceberg.

Many other rock art sites in Australia, too numerous to describe here, show astronomical designs, or putative astronomical designs.
For example, rock engravings consisting of a small circle with lines radiating outward, suggesting a sun, star, or ÒsunburstÓ motif, are  found in the Sydney basin  \citep{sim66}, although their meaning is unknown. Many  rock art sites in Kakadu 
 and the Kimberleys also appear to be astronomical, although current resources for Aboriginal astronomy research in Australia are insufficient to do them justice.


\section{Conclusion}

 Ten years ago, most published knowledge of Aboriginal astronomy referred to songs and stories, but not to practical applications, navigation, or understanding the motion of celestial objects. Very few peer-reviewed papers had been written, so that Aboriginal astronomy received little attention in mainstream anthropology.

The published literature now shows unequivocally that traditional Aboriginal Australians were careful observers of the night sky, they had a deep knowledge of the sky, and their celestial knowledge played a major role in the culture, social structure, and oral traditions. Their knowledge went far beyond simply telling stories of the constellations. Instead, they appear to have been engaged in building  self-consistent, but culturally appropriate,  models that explained the phenomena observed in the sky. This search for understanding resembles modern-day science, and so is sometimes labelled `ethnoscience'.

This knowledge also had practical applications, such as marking time and date, over scales ranging from hours to years, so that it could be used to communicate the dates of ceremonies, or mark when it was time to move camp to take advantage  of seasonal food sources. Some Aboriginal groups also used it for navigation.

\subsection{Stone Arrangements}

There is unambiguous evidence that  astronomical knowledge played a role in the construction of some stone arrangements. Many stone arrangements were aligned by the builders to cardinal directions with a precision  that implies that astronomical observations were used to determine their orientation. At least one stone arrangement is carefully aligned to astronomically significant directions on the horizon at which the solstitial and equinoctial Sun sets.  Monte Carlo tests have shown that   the probability of these distributions occurring by chance is extremely small, confirming that these alignments were deliberately chosen by the builders of the stone arrangements. We may deduce from this that traditional Aboriginal people had  a significant knowledge of astronomy, including a good understanding of the motion of the Sun, and made careful measurements to determine cardinal directions. This does not imply that the stone arrangements are ``astronomical observatories'', since  the primary function  of the stone arrangements is unknown.  Stone arrangements tend to be under-represented in the literature, and evidence suggests that there are many stone arrangements that have not yet been recorded or classified. It is important that they be examined and classified, and if appropriate  catalogued and protected, so that we may use them to understand more of the pre-contact Aboriginal cultures. 

\subsection{Relationship between Sky and Earth}
Many groups regard the sky and the Earth as `parallel planes', with counterparts in the sky to places, animals, and people on the Earth. There is also a widespread belief that they used to be more intimately connected, and some groups believe that they used to be the same thing, or that they became inverted, so what used to be the Earth is now the sky. It is widely believed that spirits, and some people such as `clever men', are still able to move between the sky and the Earth

\subsection{Navigation}
Aboriginal people used a variety of techniques to navigate the length and breadth of Australia. These included
songlines, which  are effectively oral maps of the landscape, enabling the transmission of oral navigational skills in cultures that do not have a written language, 
They extend for large distances across Australia,  were used as trading routes, and some modern Australian highways follow their paths.
Songlines on the earth are sometimes mirrored by songlines in the sky, so that the sky can be used as mnemonic to remember a route, and the sky is also used directly as a navigational tool. Other navigational techniques include using the sky as a compass, or by noting particular stars or planets as signposts, or by following the path defined by the Milky Way or the ecliptic.

\subsection{Ethnoscience}
Ethnoscience is the attempt to describe and explain natural phenomena within an appropriate cultural context, resulting in predictive power and practical applications such as navigation, timekeeping, and tide prediction. Ethnoscience is similar in its goals to modern-day  science, but is based on the limited information available to traditional Aboriginal people, and framed within the host culture.  For example, it is remarkable that the Yolngu model of tides correctly predicted how the height and timing of the tide varies with the phase of the Moon and may be contrasted with Galileo's incorrect explanation which predicted only one tide each day, and was silent on how the tides varied with the phase of the moon

\subsection{Other Outcomes}
In addition to the growth in knowledge about Aboriginal astronomy, there have been three other outcomes from these studies.

First, Aboriginal astronomy is more accessible to the general public than some other aspects of Aboriginal culture, so that Aboriginal astronomy has become a cultural bridge, giving non-Aboriginal people a glimpse of the depth and complexity of Aboriginal cultures, and promoting better understanding between cultures.

Second, an unexpected and unplanned outcome, which has been dubbed `science by stealth', is that  Aboriginal astronomy activities show how science is linked to culture in traditional Aboriginal societies, and thus illuminate how science is linked closely to our own European culture. Even reviews in the Arts media 
have applauded this approach, suggesting that combining science with culture may be an effective way of bridging the Òtwo CulturesÓ.

Third, there has  been a growth of the use of Indigenous Astronomy in the classroom, both to teach Aboriginal culture, and to encourage engagement with science, particularly amongst Indigenous students.

\subsection{The Future}
Although Aboriginal astronomy was first reported over 150 years ago  \citep{stanbridge57},  only in the last few years has there been a concerted scholarly attempt, which I have attempted to summarise here, to study the breadth and richness of Aboriginal astronomy. 
Quite apart from its scholarly value, this research can  be a powerful tool in overcoming some of the lingering prejudice that continues to permeate Australian society. The research presented here is proving valuable in building greater understanding of the depth and complexity of Aboriginal cultures.

 My colleagues and I continue to find new (to us!) information and traditional knowledge, supporting the view that research in Aboriginal astronomy is in its infancy. Many tantalising lines of evidence suggest that far more awaits us, and there are several lines of enquiry that demand attention.
For example,
\begin{itemize}
\item Only one culture (the Wardaman) has had its knowledge recorded to the depth of \cite{DS}. Presumably dozens of similar books could in principle be written on other Aboriginal cultures that are still strong, although this is probably contingent on building unusually productive partnerships like that between Bill Yidumduma Harney and Hugh Cairns.
\item Some oral traditions discussed in this review, such as the Yolngu Morning star ceremony, are barely touched. I am aware that a far greater depth of public knowledge exists, but it would require signficant ethnography to document it. Many similar examples exist, and are ripe for research.
\item I am frequently contacted by Aboriginal people wishing to tell their story. I would love to accommodate them all, but time, and the number of researchers in this area, are limited. Far more could be done with more researchers. Perhaps an online wikipedia-type tool, with appropriate safeguards against abuse, might help. 
\item Aboriginal astronomy papers tend to be written by a relatively small band of researchers with specific interests in this area. We are sometimes  criticised for focussing on this area rather than discussing it in the wider context of Aboriginal ethnography. However, most of us are not qualified to do so. I would encourage anthropologists and ethnographers to enter this field.
\item Similarly, I look forward to collaborative research that embraces and synthesises the different academic cultures. For example, a collaborative effort is required to resolve the discord between the observational evidence for complex Aboriginal number systems, cited here, and the ideas promoted in the linguistic literature, which place less emphasis on evidence.
\item Aboriginal stone arrangements are under-represented in the archaeological and anthropological literature. It seems likely that  other astronomically-aligned arrangements like Wurdi Young exist, but significant fieldwork will be required to find them.
\item Astronomers enjoy a resource that signficantly contributes to the rapid pace of developments in their field: the Astrophysics Data System (http://adsabs.harvard.edu/ ) which contains links to web-accessible copies of virtually all astronomical academic papers. Aboriginal astronomy enjoys no such resource, and my astronomy colleagues will be shocked to hear that the research for this review involved searching dusty books on obscure shelves of actual libraries! While my colleagues and I have amassed an online collection of scanned papers, it falls far short of ADS. Construction of such a resource would significantly accelerate progress in this field.
\end{itemize}

Against this optimistic view, all Aboriginal cultures are under significant threat. The elders who possess ancient knowledge grow old and pass away, often taking their knowledge with them, and not enough bright young Aboriginal people are willing to adopt the traditional lifestyle, and forego other opportunities, to continue the tradition. Even where  the tradition is strong, better education and exposure to media mean that traditional knowledge is not static, but evolves under the influence of modern education and new-age ideas. It is unclear what the future holds, but I am inspired by the  Euahlayi  and Kamilaroi elders with whom we have collaborated, who are rebuilding their culture and language. I share their hope that the Euahlayi  and Kamilaroi language will one day be taught as a second language in NSW schools, perhaps leading to a similar revitalisation as that demonstrated by the teaching of traditional language and culture in other countries, such as Wales.  I look forward to future research conducted as a collaboration between researchers and Aboriginal communities, with the twin goals of promoting better understanding of the culture while protecting the sanctity of sacred knowledge, and of  strengthening and protecting traditional knowledge and culture, so that they may thrive for countless generations to come.

\section{Acknowledgements}

I acknowledge and pay my respects to the traditional owners and elders, both past and present, of all the language groups mentioned in this paper. I particularly thank my Aboriginal colleagues and friends for sharing their traditional knowledge with us. I thank my collaborators:  Reg Abrahams, Michael Anderson, Hugh Cairns, the late John Clegg, Paul Curnow, Kristina Everett, Bob Fuller, John Goldsmith, Duane Hamacher, Bill Yidumduma Harney, Ian Maclean, Adele Pring,  and Michelle Trudgett,  all of whom have  contributed to different aspects of the research described here.  I  thank David Lee for making his work available to me prior to publication, and the late John Morieson who  introduced me to Victorian Aboriginal astronomy, and Daniel Price for drawing my attention to John Morgan's book on William Buckley. I also thank Duane Hamacher, Bob Fuller, Philip Clarke, and an anonymous referee for  helpful comments on drafts of this paper.

All photos and images in this paper are by the author unless otherwise stated.


\end{document}